\documentclass[10pt,preprint,twocolumn]{aastex63}

\usepackage{graphicx}
\usepackage{longtable}
\usepackage{mdwlist}

\usepackage{natbib}
\usepackage{multirow}
\usepackage{upgreek}
\usepackage{wasysym}
\usepackage{txfonts}
\usepackage{gensymb}
\usepackage{morefloats}
\usepackage{threeparttable}
\usepackage{rotating}
\usepackage{graphicx}

\usepackage{lineno}
\shorttitle{Extended X-ray emission around FR\,IIs}
\shortauthors{Jimenez Gallardo et al.}

\begin{document}

   \title{Extended X-ray emission around FR\,II radio galaxies: \\ hotspots, lobes and galaxy clusters}
   
   \correspondingauthor{Ana Jimenez-Gallardo}
\email{ana.jimenezgallardo@edu.unito.it}

\author[0000-0003-4413-7722]{A. Jimenez-Gallardo}
\affiliation{Dipartimento di Fisica, Universit\`a degli Studi di Torino, via Pietro Giuria 1, I-10125 Torino, Italy}
\affiliation{Istituto Nazionale di Fisica Nucleare, Sezione di Torino, I-10125 Torino, Italy}
\affiliation{INAF-Osservatorio Astrofisico di Torino, via Osservatorio 20, 10025 Pino Torinese, Italy}

\author[0000-0002-1704-9850]{F. Massaro}
\affiliation{Dipartimento di Fisica, Universit\`a degli Studi di Torino, via Pietro Giuria 1, I-10125 Torino, Italy}
\affiliation{Istituto Nazionale di Fisica Nucleare, Sezione di Torino, I-10125 Torino, Italy}
\affiliation{INAF-Osservatorio Astrofisico di Torino, via Osservatorio 20, 10025 Pino Torinese, Italy}
\affiliation{Consorzio Interuniversitario per la Fisica Spaziale, via Pietro Giuria 1, I-10125 Torino, Italy}

\author[0000-0002-5646-2410]{A. Paggi}
\affiliation{Dipartimento di Fisica, Universit\`a degli Studi di Torino, via Pietro Giuria 1, I-10125 Torino, Italy}
\affiliation{Istituto Nazionale di Fisica Nucleare, Sezione di Torino, I-10125 Torino, Italy}
\affiliation{INAF-Osservatorio Astrofisico di Torino, via Osservatorio 20, 10025 Pino Torinese, Italy}

\author[0000-0003-3073-0605]{R. D'Abrusco}
\affiliation{Center for Astrophysics $|$ Harvard \& Smithsonian, 60 Garden Street, Cambridge, MA 02138, USA}

\author{M. A. Prieto}
\affiliation{Departamento de Astrof\'isica, Universidad de La Laguna, E-38206 La Laguna, Tenerife, Spain}
\affiliation{Instituto de Astrof\'isica de Canarias (IAC), E-38200 La Laguna, Tenerife, Spain}

\author[0000-0003-4413-7722]{H. A. Pe\~na-Herazo}
\affiliation{Dipartimento di Fisica, Universit\`a degli Studi di Torino, via Pietro Giuria 1, I-10125 Torino, Italy}
\affiliation{Istituto Nazionale di Fisica Nucleare, Sezione di Torino, I-10125 Torino, Italy}
\affiliation{Instituto Nacional de Astrof\'isica, \'Optica y Electr\'onica, Puebla, Mexico}

\author[0000-0001-6305-6931]{V. Berta}
\affiliation{Dipartimento di Fisica, Universit\`a degli Studi di Torino, via Pietro Giuria 1, I-10125 Torino, Italy}

\author[0000-0001-5742-5980]{F. Ricci}
\affiliation{Instituto de Astrof\'isica and Centro de Astroingenier\'ia, Facultad de F\'isica, Pontificia Universidad Cat\'olica de Chile, Casilla 306, Santiago 22, Chile}

\author[0000-0003-1619-3479]{C. Stuardi}
\affiliation{Dipartimento di Fisica e Astronomia, Universit\`a di Bologna, via Piero Gobetti 93/2, I-40129 Bologna, Italy}
\affiliation{Istituto di Radioastronomia, INAF, via Gobetti 101, 40129, Bologna, Italy}

\author[0000-0003-1809-2364]{B. J. Wilkes}
\affiliation{Center for Astrophysics $|$ Harvard \& Smithsonian, 60 Garden Street, Cambridge, MA 02138, USA}

\author[0000-0001-6421-054X]{C. P. O'Dea}
\affiliation{University of Manitoba,  Dept. of Physics and Astronomy, Winnipeg, MB R3T 2N2, Canada}
\affiliation{School of Physics \& Astronomy, Rochester Institute of Technology, 84 Lomb Memorial Dr., Rochester, NY 14623, USA}

\author{S. A. Baum}
\affiliation{University of Manitoba,  Dept. of Physics and Astronomy, Winnipeg, MB R3T 2N2, Canada}
\affiliation{Center for Imaging Science, Rochester Institute of Technology, 84 Lomb Memorial Dr., Rochester, NY 14623, USA}

\author[0000-0002-0765-0511]{R. P. Kraft}
\affiliation{Center for Astrophysics $|$ Harvard \& Smithsonian, 60 Garden Street, Cambridge, MA 02138, USA}

\author[0000-0002-9478-1682]{W. R. Forman}
\affiliation{Center for Astrophysics $|$ Harvard \& Smithsonian, 60 Garden Street, Cambridge, MA 02138, USA}

\author{C. Jones}
\affiliation{Center for Astrophysics $|$ Harvard \& Smithsonian, 60 Garden Street, Cambridge, MA 02138, USA}

\author[0000-0001-5649-938X]{B. Mingo}
\affiliation{School of Physical Sciences, The Open University, Walton Hall, Milton Keynes, MK7 6AA, UK}

\author[0000-0003-0995-5201]{E. Liuzzo}
\affiliation{Istituto di Radioastronomia, INAF, via Gobetti 101, 40129, Bologna, Italy}

\author[0000-0002-0690-0638]{B. Balmaverde}
\affiliation{INAF-Osservatorio Astrofisico di Torino, via Osservatorio 20, 10025 Pino Torinese, Italy}

\author[0000-0003-3684-4275]{A. Capetti}
\affiliation{INAF-Osservatorio Astrofisico di Torino, via Osservatorio 20, 10025 Pino Torinese, Italy}

\author[0000-0001-8382-3229]{V. Missaglia}
\affiliation{Dipartimento di Fisica, Universit\`a degli Studi di Torino, via Pietro Giuria 1, I-10125 Torino, Italy}
\affiliation{Istituto Nazionale di Fisica Nucleare, Sezione di Torino, I-10125 Torino, Italy}
\affiliation{INAF-Osservatorio Astrofisico di Torino, via Osservatorio 20, 10025 Pino Torinese, Italy}

\author[0000-0003-4223-1117]{M. J. Hardcastle}
\affiliation{Centre for Astrophysics Research, School of Physics, Astronomy and Mathematics, University of Hertfordshire, College Lane, Hatfield AL10 9AB, UK}

\author[0000-0002-1824-0411]{R. D. Baldi}
\affiliation{Dipartimento di Fisica, Universit\`a degli Studi di Torino, via Pietro Giuria 1, I-10125 Torino, Italy}
\affiliation{Istituto di Radioastronomia, INAF, via Gobetti 101, 40129, Bologna, Italy}
\affiliation{Department of Physics \& Astronomy, University of Southampton, Hampshire SO17 1BJ, Southampton, UK}
\affiliation{INAF - Istituto di Astrofisica e Planetologia Spaziali, via Fosso del Cavaliere 100, I-00133 Roma, Italy}

\author[0000-0002-1824-0411]{L. K. Morabito}
\affiliation{Centre for Extragalactic Astronomy, Department of Physics, Durham University, Durham, DH1 3LE, UK}

   \date{\today}
   \begin{abstract}
   
We present a systematic analysis of the extended X-ray emission discovered around 35 FR\,II radio galaxies from the revised Third Cambridge catalog (3CR) $\textit{Chandra}$ Snapshot Survey with redshifts between 0.05 to 0.9.

We aimed to (i) test for the presence of extended X-ray emission around FR\,II radio galaxies, (ii) investigate if the extended emission origin is due to Inverse Compton scattering of seed photons arising from the Cosmic Microwave Background (IC/CMB) or to thermal emission from an intracluster medium (ICM) and (iii) test the impact of this extended emission on hotspot detection.

We investigated the nature of the extended X-ray emission by studying its morphology and compared our results with low-frequency radio observations (i.e., $\sim$150 MHz), in the TGSS and LOFAR archives, as well as with optical images from Pan-STARRS. In addition, we optimized a search for X-ray counterparts of hotspots in 3CR FR\,II radio galaxies.

We found statistically significant extended emission ($>$3$\sigma$ confidence level) along the radio axis for $\sim$90\%, and in the perpendicular direction for $\sim$60\% of our sample. We confirmed the detection of 7 hotspots in the 0.5 - 3 keV.

In the cases where the emission in the direction perpendicular to the radio axis is comparable to that along the radio axis, we suggest that the underlying radiative process is thermal emission from ICM. Otherwise, the dominant radiative process is likely non-thermal IC/CMB emission from lobes. We found that non-thermal IC/CMB is the dominant process in $\sim$70\% of the sources in our sample, while thermal emission from the ICM dominates in $\sim$15\% of them.
   \end{abstract}

\section{Introduction}

The Third Cambridge Radio Catalog (3C) is an astronomical catalog of radio sources detected originally at 159 MHz and published in 1959 by \citet{Edge1959}. In 1962, Bennett created the first revised version of the catalog (3CR), using observations at 178 MHz. The 3CR catalog was considered as the definitive listing of the brightest radio sources in the Northern Hemisphere at flux density higher than 9 Jy and declinations above 5\degr\ until Laing, Riley and Longair (1983) published a further revision in 1983, called 3CRR. This last revision includes galaxies which were not detected in the original catalogue due to shortcomings of the original observations, but which otherwise meet the flux and declination limits. The 3CRR catalogue includes all extragalactic radio sources on the Northern Hemisphere with 178 MHz flux density greater than 10.9 Jy lying at declination greater than 10\degr\ and Galactic latitude $|b|>$10 \degr. This is formally a complete sample of radio galaxies and radio loud quasars observed at low radio frequencies and is among the most studied catalogs of radio-loud active galactic nuclei (AGNs). An additional revision of the catalog created by \citet{Bennet62} was carried out by Spinrad et al. in 1985, including new optical identifications of the sources in the 3CR catalog.

The 3C catalog and its revised versions have proven to be excellent surveys to understand the properties of radio-loud sources (e.g., \citealt{Schmidt1963}, \citealt{Shields1999}, \citealt{Tadhunter2016}), mainly due to their wide radio frequency coverage (see e.g., \citealt{Hardcastle2000}, \citealt{Giovannini2005}, \citealt{Madrid2006}, \citealt{Privon2008}, \citealt{Hilbert2016}, \citealt{Kotyla2016} and \citealt{Balmaverde2019}). In addition, almost all 298 sources listed in the 3CR have been observed in the infrared (see e.g., \citealt{Baldi2010}; \citealt{Werner2012} and \citealt{Dicken2014}), optical (see e.g., \citealt{Hiltner1991}, \citealt{deKoff1996}, \citealt{Chiaberge2000}, \citealt{Buttiglione2009,Buttiglione2011}, \citealt{Tremblay2009} and \citealt{Baldi2019}), and UV (\citealt{Allen2002} and \citealt{Baldi2008}) bands in the last decades. More recently, the 3CR catalog wavelength coverage was increased up to the X-rays (see e.g., \citealt{Prieto1996}, \citealt{Evans2006}, \citealt{Hardcastle2006}, \citealt{Balmaverde2012}, \citealt{Wilkes2013}, \citealt{Maselli2016} and \citealt{Kuraszkiewicz2020}).

However, despite this large suite of available multifrequency observations, before $Chandra$ Cycle 9 only $\sim60\%$ of the 3CR sources had observations with $Chandra$ and only $\sim$30\% were covered with $XMM-Newton$ (see e.g., \citealt{Massaro2018} for a recent summary). Therefore, in 2007, we began the 3CR $Chandra$ snapshot survey to complete the catalog X-ray coverage (\citealt{Massaro2010a}). 

Several results have been achieved to date thanks to this snapshot survey, including several X-ray follow up observations of interesting targets (such as 3CR\,171.1 by \citealt{Hardcastle2010} and \citealt{Balmaverde2012}, 3CR\,305 by \citealt{Hardcastle2012}, 3CR\,105 and 3CR\,445 by \citealt{Orienti2012} and 3CR\,227 and 3CR\,295 by \citealt{Migliori2020}). For instance, this survey led to the discovery of X-ray counterparts of radio jet knots and hotspots (see e.g., \citealt{Massaro2015} for 3CR archival observations) as well as diffuse X-ray emission around several radio sources (\citealt{Massaro2009a}, \citealt{Massaro2018} and \citealt{Stuardi2018}) due to the presence of a hot intracluster medium (ICM; see e.g., 3CR\,89, 3CR\,196.1 and 3CR\,320 by \citealt{Dasadia2016}, \citealt{Ricci2018} and \citealt{Vagshette2019}, respectively, and 3CR\,17 by \citealt{Massaro2009b} and \citealt{Madrid2018}, to name a few examples) or to Inverse Compton scattering (IC) of photons from the Cosmic Microwave Background (CMB) in their lobes (IC/CMB; see e.g., 3CR\,459 by \citealt{Maselli2018}). As defined by \citet{Leahy1993}\footnote{http://www.jb.man.ac.uk/atlas/anatomy.html}, we call ``hotspots'' those surface brightness peaks at the jet termination with no significant diffuse radio emission beyond them and detected above ten times the rms noise level at frequencies above $\sim$1 GHz. Hotspots are also relatively ``compact'' (i.e., enclosed in circular regions of $\sim$2-4 arcsec radius) and distinct from radio lobes (i.e., extended regions of diffuse radio emission, on scales of tens of kpc scales, whose perimeter is mostly well-defined and with intensity tending to zero towards their perimeter). In addition, several radio galaxies with FR\,II morphology (i.e., edge-brightened radio sources, as defined by \citealt{Fanaroff1974}), observed during the 3CR Snapshot Survey, show diffuse X-ray emission with no apparent radio counterpart at GHz frequencies (see e.g., \citealt{Massaro2013c} and \citealt{Stuardi2018}).

In the last two decades, examples of radio sources showing diffuse X-ray emission extending, well beyond the radio structure, were discovered at high redshifts (i.e., $1.7<z<3.8$) (e.g., 3CR\,294, HDF\,130, 3CR\,191, 3CR\,432, 4C\,60.07, 4C\,03.24 and 4C\,19.71 by \citealt{Fabian2003,Fabian2009}, \citealt{Erlund2006}, \citealt{Smail2009,Smail2012}, respectively). This X-ray emission could be due to either (i) the X-ray counterpart of extended steep-spectrum radio structures observable at MHz but not at GHz frequencies, or (ii) the ICM whenever the source lies in a group/cluster of galaxies, or (iii) a mixture of both processes. \citet{Croston2005} studied X-ray lobe emission of 33 FR II radio sources, finding a high rate of lobe detection (at least one per source in 20 sources of their sample) and concluding that the main X-ray emission mechanism from lobes is IC/CMB. Following this result, \citet{Ineson2013,Ineson2015,Ineson2017} differentiated between thermal emission from the ICM and non-thermal emission from IC/CMB in a sample of radio galaxies based on their radio morphology. Thus, they considered that X-ray emission spatially associated with lobes was due to IC/CMB, while the rest of the extended X-ray emission was thermal emission from the ICM. Additionally, using a complete subset of 2 Jy sources (at $0.05<z<0.2$), \citet{Mingo2017} found lobe emission in FR II radio galaxies consistent with IC/CMB emission, confirming previous results by \citet{Croston2005}.

Carrying out the 3CR $Chandra$ Snapshot Survey, extended X-ray emission was discovered around many of the 3CR sources ($\sim$50 sources out of 262 observed to date, \citealt{Massaro2013c}). However, previous analyses were mainly focused on searching for X-ray counterparts of hotspots rather than ICM signatures. Thus, here, we present a refined analysis of a selected sample of FR\,II radio galaxies observed in the last decade during the 3CR $\textit{Chandra}$ Snapshot Survey with the main goals of:

\begin{enumerate}
    \item testing for the presence of extended X-ray emission around FR II radio galaxies and comparing the structure of this emission with the radio morphology at GHz and at $\sim$150 MHz frequencies (for those having available MHz observations);
    
    \item investigating the origin of this extended emission (either as IC/CMB emission from lobes or as thermal emission from the ICM);
    
    \item verifying previous claims of X-ray detected hotspots and checking if they could be related to fluctuations of the $\textit{local}$ X-ray background by refining the background regions and energy range used to claim their detection.
\end{enumerate}

This last point arose since, although a significant number of X-ray counterparts were discovered at the locations of radio hotspots ($\sim$40; \citealt{Massaro2013c}), previous analyses of 3CR \textit{Chandra} Snapshot Survey observations (reported in \citealt{Massaro2010a, Massaro2012, Massaro2013c, Massaro2018} and \citealt{Stuardi2018}) did not take into account the possibility that detected hotspots could be fluctuations of the extended X-ray emission due to (i) radiation arising from radio lobes and/or to (ii) the presence of ICM in those cases where the radio galaxy belongs to a galaxy cluster. 

To tackle our goals, we selected a sample of radio galaxies with clear FR\,II radio morphology, observed as part of the 3CR \textit{Chandra} Snapshot Survey to create a uniformly observed sample of radio sources (see \S~\ref{sec:sample} for all details about the sample selection criteria).

We additionally computed surface brightness profiles of FR\,II radio galaxies with clearly detected extended X-ray emission along their radio axes and verified the presence of galaxy clusters by using archival optical observations to test for the presence of red sequences (i.e., a color-magnitude relation for galaxies belonging to the same galaxy cluster; see \citealt{Visvanathan1977}) in the fields of those galaxies with diffuse X-ray emission not spatially associated with the radio structure. Finally, we performed a comparison with low radio frequencies (i.e., $\sim$150 MHz) available in the Tata Institute of Fundamental Research (TIFR) Giant Metrewave Radio Telescope (GMRT) Sky Survey (TGSS\footnote{http://tgssadr.strw.leidenuniv.nl/doku.php}) and in the LOw-Frequency ARray (LOFAR\footnote{http://www.lofar.org}) Two-meter Sky Survey (LoTSS) archives. 

The paper is organized as follows. In \S~\ref{sec:sample} we present the sample selection. The $Chandra$ data reduction is described in \S~\ref{sec:reduction} while \S~\ref{sec:analysis} is dedicated to the data analysis. Results are described in \S~\ref{sec:results}, and a comparison with observations carried out at radio, infrared and optical frequencies is shown in \S~\ref{sec:multifreq}. \S~\ref{sec:summary} is devoted to our discussion and conclusions. Tables with properties for each image of all selected radio sources, as well as images for all sources are collected in Appendix \ref{app}. Finally, results from the serendipitous discovery of hotspots made with the Wide-field Infrared Survey Explorer (WISE\footnote{https://irsa.ipac.caltech.edu/Missions/wise.html}) are shown in Appendix \ref{app:wise}.

Unless otherwise stated we adopt cgs units for numerical results and we also assume a flat cosmology with $H_0=69.6$ km s$^{-1}$ Mpc$^{-1}$, $\Omega_{M}=0.286$ and $\Omega_{\Lambda}=0.714$ \citep{bennett14}. Spectral indices, $\alpha$, are defined by flux density, S$_{\nu}\propto\nu^{-\alpha}$. Optical magnitudes obtained from the Panoramic Survey Telescope and Rapid Response System (Pan-STARRS\footnote{https://catalogs.mast.stsci.edu/panstarrs/}; \citealt{Chambers2016}) catalog are in a photometric system close to AB magnitudes described in \citet{Tonry2012}, with uncertainties between $\sim10^{-1}$ to $10^{-3}$ magnitudes. The average background level of the X-ray images is $\sim0.04$ photons arcsec$^{-1}$ in the energy band of 0.5 - 3 keV.

\section{Sample selection}
\label{sec:sample}

We initially selected all radio galaxies with a classical FR\,II radio morphology (i.e., edge-brightened, \citealt{Fanaroff1974}) out of those listed in the 3CR catalog and observed during the 3CR $Chandra$ Snapshot Survey before Cycle 20. Radio sources belonging to the 3CRR sample were not included since a dedicated paper is already in preparation (\citealt{Wilkes2020}). Criteria to select FR\,II radio sources on the basis of their radio morphology are as adopted in \citet{Capetti2017}. Additionally, we expect the physical mechanisms that give rise to potential X-ray emission in FR IIs to be different from those of FR Is (i.e., edge-darkened), because of differences in their particle content (see \citealt{Croston2018}).

Our initial sample includes 72 sources. Then, we excluded (i) sources with angular size (measured as the angular separation of the radio position of both hotspots) smaller than 5 arcsec measured at GHz frequencies in radio maps obtained from the Very Large Array (VLA\footnote{http://archive.nrao.edu/nvas/}) and (ii) three sources, namely: 3CR\,187, 3CR\,196.1 and 3CR\,320 for which deeper investigations have been already published or are in preparation (see \citealp{Paggi2020}, \citealp{Ricci2018} and \citealp{Vagshette2019}, respectively). We chose to exclude sources with angular sizes below 5 arcsec since, at the redshift range of our sample, it corresponds to 5 - 15 kpc and, therefore, we exclude compact radio sources. In addition, $\sim$ 90\% of the \textit{Chandra} point-spread function (PSF) is enclosed in a circular region of 2 arcsec radius. Thus, selecting sources that extend beyond 5 arcsec allows us to detect extended X-ray emission, beyond the relatively bright unresolved core present in a large fraction of FR II radio galaxies (see e.g.  \citealt{Massaro2011b}).

Our final sample thus includes 35 FR\,II radio galaxies, 24 optically classified as High-Excitation Radio Galaxies (HERGs), 7 as Low-Excitation Radio Galaxies (LERGs), 2 as Broad-Line Radio Galaxies (BLRGs; see e.g., \citealt{Laing1994}, \citealt{Buttiglione2010}, \citealt{Baldi2019}, for works on optical classification of radio galaxies) and 2 with no clear optical identification (3CR\,103 and 3CR\,435B). In total, these FR\,II radio galaxies have 71 hotspots since 3CR\,133 shows a double hotspot. 

Table \ref{tab:sample} gathers information about all the galaxies in our sample including their coordinates, redshift, kiloparsec scale, angular size according to the radio morphology, and their optical classification, as well as information about the X-ray and radio observations used in the analysis.

\begin{table*}[]
   \centering
    \caption{Source parameters for the selected sample of FR\,II radio galaxies}
    \begin{tabular}{|l|r|r|r|r|r|l|r|r|r|}
  \hline
  \multicolumn{1}{|c|}{3CR} &
  \multicolumn{1}{c|}{R.A.} &
  \multicolumn{1}{c|}{Dec.} &
  \multicolumn{1}{c|}{$z$} &
  \multicolumn{1}{c|}{kpc scale} &
  \multicolumn{1}{c|}{Ang. size} &
  \multicolumn{1}{c|}{Opt.} &
  \multicolumn{1}{c|}{$\textit{Chandra}$} &
  \multicolumn{1}{c|}{Radio freq.}&
  \multicolumn{1}{c|}{Beam size}\\
  \multicolumn{1}{|c|}{Name} &
  \multicolumn{1}{c|}{(J2000)} &
  \multicolumn{1}{c|}{(J2000)} &
  \multicolumn{1}{c|}{} &
  \multicolumn{1}{c|}{(kpc/arcsec)} &
  \multicolumn{1}{c|}{(arcsec)} &
  \multicolumn{1}{c|}{class} &
  \multicolumn{1}{c|}{Obs. Id} &
  \multicolumn{1}{c|}{(GHz)}&
  \multicolumn{1}{c|}{(arcsec)}\\
  \hline

18   & 00 40 50.53 &  +10 03 26.65  & 0.188  &   3.167  &   55  &   BLRG    &  9293   &   1.4 &  1.70   \\    
44$^*$   & 01 31 21.65 &  +06 23 43.14  & 0.66   &   7.059  &   64  &   HERG    &  16048  &   --  &   --    \\ 
52   & 01 48 28.91 &  +53 32 28.04  & 0.29   &   4.389  &   55  &   HERG    &  9296   &   8.0 &  0.38   \\ 
54   & 01 55 30.26 &  +43 45 59.05  & 0.827  &   7.694  &   53  &   HERG    &  16049  &   8.0 &  0.40   \\  
63   & 02 20 54.30 &  -01 56 50.65  & 0.175  &   2.990  &   18  &   HERG    &  12722  &   8.0 &  3.36   \\ 
69   & 02 38 02.66 &  +59 11 50.50  & 0.458  &   5.878  &   47  &   HERG    &  18092  &   4.8 &  5.00   \\ 
103  & 04 08 03.22 &  +43 00 33.93  & 0.33   &   4.796  &   82  &           &  13874  &   1.4 &  3.27   \\ 
107  & 04 12 22.62 &  -00 59 32.69  & 0.785  &   7.556  &   15  &   HERG    &  16052  &   4.8 &  0.46   \\  
114  & 04 20 22.23 &  +17 53 56.86  & 0.815  &   7.655  &   53  &   LERG    &  16053  &   4.8 &  1.47   \\ 
133  & 05 02 58.47 &  +25 16 25.27  & 0.2775 &   4.254  &   12  &   HERG    &  9300   &   1.4 &  0.98   \\ 
135  & 05 14 08.36 &  +00 56 32.48  & 0.1273 &   2.294  &   121 &   HERG    &  9301   &   8.0 &  0.66   \\ 
165$^*$  & 06 43 07.40 &  +23 19 02.60  & 0.2957 &   4.449  &   76  &   LERG    &  9303   &   --  &   --    \\
166  & 06 45 24.10 &  +21 21 51.27  & 0.2449 &   3.883  &   39  &   LERG    &  12727  &   1.4 &  1.37   \\
169.1  & 06 51 14.83 &  +45 09 28.48  & 0.633  &   6.931  &   46  &   HERG    &  16056  &   8.0 &  0.40   \\ 
180  & 07 27 04.88 &  -02 04 30.34  & 0.22   &   3.581  &   107  &   HERG    &  12728 &   8.0 &  0.36   \\ 
197.1  & 08 21 33.60 &  +47 02 37.15  & 0.1282 &   2.307  &   15  &   HERG    &  9306   &   4.8 &  1.32   \\ 
198$^*$  & 08 22 33.58 &  +05 56 30.82  & 0.0814 &   1.545  &   283 &   HERG    &  12730  &   --  &  --     \\ 
223.1  & 09 41 24.02 &  +39 44 41.71  & 0.1075 &   1.979  &   78  &   HERG    &  9308   &   1.4 &  1.67   \\ 
268.2  & 12 00 58.73 &  +31 33 21.55  & 0.362  &   5.096  &   97  &   HERG    &  13876  &   4.8 &  0.46   \\ 
272$^*$  & 12 24 28.44 &  +42 06 36.51  & 0.944  &   8.005  &   56  &   HERG    &  16061  &   --  &   --    \\
287.1  & 13 32 53.27 &  +02 00 45.86  & 0.2156 &   3.526  &  112  &   HERG    &  9309   &   1.4 &  5.00   \\
293.1  & 13 54 40.52 &  +16 14 43.15  & 0.709  &   7.272  &   44  &   HERG    &  16066  &   4.8 &  0.42   \\
306.1  & 14 55 01.41 &  -04 20 59.94  & 0.441  &   5.751  &   90  &   HERG    &  13885  &   4.8 &  1.75   \\
313  & 15 11 00.04 &  +07 51 50.15  & 0.461  &   5.900  &   133  &   HERG    &  13886 &   8.0 &  2.43   \\
332  & 16 17 42.54 &  +32 22 34.39  & 0.151  &   2.649  &   69  &   HERG    &  9315   &   1.4 &  4.40   \\
357  & 17 28 20.11 &  +31 46 02.55  & 0.166  &   2.866  &   76  &   LERG    &  12738  &   4.8 &  1.92   \\
379.1  & 18 24 33.00 &  +74 20 58.87  & 0.256  &   4.013  &   76  &   HERG    &  12739  &   1.4 &  1.49   \\
403.1$^*$  & 19 52 30.44 &  -01 17 22.35  & 0.0554 &   1.083  &   107 &   LERG    &  12741  &   --  &   --    \\   
411  & 20 22 08.43 &  +10 01 11.38  & 0.467  &   5.943  &   26  &   HERG    &  13889  &   1.4 &  1.30   \\
430  & 21 18 19.10 &  +60 48 07.68  & 0.0555 &   1.086  &   85  &   LERG    &  12744  &   4.8 &  1.34   \\
434$^*$  & 21 23 16.24 &  +15 48 05.80  & 0.322  &   4.718  &   13  &   LERG    &  13878  &   --  &   --    \\
435A$^*$  & 21 29 05.45 &  +07 32 59.75  & 0.471  &   5.971  &  24  &   BLRG    &  13890  &   --  &   --    \\
435B  & 21 29 06.10 &  +07 32 54.80  &  0.865  &   7.804  &  44  &           &  13890  &   4.8&  1.76   \\
456$^*$  & 23 12 28.08 &  +09 19 26.39  & 0.233  &   3.741  &   7  &   HERG    &  12746   &   --  &   --    \\
458  & 23 12 52.08 &  +05 16 49.80  & 0.289  &   4.378  &   197 &   HERG    &  12747  &   4.8 &  2.33   \\    

\hline

    \end{tabular}
    
    Column description: (1) source name; (2) right ascension (R.A.); (3) declination (dec.); (4) redshift; (5) kiloparsec scale; (6) estimation of the radio emission angular size, measured as the angular separation between the hotspots radio position; (7) optical classification where HERG stands for High-Excitation Radio Galaxies, LERG stands for Low-Excitation Radio Galaxies and BLRG, for  Broad-Line Radio Galaxies ; (8) $Chandra$ observation identification number; (9) frequency of radio maps (all obtained from the VLA archive) used for the registration (see \S~\ref{sec:reduction}); (10) major axis beam size of the radio maps. Sources marked with an asterisk are those that could not be registered.
\label{tab:sample}
\end{table*}

\section{\textit{Chandra} data reduction}
\label{sec:reduction}

Data reduction for all selected observations was carried out following the same procedures described in the $Chandra$ Interactive Analysis of Observations (CIAO; \citealt{Fruscione2006}) threads\footnote{http://cxc.harvard.edu/ciao/threads/} and using CIAO v4.11 and $Chandra$ Calibration Database v4.8.2. Here, we report basic details and differences, specifically adopted to achieve our goals, with respect to our previous analyses (see e.g., \citealt{Massaro2010a} and \citealt{Massaro2011b} for additional information).

We binned images in energy, in the 0.5 - 3 keV energy range for all radio galaxies in our sample. We chose to restrict our analysis to the  0.5 - 3 keV energy range since extended X-ray radiation is emitted predominantly in the soft band. This could only affect our calculations for a source in a very dense environment. However, we do not expect that to be the case for sources in our sample because all sources were already inspected in the original data papers (see \citealt{Ineson2013,Ineson2015,Ineson2017} and \citealt{Croston2017}).

Astrometric registration between radio and X-ray images was performed by adopting the same procedure used by previous papers on the 3CR \textit{Chandra} Snapshot Survey (see \citealt{Massaro2010a}; \citealt{Massaro2012}; \citealt{Massaro2013c}; \citealt{Massaro2018}; \citealt{Stuardi2018}) and deviating from those results by $<$ 1 arcsec. Frequencies and beam sizes of radio maps, obtained from the VLA archive, used in each case, are reported in Table \ref{tab:sample}, where sources with no radio frequency, indicated in the table, are those for which no astrometric registration was possible (due to the lack of core detection in radio observations), namely, 3CR\,44, 3CR\,165, 3CR\,198, 3CR\,272, 3CR\,403.1, 3CR\,434, 3CR\,435A and 3CR\,456.

\begin{figure}
    \centering
\includegraphics[width=9.5cm]{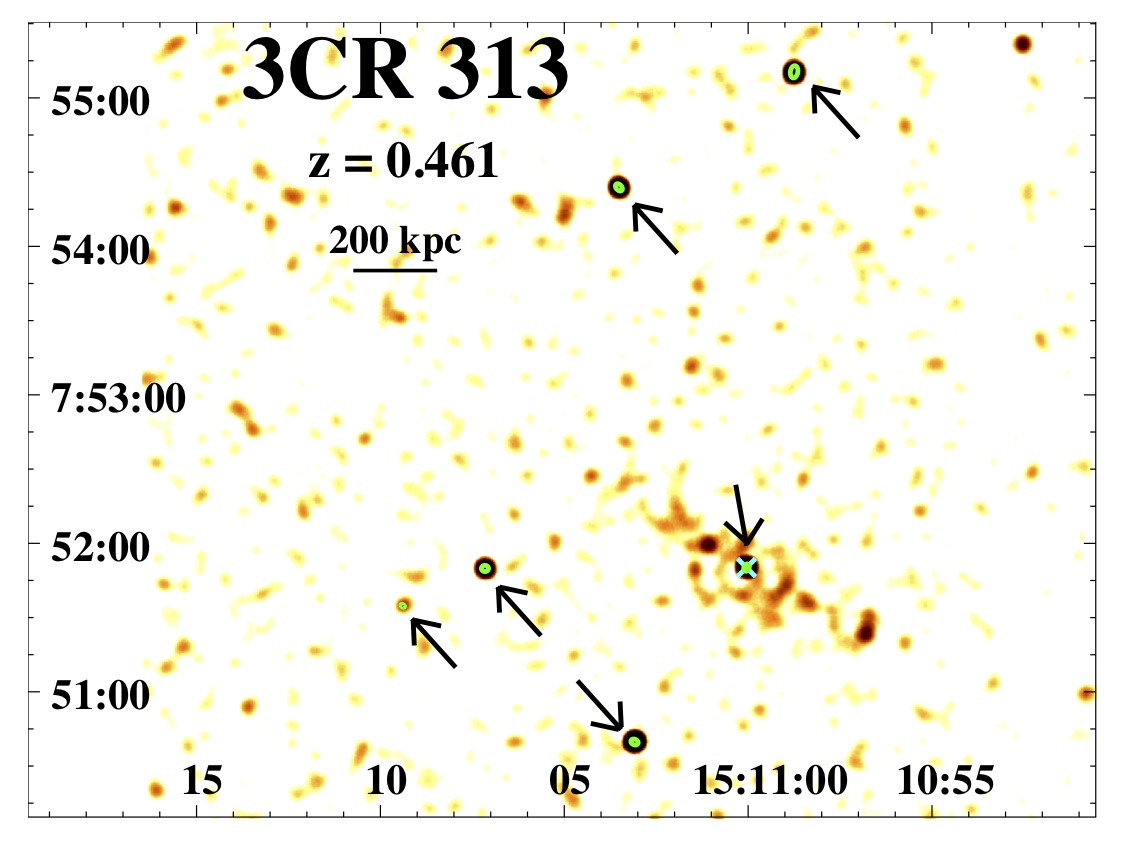}
\caption{0.5 - 3 keV $Chandra$ image of 3CR\,313. Point-like sources detected in the 3CR\,313 field using \textsc{wavedetect} are marked with black arrows. The image has been smoothed using a Gaussian kernel of 4.92 arcsec. The position of the nucleus is marked with a cross. Point sources marked are removed during the point source subtraction to obtain the images in Fig. \ref{fig:sample}. This image shows how effective \textsc{wavdetect} is at detecting point sources with the parameters we selected.}
\label{fig:pointsource}
\end{figure}

In Fig. \ref{fig:sample}, we show $Chandra$ 0.5 - 3 keV images for all sources in our sample, with different pixel sizes and smoothed with different Gaussian kernels, as reported in Table \ref{tab:parameters}. Overlaid contours correspond to the radio emission at different frequencies, as reported in Table \ref{tab:parameters}. Radio maps were obtained from the VLA archive. We removed point-like sources from these images (including the X-ray nuclei of the radio galaxies but not their hotspots) to highlight the presence of X-ray extended emission, although we only used the point-source subtracted images for visualization purposes (see \ref{fig:sample}) and to create X-ray surface brightness profiles (see \S~\ref{sec:analysis}). The procedure adopted is reported in the following. We detected point-like sources in the range 0.5 - 7 keV using the \textsc{wavdetect} task, available in CIAO, with a sequence of $\sqrt{2}$ wavelet scales, from 1 to 16 to cover different size sources, and a false-positive probability threshold set to the value of $10^{-6}$, which is the value recommended for a 1024 $\times$ 1024 image in the CIAO threads\footnote{https://cxc.harvard.edu/ciao/threads/wavdetect/} to make sure we do not over-subtract point sources. Next, we generated corresponding elliptical regions using the \textsc{roi} task and with the \textsc{dmfilth} task we built the final point-like source subtracted images. \textsc{dmfilth} replaces counts in each region where a point-like source is detected, defined by \textsc{roi}, by sampling the Poisson distribution of the pixel values in concentric background regions. As an example, in Fig. \ref{fig:pointsource} we show the field of 3CR\,313 marking the point-like sources detected by \textsc{wavdetect}. 

Since some of the background and foreground X-ray sources could be active galaxies (see e.g., \citealt{Horst2008}, \citealt{Eckart2010}, \citealt{Massaro2011a} and \citealt{Assef2013}), we tested if the X-ray sources detected with {\it wavdetect} have a mid-infrared counterpart in the recent release of the Wide-field Infrared Survey Explorer (\textit{WISE}) catalog (i.e., \textit{AllWISE}\footnote{http://wise2.ipac.caltech.edu/docs/release/allwise/}). Within a circle of radius of 30 arcsec centered in the sources, in our sample, all point sources from the \textit{AllWISE} catalog with X-ray counterparts were detected by \textsc{wavdetect}. Therefore, this suggests that the thresholds chosen to run 
\textsc{wavdetect} are reliable.


\section{\textit{Chandra} data analysis}
\label{sec:analysis}

\subsection{Detection of hotspots and extended X-ray emission}

\begin{figure*}
    \centering
\includegraphics[width=8.5cm]{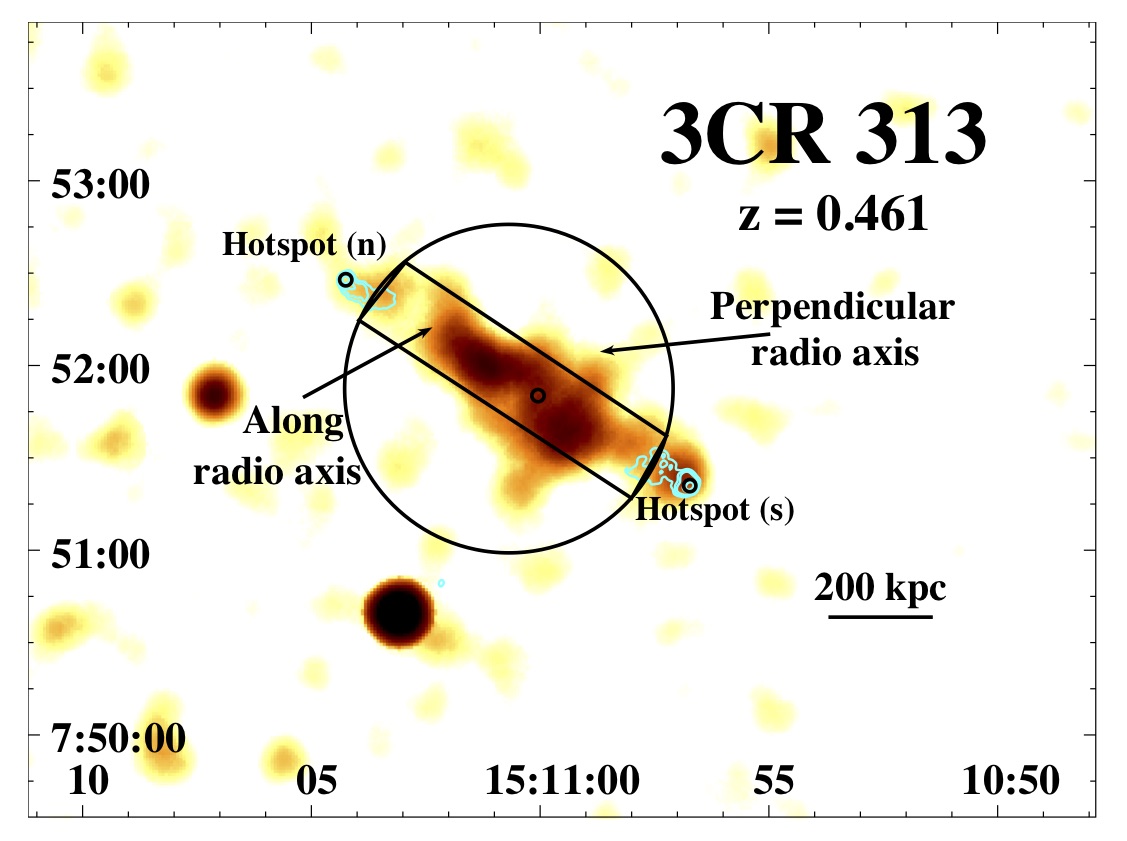}
\includegraphics[width=8.5cm]{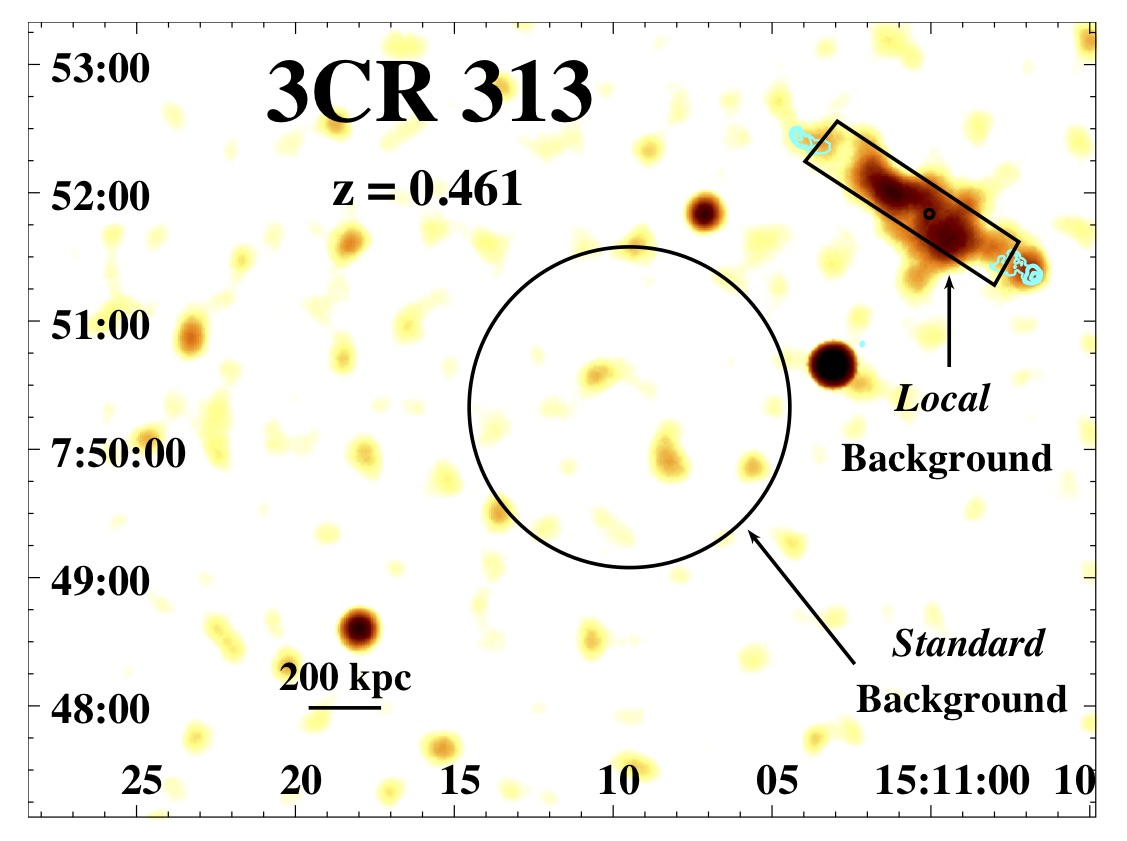}
\caption{Left panel: Example of regions chosen for the detection of hotspots, as well as extended X-ray emission along the radio axis and perpendicular to the radio axis. The 0.5 - 3 keV X-ray emission was smoothed with a 11.81 arcsec Gaussian kernel and with pixels of 0.984 arcsec. The 8 GHz radio emission is shown as blue contours whose levels are reported on Table \ref{tab:parameters}. Similar regions to the ones shown were chosen for the other sources. Right panel: Example of regions chosen as background for 3CR\,313 (in black) in the 0.5 - 3 keV band $Chandra$ observation. The X-ray emission and radio contours are the same as those in the left panel. These background regions (and similar ones for the other sources) were used while carrying out our extended emission and hotspot detection analysis.}
\label{fig:detreg}
\end{figure*}

Throughout this analysis, we used unbinned and unsmoothed images with photons restricted to the 0.5 - 3 keV band. To search for extended X-ray emission surrounding radio galaxies in our sample as well as the X-ray counterparts of radio hotspots, we considered three different regions (as shown in Fig. \ref{fig:detreg}):
\begin{enumerate}
\item hotspot regions: circular regions of 2 arcsec radius centered on the radio position of all hotspots and radio cores;
\item a rectangular region, defined along the radio axis on the basis of radio contours, excluding the regions corresponding to both the radio core and the hotspots;
\item a circular region, centered on the radio core position and extended as the previous one, but excluding the rectangular region along the radio axis, radio core and hotspots (circles of 2 arcsec radius each).
\end{enumerate}

The starting level of radio contours to select these regions was set to 5 times the root mean square of the radio map as reported in Table \ref{tab:parameters}, which is usually $\sim$1 mJy/beam.

X-ray detection significance for hotspots was estimated using the following procedure:

\begin{enumerate}
    \item we computed the background, choosing an appropriate region as described below;
    \item we identified a region for the X-ray hotspot (a 2 arcsec radius circular region centered at the location of the radio hotspot);
    \item we computed the number of photons in the hotspot region;
    \item then, assuming the number of photons in the background follows a Poissonian distribution with mean the number of photons that we measured for the background region, we compute the probability of detecting the observed number of photons in the hotspot region. Detection probability is indicated using the Gaussian equivalent $\sigma$.

\end{enumerate}
For each hotspot region selected, we assume two different background regions: (i) the {\it standard} X-ray background measured in a circular region as large as the region perpendicular to the radio axis and located on the same charge-coupled device (CCD) chip, far enough from the radio galaxy (i.e., at least a few tens of arcsec) to avoid the smearing of the PSF on CCD borders and contamination from the source, rescaled to the hotspot region size; and (ii) a so-called {\it local} background, defined as the region along the radio axis where IC/CMB could be present, also rescaled to the size of the hotspot region. In Table \ref{tab:dets}, we show the detection significance obtained for all features and in Table \ref{tab:counts}, we show the background-subtracted number of photons in each region, using the {\it standard} background.  

As an example, in the left panel of Fig. \ref{fig:detreg}, regions selected to carry out our analysis for 3CR\,313 are shown over its 0.5 - 3 keV X-ray emission. On the other hand, in the right panel of Fig. \ref{fig:detreg} we show the {\it standard} X-ray background compared to the {\it local} background, also for 3CR\,313.  

We chose to label the region along the radio axis as {\it local} background only when this region was used as background. We used the number of photons in the {\it standard} X-ray background to also compute the detection significance of the extended X-ray emission along the radio axis and perpendicular to the radio axis.

\subsection{X-ray surface brightness profiles}

We computed 0.5 - 3 keV, exposure corrected X-ray surface brightness profiles in two directions: (i) along the radio axis, and (ii) perpendicular to the radio axis. To estimate the background, we used blank-sky files available in the $Chandra$ Calibration Database. Final background event files were obtained by re-projecting blank-sky files to the same tangent planes as observations. Next, we renormalized exposure times of background event files, making observation count rates in the 9 - 12 keV band match background files count rates in the same band, where we expect emission to come mainly from particle background (same procedure adopted and described in \citealt{Hickox2006}).

\begin{figure}
    \centering
    \includegraphics[width=8.5cm]{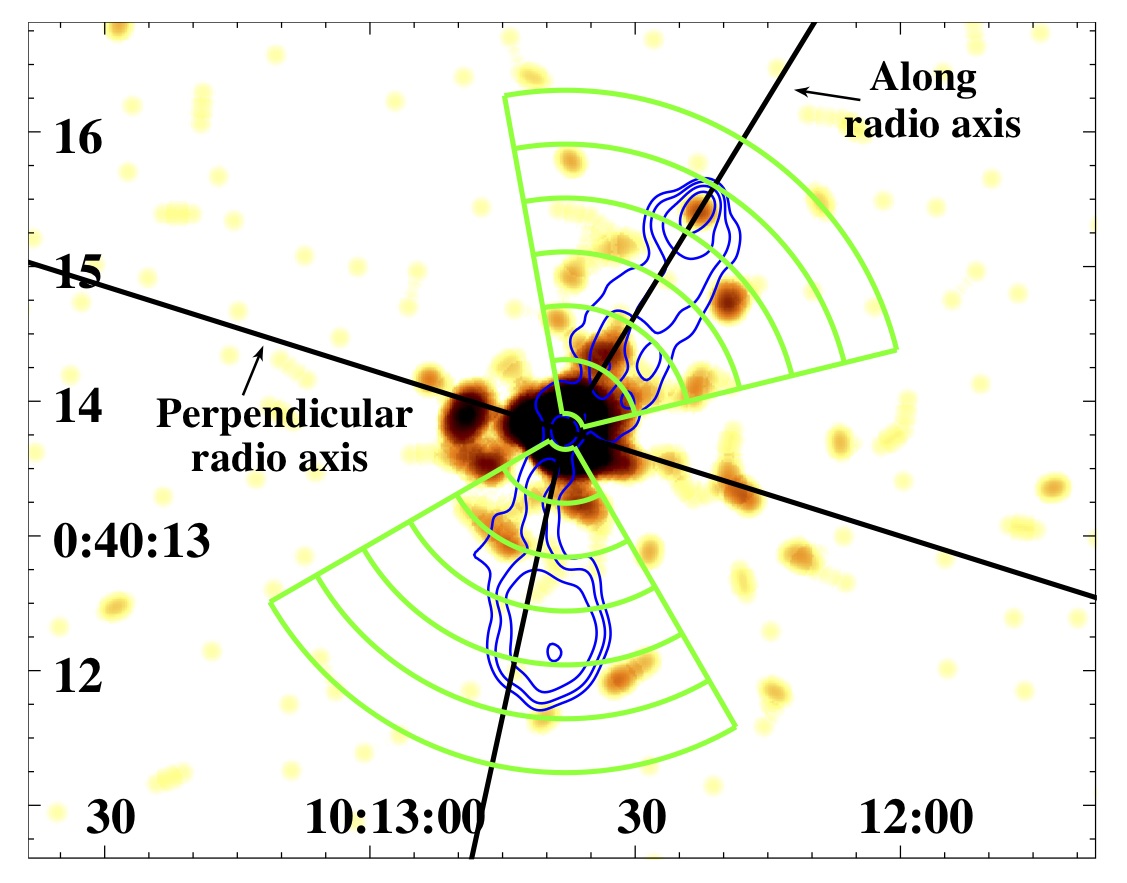}
\caption{Directions along and perpendicular to the radio axis chosen for 3CR\,18, as well as bins along the radio axis. The image shows the X-ray emission between 0.5 and 3 keV with 1.4 GHz VLA contours overlaid in blue. Parameters of the X-ray image and radio contour levels are given in Table \ref{tab:parameters}. These directions and bins were chosen based on the radio morphology and used to make surface brightness profiles.}
\label{conedirec}
\end{figure}

We built these profiles for the four sources in our sample with more than 100 photons along the radio axis, namely, 3CR\,18, 3CR\,198, 3CR\,287.1 and 3CR\,332. Fig. \ref{conedirec} shows an example for the bins chosen for 3CR\,18. We chose regions along the radio axis on the basis of the radio emission (as shown in Fig. \ref{conedirec}). Once these regions were chosen, we built radial bins with fixed width of 4 arcsec to 8 arcsec, depending on each source, up to the maximum extent of the radio contours. Beyond the radio structure, we extended the radial profiles until we reached a signal-to-noise ratio (SNR) of three was achieved or once a maximum radius (set to be inside the CCD chip) was reached.



\section{Results}
\label{sec:results}

\subsection{Detecting X-ray extended emission}

In comparison with the previous analyses carried out in the data papers of the 3CR $Chandra$ Snapshot Survey, we restricted our analysis to the energy range 0.5 - 3 keV because the extended X-ray emission, both coming from the thermal ICM or from IC/CMB, peaks in the soft band. We also chose to use the soft band to detect X-ray counterparts of hotspots since these observations were taken during \textit{Chandra} Cycles 9, 12 and 13, when the sensitivity of the soft band was not significantly degraded. Additionally, the hard band generally presents a higher background than the soft band. This is supported by the fact that during the XJET project, carried out by \citet{Massaro2011b}, only two hotspots out of the 32 detected in radio galaxies and quasars presented more counts in the hard band than in the soft band. In this work, the detection analysis was carried out using the full band (0.5 - 7 keV), observations had exposure times of less than 35 ks and the soft band was defined as 0.5 - 2 keV. However, we also checked the influence of carrying out this analysis using the full band.

Table \ref{tab:dets} shows the results of the detection analysis for regions along the radio axis and perpendicular to the radio axis and hotspots, where $\sigma_{\parallel}$ and $\sigma_{\bot}$ are the detection significance of the emission along and perpendicular to the radio axis and $\sigma_{i}$ and $\sigma^{loc}_{i}$ are the detection significance of hotspots against the \textit{standard} background and against the \textit{local} background, respectively. Sources and hotspots labelled with asterisks are those previously detected in the survey papers.

We detected extended X-ray emission (in the 0.5 - 3 keV range) along the radio axis against the \textit{standard} background, for 32 out of 35 radio galaxies (i.e., $\sim90\%$) in our sample above $3\sigma$ level of confidence, while this number decreases to 24 (i.e., $\sim70\%$) when considering a threshold of 5$\sigma$. 

On the other hand, we detected extended X-ray emission in the direction perpendicular to the radio axis, using the \textit{standard} background, in 22 radio galaxies out of 35 above 3$\sigma$ confidence level ($\sim60\%$), 15 ($\sim40\%$) when setting the level of confidence above 5$\sigma$. 

\begin{table}[]
   \begin{center}
  
    \caption{Detection significance for different radio components}
    \begin{tabular}{|l|c|c|c|c|c|c|c|c|}
  \hline
3CR Name & $\sigma_{\parallel}$ & $\sigma_{\bot}$& id$_{1}$ & $\sigma_{1}$ & $\sigma^{loc}_{1}$ & id$_{2}$ & $\sigma_{2}$ & $\sigma^{loc}_{2}$\\
\hline
   18      & $>5$ & $>5$ & n29     & 4    & --   & s26     & --   & --\\
   44      & $>5$ & 4   & n45$^{C}$     & 3    & --   & s19     & 3    & --\\ 
   52      & $>5$ & 5   & n29$^{*,C}$ & --   & --   & s26     & --   & -- \\
   54      & 4    & --   & n34     & --   & --   & s19$^*$ & 4    & 3 \\
   63      & $>5$ & $>5$   & n11     & --   & --   & s7      & $>5$ & --  \\
   69      & 5    & --   & n22$^{W}$     & --   & --   & s25     & 3    & --  \\
   103     & $>5$ & --   & n45     & --   & --   & s37     & --   & --  \\
   107$^\dagger$ & $>5$ & --   & n8      & --   & --   & s7      & --   & --  \\
   114     & 5    & 3     & n29     & --   & --   & s24     & --   & --  \\
   133     & $>5$ & $>5$ & w4      & --    & --   & e5$^{C}$      & $>5$    & -- \\
   135     & --   & --   & e77     & --   & --   & w43     & --   & --  \\
   165     & $>5$ & $>5$ & n29     & --   & --   & s48     & --   & --  \\
   166     & $>5$ & 5    & n15     & 4    & --   & s24$^{W}$     & --   & --  \\
   169.1     & 3    & --    & n25     & --   & --   & s22     & 3    & --  \\
   180     & 3    & --   & n43     & --   & --   & s63     & --   & --  \\
   197.1     & $>5$ & $>5$ & n6      & 3    & --   & s10     & 3    & --  \\
   198     & $>5$ & 4    & n107    & --   & --   & s176    & --   & --  \\
   223.1     & 3    & --   & n40     & --   & --   & s38     & --   & --  \\
   268.2     & 5    & --  & n42     & --   & --   & s55$^*$ & 3    & 3   \\
   272     & 4    & --  & n24     & 3    & --   & s32     & 3    & --  \\
   287.1     & $>5$ & 5     & w65$^{C}$     & 4   & --   & e47     & 3    & --  \\
   293.1     & 5    & 4    & n15     & --   & --   & s29     & --   & --  \\
   306.1     & $>5$ & --   & n44     & 3    & --   & s47$^{C}$     & --   & --  \\
   313 & $>5$ & 4    & n90$^*$ & 4    & 3    & s43$^*$ & 4    & 3   \\
   332     & $>5$ & 5    & n34$^{W}$     & 3    & --   & s34     & --   & --  \\
   357     & $>5$ & --   & w32     & --   & --   & e44     & --   & --  \\
   379.1     & --   & 3    & n44     & --   & --   & s32     & --   & --  \\
   403.1     & 3    & $>5$  & e35     & --   & --   & w72     & --   & --  \\
   411     & $>5$ & $>5$ & w13     & 3    & --   & e13     & --   & --  \\
   430     & --   & $>5$ & n44     & --   & --   & s41     & --   & --  \\
   434     & 3    & --   & w8      & --   & --   & e5      & 4    & 3   \\
   435A    & $>5$ & 4    &n9      & 3    & --   & s14     & 4    & --  \\
   435B    & $>5$ & $>5$ &e22     & 3    & --   & w22     & 4    & --  \\
   456     & 4    & $>5$ & n4      & $>5$ & 4    & s3      & 4    & --  \\
   458     & 5    & --   & e75$^*$ & 4    & 4    & w122    & --   & --  \\

\hline
    \end{tabular}
     \end{center}  
    Column description: (1) source name; (2) detection significance of the emission along the radio axis; (3) detection significance of the emission perpendicular to the radio axis; (4) first hotspot label; (5) detection significance of the first hotspot using the \textit{standard} X-ray background; (6) detection significance of the first hotspot using the \textit{local} background; (7) second hotspot label; (8) detection significance of the second hotspot using the \textit{standard} X-ray background; (9) detection significance of the second hotspot using the \textit{local} background.\\
    Notes:\\
    ($^\dagger$) Sources with extended X-ray emission previously reported in the survey papers.\\ 
    ($^*$) Hotspots detected in previous survey papers.\\
    ($^C$) Hotspots detected in the $Chandra$ Source Catalog.\\
    ($^W$) Hotspots detected in $WISE$.\\
    3CR\,133 presents a double hotspot, e6, detected only over the \textit{standard} background at 5$\sigma$ level of confidence.
\label{tab:dets}
\end{table}

\subsection{Detecting X-ray counterparts of hotspots}

Using the background estimated in the \textit{standard} region, we detected 30 hotspots out of 35 sources in our sample, all above the 3$\sigma$ confidence level and 4 of them above 5$\sigma$. If, instead, we consider the background in the \textit{local} region to guarantee that detected hotspots are not just due to background fluctuations of any diffuse emission, the number of detected hotspot decreases to 7 above the 3$\sigma$ confidence level ($\sim10$\% of the total number of hotspots). Therefore, $\sim$75\% of hotspots detected, without taking into account the \textit{local} background, could be fluctuations of the extended X-ray emission along the radio axis. Using the full band (0.5 - 7 keV) in our detection analysis, we detected ($>3\sigma$) only three hotspots out of those not detected using the soft band, namely s7 in 3CR\,107, s63 in 3CR\,180 and s47 in 3CR\,306.1; while the significance level of $\sim$80\% (i.e., 56) of the hotspots in our sample decreases when we consider the full band instead of the soft band, with the significance of 9 decreasing below 3 $\sigma$. 

Results of the detection analysis are shown in Table \ref{tab:dets}. We labelled hotspots using the same notation as in the XJET project\footnote{http://hea-www.harvard.edu/XJET/} (\citealt{Massaro2011b}): their orientation (north, n, south, s, east, e, or west, w) followed by their angular separation (in arcsec) from the core of the radio galaxy.

In Table \ref{tab:counts}, we show a summary of the sources with newly detected hotspots compared with previous works on the 3CR $Chandra$ Snapshot Survey. Previous works detected only 6 hotspots ($\sim$10 \%) for the same sources/observations analyzed here, while only five of them are detected in the second version of the $Chandra$ Source Catalog (CSC2\footnote{https://cxc.harvard.edu/csc/}; see \citealt{Evans2010} for the first version of the catalog and \citealt{Evans2019, Evans2020} for the second release). However, the detection of these hotspots in the CSC2 is only marginal (below a $3\sigma$ confidence level) and, therefore, they could be spurious detections instead of X-ray counterparts of the radio hotspots. Using the \textit{standard} background and restricting the energy to the soft band (in contrast with previous analyses, including that of CSC2, that used the full band), we detected new hotspots in 16 sources (see Table \ref{tab:counts}); while using the \textit{local} background, we detected new hotspots in 2 sources and confirmed hotspots from previous analyses in 4 sources (see Table \ref{tab:counts}), discarding only one hotspot whose detection was claimed in previous survey papers (i.e., n29 in 3CR\,52).

Photons in both hotspot regions as well as along the radio axis and in the region perpendicular to the radio axis (see \S~\ref{sec:analysis} for the definition of each region) and their corresponding background levels are reported in Table \ref{tab:counts}, together with an estimate of the ratio of photons along and perpendicular to the radio axis ($\rho$), defined as follows:

\begin{equation}
    \rho=\frac{N_{ph,\parallel}}{N_{ph,\bot}}\cdot\frac{A_{\bot}}{A_{\parallel}}
    \label{eq:ratio}
\end{equation}
where $N_{ph,\parallel}$ and $N_{ph,\bot}$ are the number of background-subtracted photons along the radio axis and in the region perpendicular to the radio axis, respectively, and $A_{\parallel}$ and $A_{\bot}$ are the areas of the regions along and perpendicular to the radio axis. This parameter gives us an estimate of the importance of the extended X-ray emission perpendicular to the radio axis with respect to the emission along the radio axis. In this table, we also compiled information about which sources present newly detected hotspots, which ones are in known optical galaxy clusters, and the dominant emission process responsible for the extended X-ray emission in each case, according to the detection significance of the extended emission along and perpendicular to the radio axis.

\subsection{Source details}
For a few targets, we found that the extended X-ray emission here discovered presents intriguing structures, namely 3CR\,103, 3CR\,268.2, 3CR\,403.1, 3CR\,430 and 3CR\,458, with no associated Planck galaxy clusters (see \citealt{Ade2016} and \citealt{Piffaretti2011}). These structures are discussed below.

\textbf{3CR\,103} ($z=0.33$). We did not detect significant emission perpendicular to the radio axis arising from this galaxy, but we detected extended emission at 93 arcsec (446 kpc at the source redshift) above $5\sigma$ confidence level to the south-west of the galaxy that could indicate the presence of a galaxy cluster. However, there is a bright star close to the position of the extended emission that  prevented us from obtaining precise redshift estimates of nearby optical sources, necessary to verify the presence of a galaxy cluster/group and deserving future follow up observations. 

\begin{figure*}
    \centering
    \includegraphics[width=8.5cm]{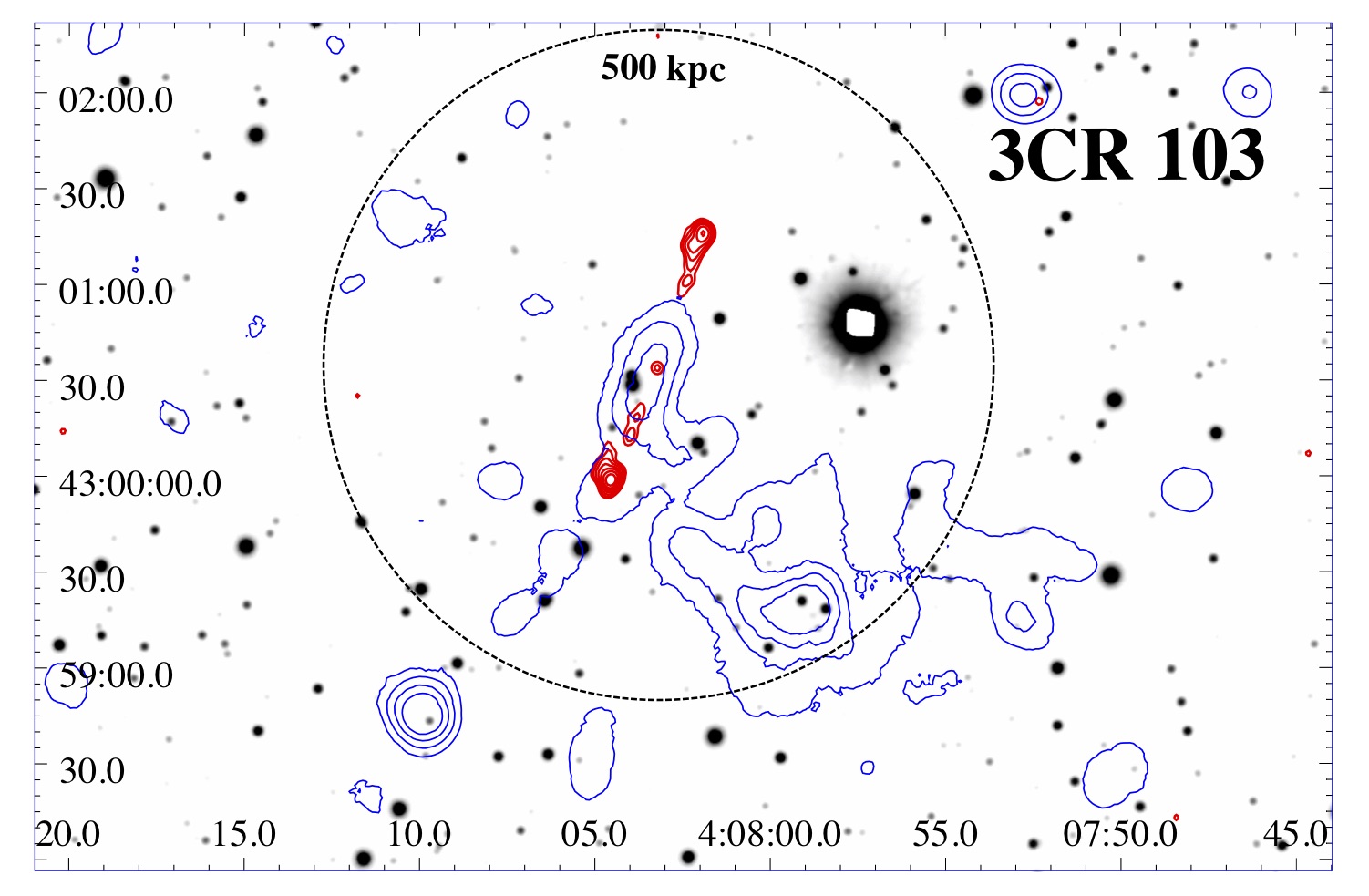}
    \includegraphics[width=8.5cm]{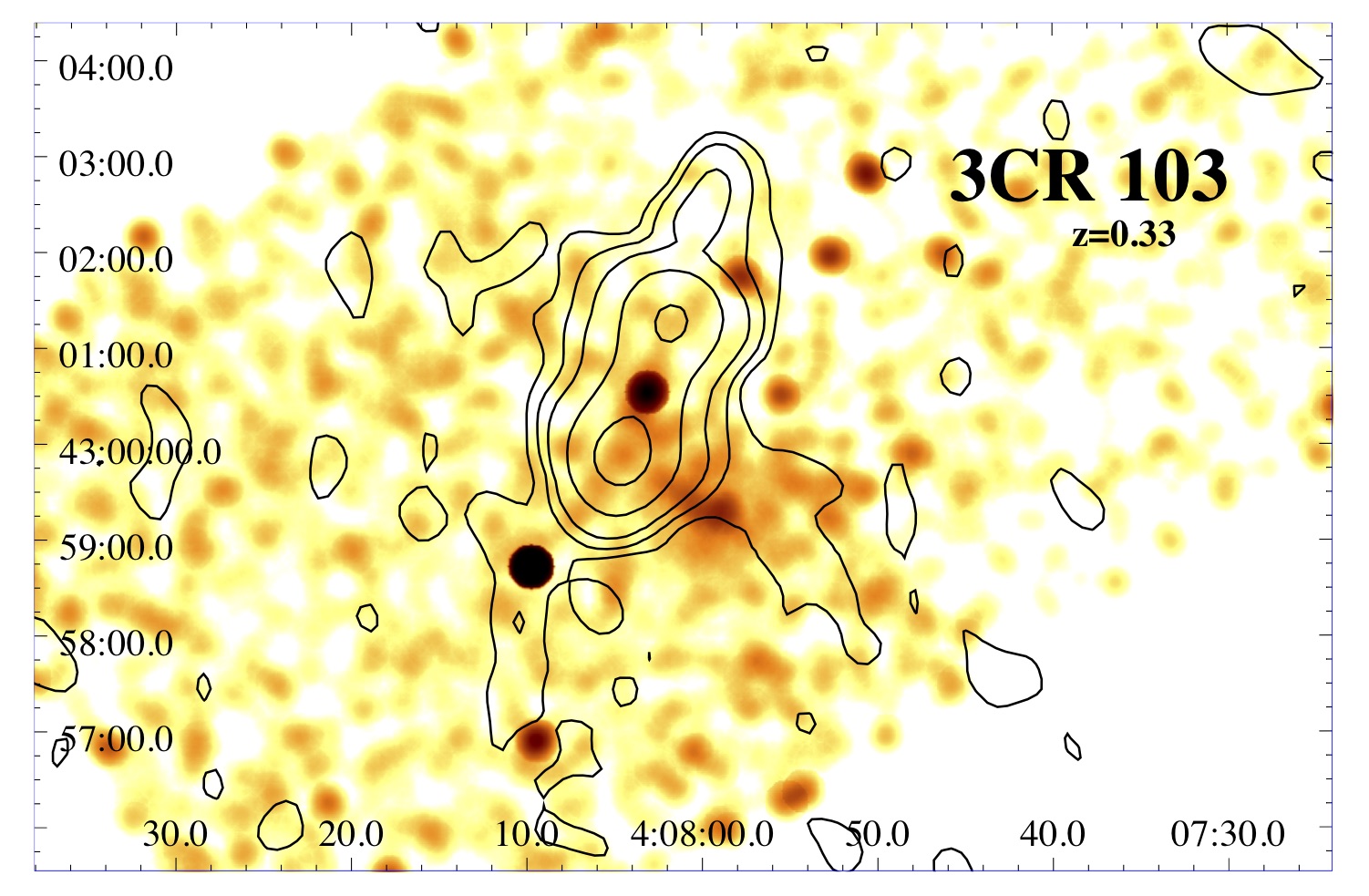}
\caption{Left: Pan-STARSS optical image in the $R$ band with VLA 1.4 GHz (in red) and $Chandra$ 0.5 - 3 keV (in blue) contours overlaid of 3CR\,103. Radio contours were drawn starting at 8 mJy/beam and increasing by a factor of 2. X-ray contours were drawn from the Chandra event file in the 0.5 - 3 keV energy range, smoothed with a 19.68 arcsec Gaussian kernel at levels of 0.06, 0.08, 0.12, 0.24, 0.48 and 1.20 photons and with pixels of 1.968 arcsec. The dashed black circle has a 500 kpc radius. The X-ray contours show diffuse X-ray emission at 446 kpc to the south-west of 3CR\,103 that could indicate that 3CR\,103 belongs to a galaxy cluster. Right: $Chandra$ image (exposure corrected) in the 0.5 - 3 energy range at center-band energy of 2.3 keV and smoothed with a Gaussian kernel of 13.78 arcsec with TGSS contours overlaid in black drawn at 0.025, 0.05, 0.125 and 0.75 Jy/beam. TGSS radio map has a beam size of 25 arcsec. The 150 MHz emission coincident with the extended X-ray emission hints at the present of a radio halo in a galaxy cluster.}
\label{fig:cluster1}
\end{figure*}

The left panel of Fig. \ref{fig:cluster1} shows an optical image from Pan-STARRS of the field of 3CR\,103 with radio contours (at 1.4 GHz) and X-ray contours (in the 0.5 - 3 keV band). The extended X-ray emission appears to be associated with the radio galaxy and seems to correspond to an overdensity of galaxies. Right panel of Fig. \ref{fig:cluster1} shows the 0.5 - 3 keV X-ray emission of 3CR\,103 with TGSS contours overlaid. Some radio emission can be seen matching the position of the extended X-ray emission, which could hint at the presence of a radio halo in a galaxy cluster (see e.g. \citealt{Giovannini1993}, \citealt{Burns1995}, \citealt{Feretti1997a,Feretti1997b}, \citealt{Giovannini1999}, \citealt{Govoni2001}, \citealt{Feretti2012} and \citealt{vanWeeren2019}).

\begin{figure}
    \centering
\includegraphics[width=9.cm]{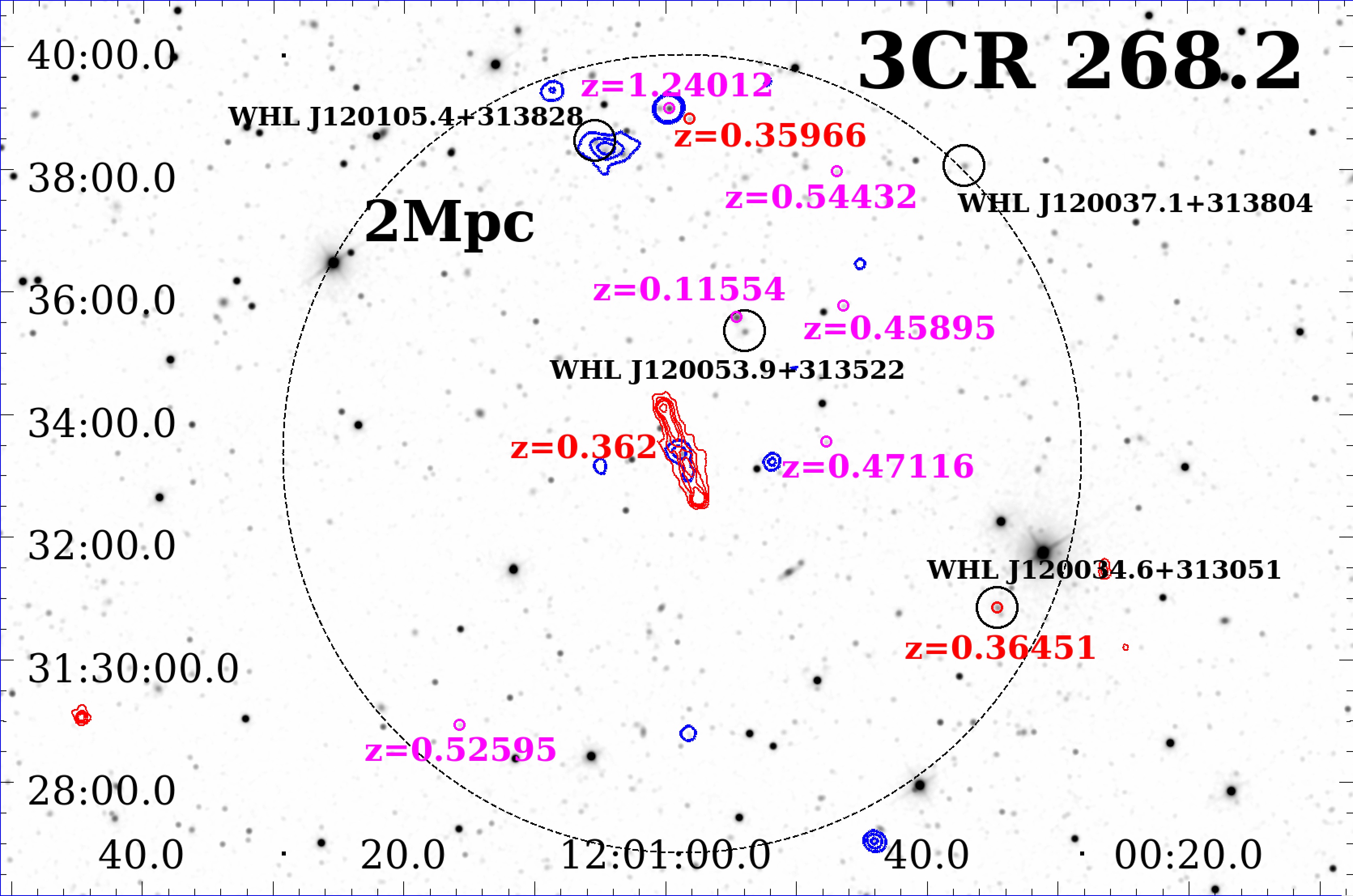}
\caption{The SDSS $R$ band image of the field around 3CR\,268.2 (\citealt{ahn12}). Radio contours of the LoTSS dataset are marked in red as in Figure \ref{fig:lofar} to indicate the location of 3CR\,268.2. Its redshift is also reported in red. The large dashed circle has 2\,Mpc radius computed at the central source redshift of 0.362 and centered on its position. Unrelated optical sources with spectroscopic redshifts are marked with magenta circles and two with redshift difference less than 0.003 with respect to 3CR\,268.2 are marked in red. Finally, black circles mark the location of the four candidate galaxy clusters found using photometric redshifts by \citet{Wen2012}. Blue X-ray contours are drawn from the {\it Chandra} event file with pixel size of 0.984 arcsec and smoothed with a Gaussian kernel of 13.78 arcsec at levels of 0.03, 0.05, and 0.07 photons. The candidate galaxy cluster WHL J120105.4+313828, at $z_{ph}=0.3082$, in the northern direction appears to have an extended X-ray counterpart. The presence of sources at similar redshifts, as the one of 3CR\,268.2, suggests the presence of a galaxy group surrounding it.}
\label{fig:3c268p2}
\end{figure}

\textbf{3CR\,268.2} ($z=0.362$). We found no significant emission perpendicular to the radio axis. Nevertheless, we detected extended X-ray emission 302 arcsec (1.5 Mpc at the source redshift) above 5$\sigma$ confidence level to the north-east of this source that could be due to the presence of an unrelated galaxy cluster (see Fig. \ref{fig:3c268p2}).
We explored the optical data available in the SDSS archive (see e.g., \citealt{ahn12}) and in Figure~\ref{fig:3c268p2} we show the $R$ band image of the field around 3CR\,268.2. There are 8 sources with spectroscopic redshift estimates, two of them, marked in blue in Figure~\ref{fig:3c268p2}, have a velocity dispersion of $\sim$725 km s$^{-1}$, thus corresponding to the typical velocity dispersion in groups and clusters of galaxies (see e.g., \citealt{moore93}, \citealt{eke04}, \citealt{berlind06} and \citealt{massaro19}). This suggests the possible presence of a galaxy group surrounding 3CR\,268.2. Moreover, \citet{Wen2012}, using the photometric redshifts of the SDSS database, also claimed the presence of four candidate galaxy clusters highlighted with black circles in the same figure, namely: WHL J120053.9+313522 at $z_{ph}=0.2993$, WHL J120105.4+313828 at $z_{ph}=0.3082$ (which coincides with the position of the extended X-ray emission at 302 arcsec from 3CR\,268.2), WHL J120034.6+313051 at $z_{ph}=0.3528$ (with a tentative association with a source having spectroscopic redshift 0.3645) and WHL J120037.1+313804 at $z_{ph}=0.3366$, two of them marginally consistent with that of 3CR\,268.2 lying at $z=0.362$.

\begin{figure}
    \centering
\includegraphics[width=9.cm]{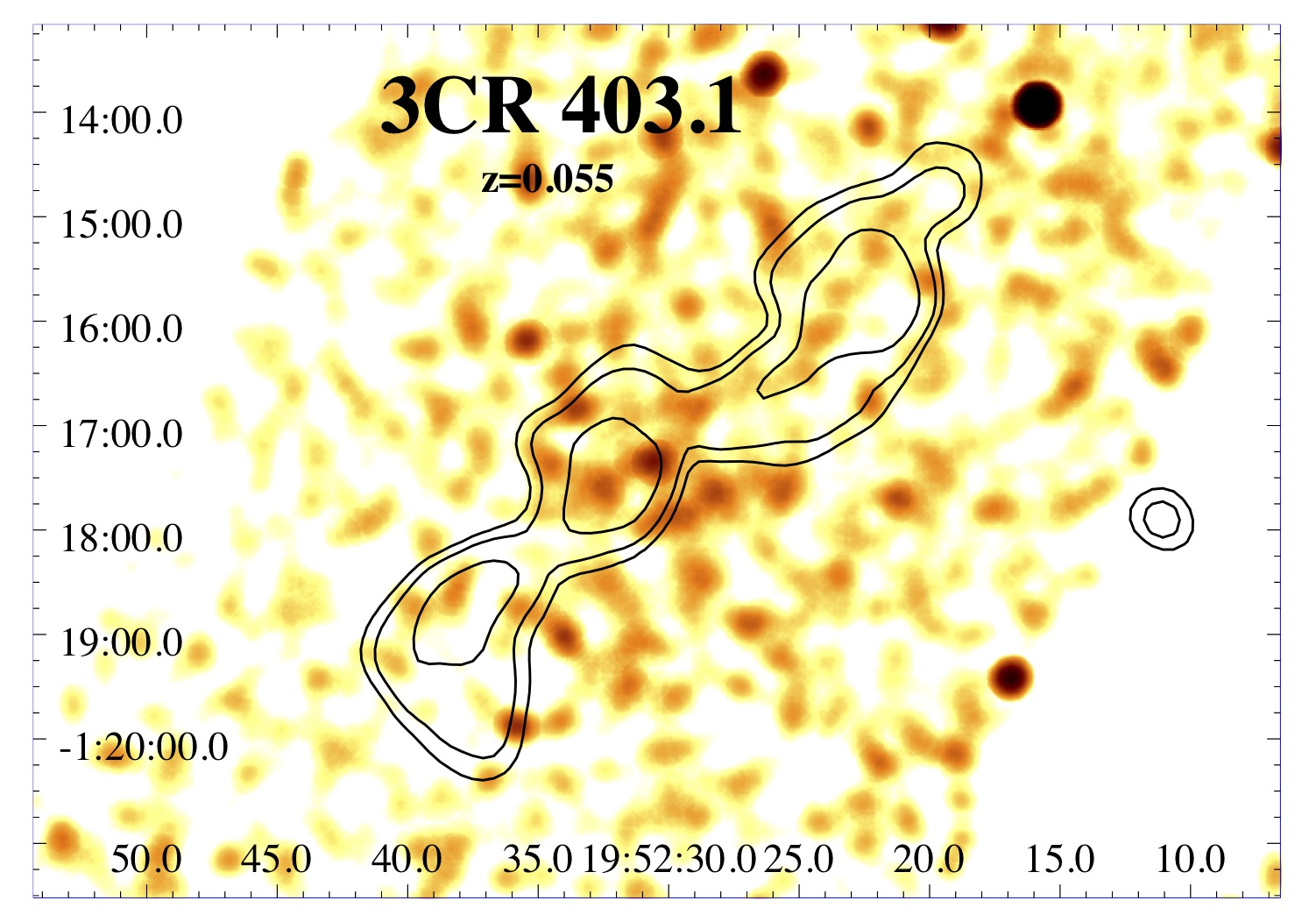}
\caption{$Chandra$ image (exposure corrected) in the 0.5 - 3 energy range and center-band energy of 2.3 keV with TGSS contours overlaid in black drawn at 0.025, 0.05, 0.125 and 0.75 Jy/beam. TGSS radio map has a beam size of 25 arcsec. The fact that the low-frequency radio emission is not co-spatial with the X-ray emission suggests the presence of a radio relic.}
\label{fig:3c403p1}
\end{figure}

\textbf{3CR\,403.1} ($z=0.0554$). We detected significant emission along and in the perpendicular direction to the radio axis. The extended X-ray emission is aligned with the GHz radio structure (see Fig. \ref{fig:sample}); however, the TGSS structure is perpendicular to the X-ray extended emission as shown in Fig. \ref{fig:3c403p1}. This could indicate the presence of a radio relic (see e.g., \citealt{Rottgering1994,Rottgering1997}, \citealt{Bagchi1998}, \citealt{Feretti1998}, \citealt{Giovannini1999}, \citealt{Feretti2012} and \citealt{vanWeeren2019}).

\textbf{3CR\,430} ($z=0.0555$). We detected significant emission perpendicular to the radio axis (above $5 \sigma$ level of confidence) that could be due to ICM from a galaxy cluster at the same position as the radio galaxy. We additionally detected extended X-ray emission at 118 arcsec (128 kpc at the source redshift) above 5$\sigma$ confidence level to the south-east of this source that could hint at the presence of another galaxy cluster, possibly undergoing a merging process with the former. Assuming this emission comes from ICM from a galaxy cluster, the brightest cluster galaxy (BCG) would be J211836.50+604704.2 whose redsift is $z = 0.05593 \pm 0.00026$, which supports the hypothesis of these two galaxy clusters undergoing a merging process.

\begin{figure*}
    \centering
\includegraphics[width=9.cm]{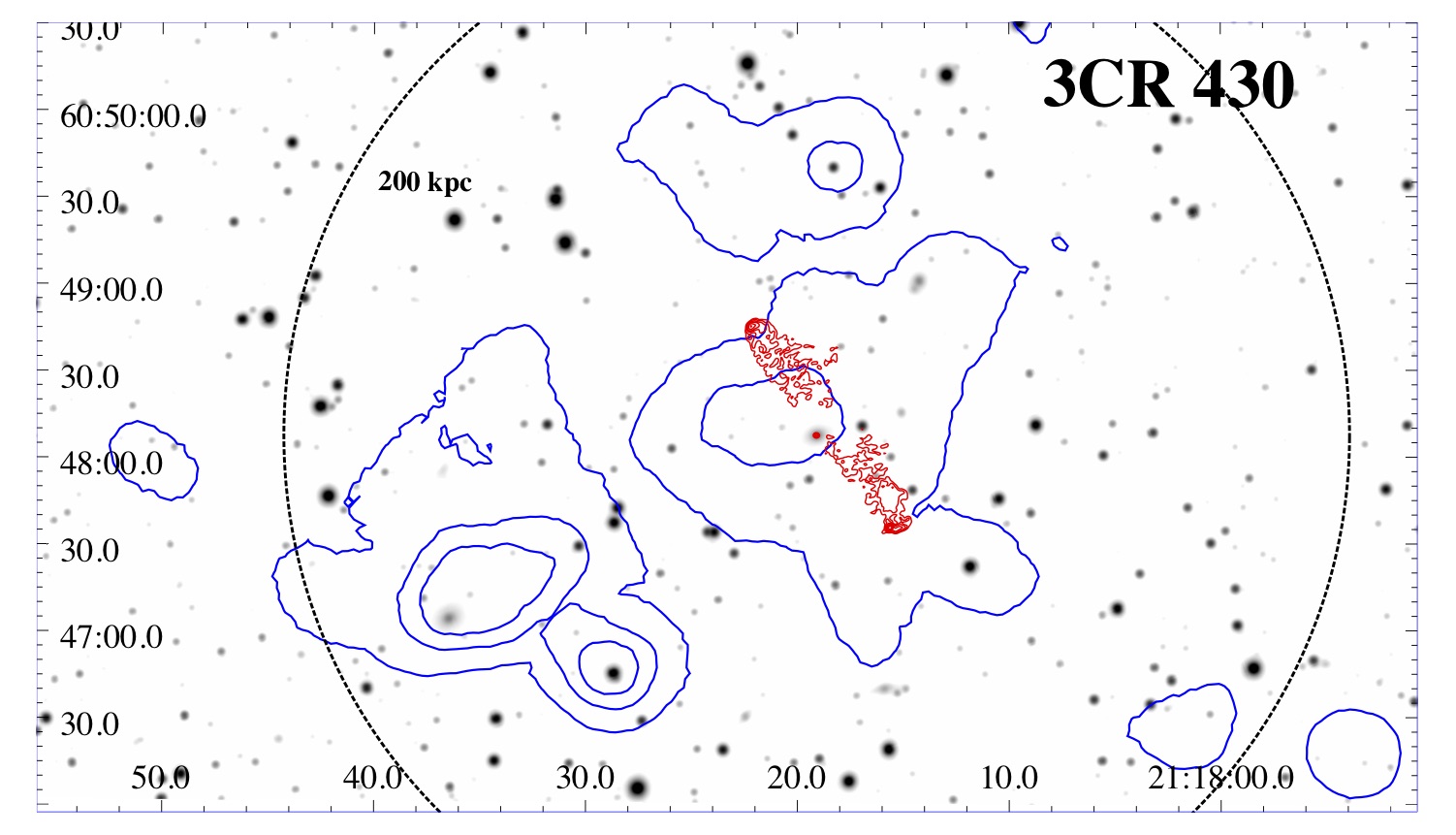}
\includegraphics[width=8.cm]{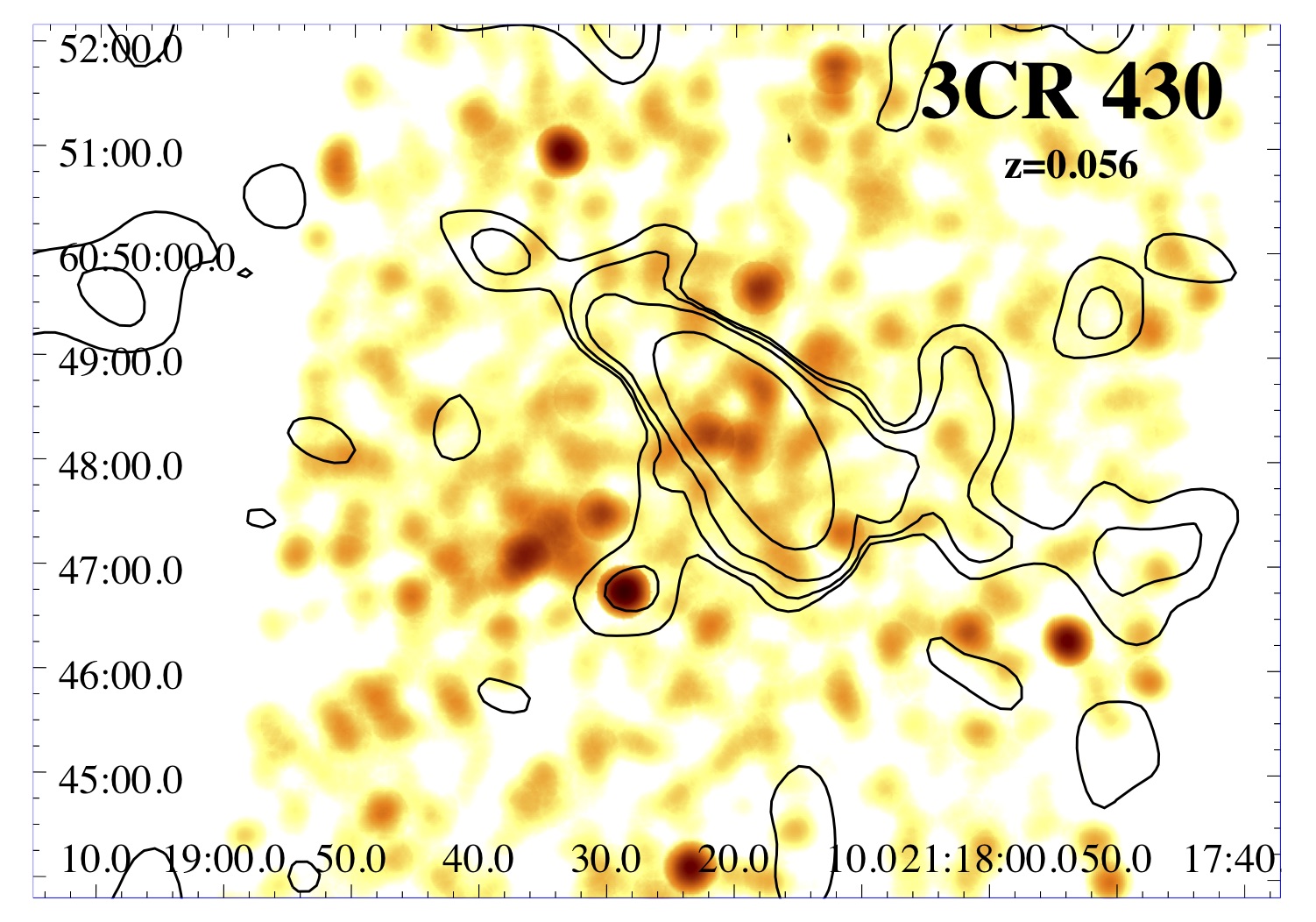}
\caption{Left: Pan-STARSS optical image in the $R$ band with VLA 1.4 GHz (in red) and Chandra 0.5 - 3 keV (in blue) contours overlaid of 3CR\,430. Radio contours were drawn starting at 0.2 mJy/beam and increasing by a factor of 2. X-ray contours were drawn from the $Chandra$ event file in the 0.5 - 3 keV energy range, smoothed with a 29.52 arcsec Gaussian kernel at levels of 0.005, 0.08, 0.1 photons and with 1.968 arcsec pixels. The dashed black circle has a 200 kpc radius. The two different extended X-ray emission structures could indicate the presence of two galaxy clusters undergoing a merging process. Right: $Chandra$ image with 0.984 arcsec pixels, exposure corrected in the 0.5 - 3 energy range at center-band energy of 2.3 keV and smoothed with a Gaussian kernel of 15.74 arcsec  with TGSS contours overlaid in white drawn at 0.025, 0.05, 0.125 and 0.75 Jy/beam. TGSS radio map has a beam size of 25 arcsec. TGSS contours mark the presence of radio emission unrelated to the extended X-ray emission, which could indicate the presence of a radio relic.}
\label{fig:cluster2}
\end{figure*}

The left panel of Fig. \ref{fig:cluster2} shows an optical image of the field of 3CR\,430 with radio contours (at 4.9 GHz) and X-ray contours (in the 0.5 - 3 keV band). The right panel of Fig. \ref{fig:cluster2} shows the 0.5 - 3 keV $Chandra$ image of 3CR\,430 with TGSS contours overlaid. Diffuse radio emission lying beyond the extended X-ray emission could indicate the presence of a radio relic if the X-ray emission arises indeed from two merging galaxy clusters.

\textbf{3CR\,458} ($z=0.289$). We detected extended X-ray emission 105 arcsec (460 kpc at the source redshift) above a 5$\sigma$ confidence level to the north-west of this source that could be due to the presence of another galaxy cluster maybe related to the radio galaxy (see Fig. \ref{fig:sample}). \citet{Wen2015} reported the presence of a galaxy cluster at $z=0.407$ at the position of the extended emission, which would mean that this extended emission would be unrelated to the radio galaxy (at $z=0.289$). However, there is also another galaxy for which the photometric redshift is known at the position of this extended emission (\citealt{Gorshkov1995}). This galaxy, at $z=0.29$, would be consistent with a source belonging to the same galaxy cluster as 3CR\,458.

\subsubsection{X-ray surface brightness profiles}

\begin{figure*}[h]
\includegraphics[width=9.5cm]{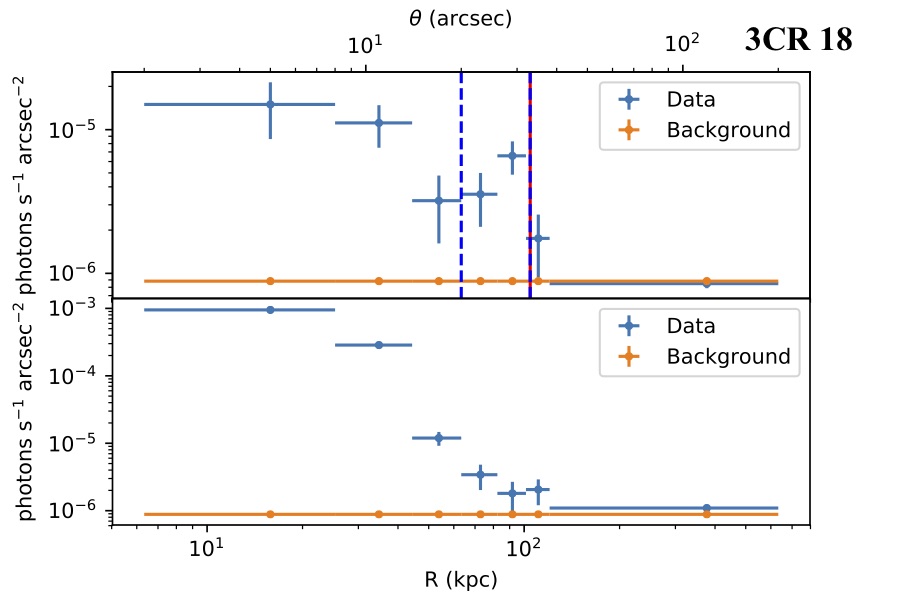}
\includegraphics[width=9.5cm]{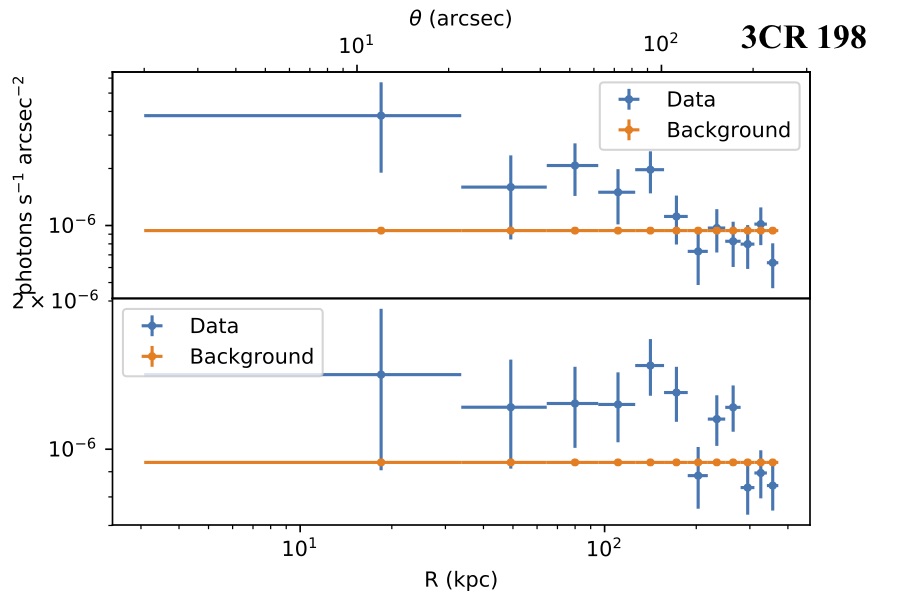}
\includegraphics[width=9.5cm]{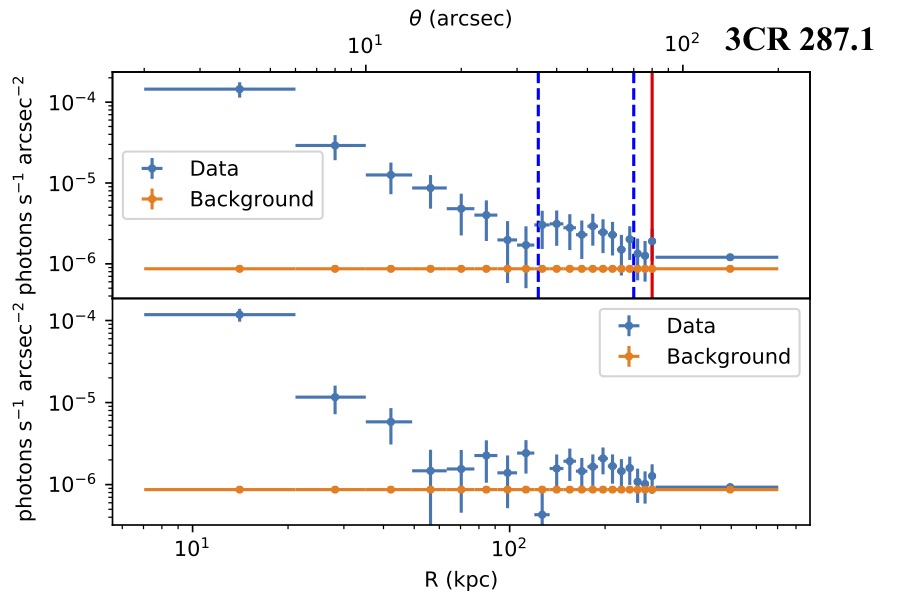}
\includegraphics[width=9.5cm]{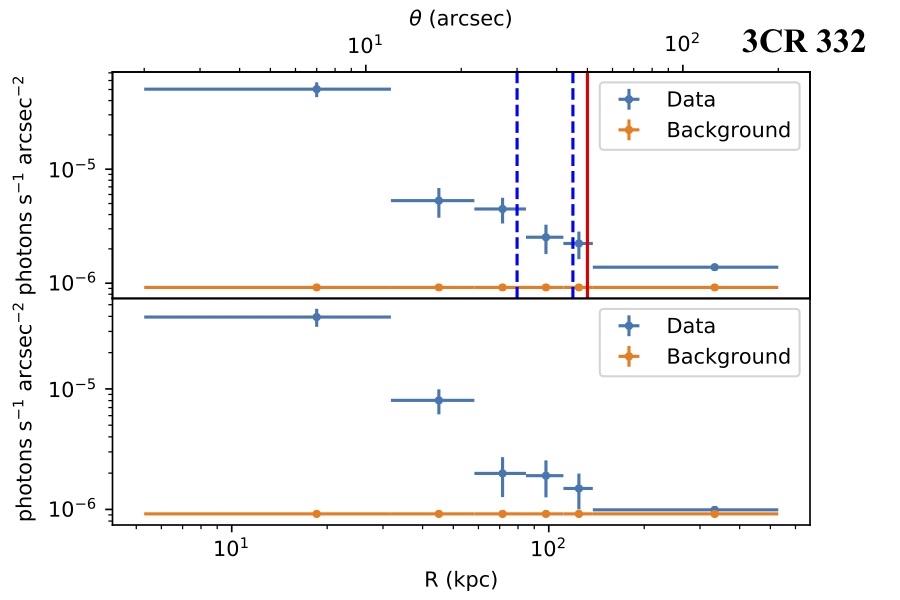}

\caption{Surface brightness profiles for 3CR\,18, 3CR\,198, 3CR\,287.1 and 3CR\,332. The top profile in each figure corresponds to the profile along the radio axis, while the bottom profiles correspond to the profiles in the direction perpendicular to the radio axis. The orange points show the background level in each case. The red vertical line marks the extension of the radio emission along the radio axis while the dashed blue lines mark the position of the hotspots. All sources, except for 3CR\,198 seem to have X-ray emission extended at least as much as the radio emission.}
\label{fig:profs}
\end{figure*}

For sources with more than 100 photons along the radio axis (see Table \ref{tab:counts}), 3CR\,18, 3CR\,198, 3CR\,287.1 and 3CR\,332, we derived X-ray surface brightness profiles along the radio axis and in the direction perpendicular to the radio axis. Images are shown in Fig. \ref{fig:profs} where the profile along the radio axis is represented on top of the one perpendicular to the radio axis, the extension of the radio emission is marked by a red vertical line, the blue dotted lines represent the hotspot region and the background level is indicated in orange. Although we only considered sources with the largest number of photons along the radio axis, the number of photons in each direction was still too low to obtain reliable results from the $\beta$-profile fit (\citealt{Cavaliere1976,Cavaliere1978}). Although we cannot distinguish thermal and non-thermal emission using $\beta$-profile fits, the X-ray surface brightness profiles show that the X-ray emission is extended at least up to the radio structure extent in all cases but for 3CR\,198.

\section{A multifrequency perspective}
\label{sec:multifreq}

\subsection{Low radio frequency observations}

We compared our $Chandra$ observations with low radio frequency observations from TGSS and LOFAR at $\sim$150 MHz. LOFAR observations were only available for 11 of our targets. 

LOFAR observations are part of the forthcoming Data Release 2 (DR2) of LoTSS\footnote{The DR2 v2.2 was run as part of the ddf pipeline (https://github.com/mhardcastle/ddf-pipeline) and the LoTSS DR1 consists of images at 6 arcsec resolution and and $\sim$70 $\mu$Jy/beam sensitivity covering an area of $\sim$400 square degrees while the footprint of the DR2 will cover an area of approximately 5700 square degrees, both performed in the northern hemisphere.}, i.e, a deep 120-168 MHz imaging survey. These datasets were processed by the international LOFAR collaboration as part of the LOFAR Data Release 1 and 2 (\citealt{Shimwell2017,Shimwell2019} and \citealt{Tasse2020}, respectively). 

Electrons responsible for IC/CMB emission from lobes in the soft X-rays (0.5 - 3 keV) are also responsible for the emission of synchrotron radiation at a frequency given by:

\begin{equation}
\nu_{syn}^{obs} = 3.72\times10^{-5} \nu_{IC}^{obs} \frac{B}{1+z}\, Hz
\end{equation}
where $\nu_{syn}^{obs}$ is the observed frequency of the synchrotron radiation of the same electrons scattering the CMB at $\nu_{IC}^{obs}$ in the Thomson regime, while $B$ is the magnetic field in Gauss and $z$ the source redshift. Then, considering $\nu_{IC}^{obs}\sim$10$^{17}$ Hz, and $B\sim$10 $\mu$G (as found in the lobes of several radio galaxies by \citealt{Ineson2017}), we could detect synchrotron radiation from the same electrons emitting via IC/CMB in X-rays at radio frequencies of tens of MHz. Therefore, using radio observations at $\sim$150 MHz allows us to get closer to the low-frequency regime needed to produce the IC/CMB radiation in the $Chandra$ energy band. Thus, when we find low radio frequency lobe-emission co-spatial with the extended X-ray structure, we tend to favor a possible IC/CMB interpretation, if fainter or null cross-cone emission is present.

Sources for which LoTSS data were available are: 3CR\,18, 3CR\,54, 3CR\,198, 3CR\,223.1, 3CR\,268.2, 3CR\,272, 3CR\,313, 3CR\,332 and 3CR\,357. LoTSS images are shown in Figure \ref{fig:lofar}, where the X-ray emission, in the 0.5 - 3 keV band, can be seen with VLA contours in blue and 150 MHz LoTSS contours in black overlaid. X-ray and VLA image parameters are reported in Table \ref{tab:parameters}, while LoTSS image parameters are collected in the caption of Fig. \ref{fig:lofar}. In general, 150 MHz emission traces radio lobes along the whole radio axis, even where no high-frequency radio emission can be seen. The emission at 150 MHz matches the extent of the X-ray emission which is in line with IC/CMB being the dominant emission process occurring co-spatially with radio lobes.

\begin{figure*}[h]
\includegraphics[width=9.cm]{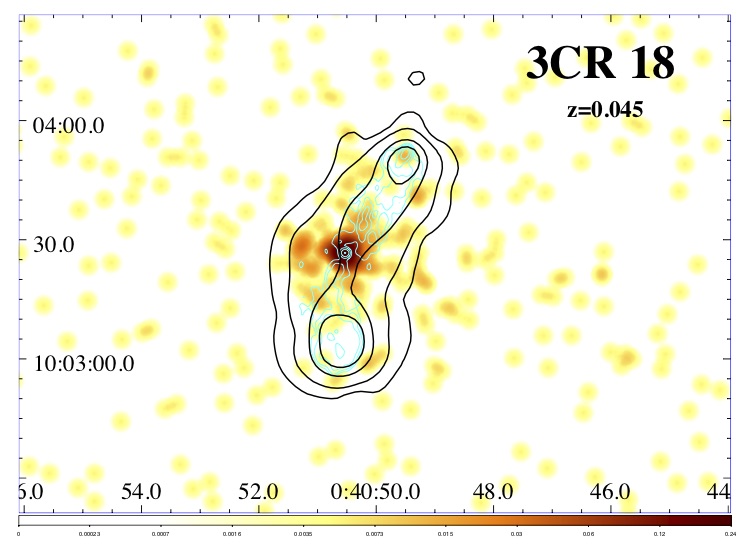}
\includegraphics[width=9.cm]{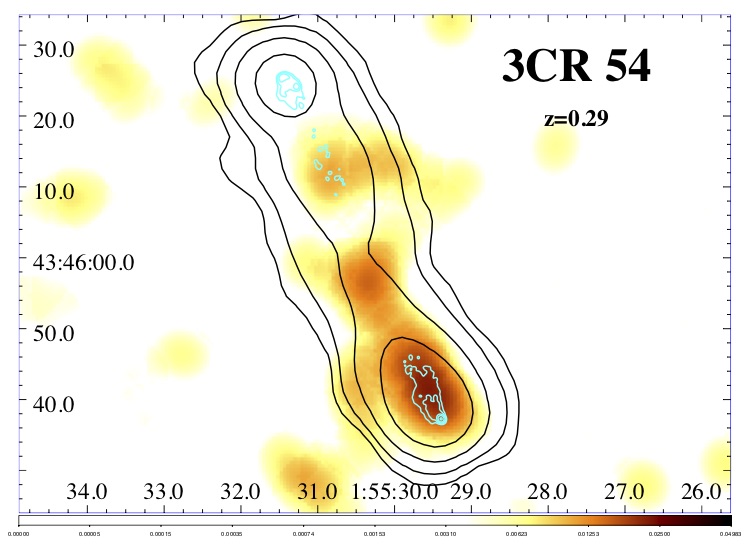}
\includegraphics[width=9.cm]{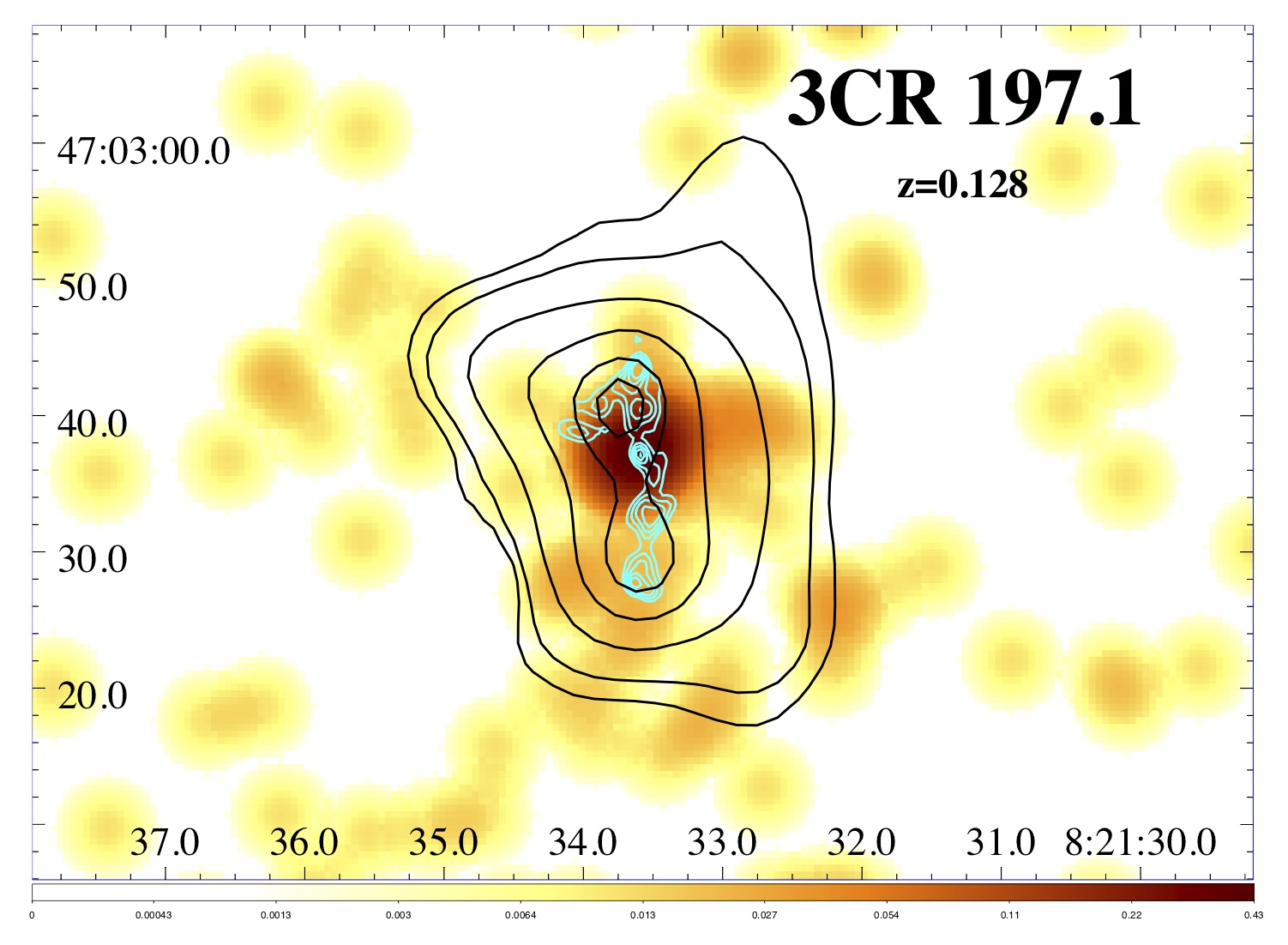}
\includegraphics[width=9.cm]{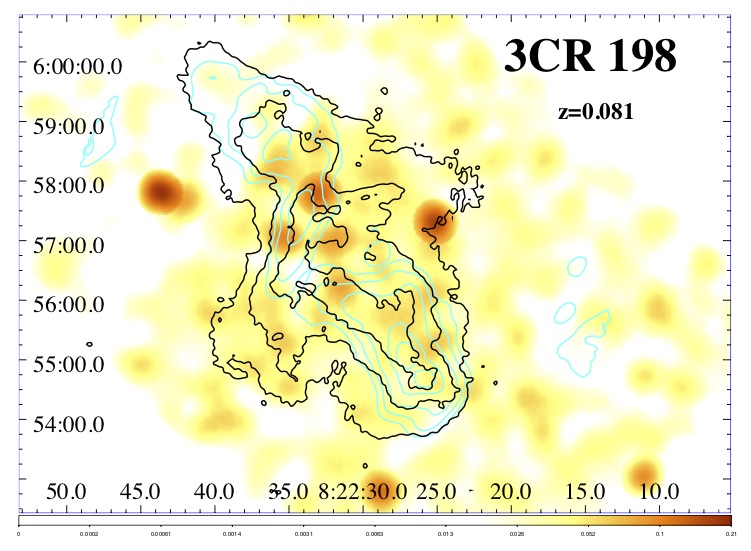}
\includegraphics[width=9.cm]{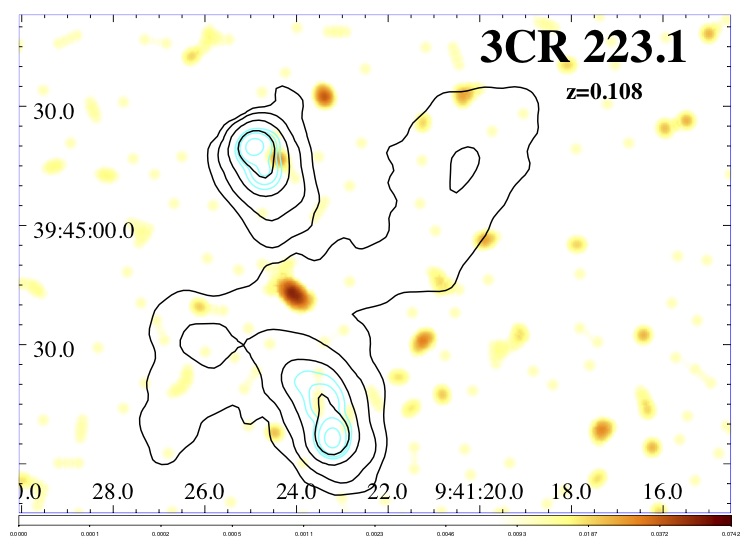}
\includegraphics[width=9.cm]{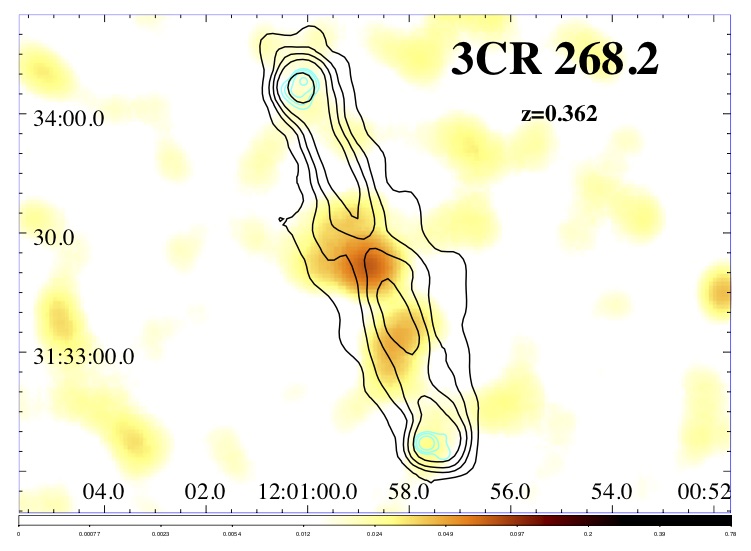}
\caption{X-ray image, VLA contours (cyan) and 150 MHz contours (black). X-ray and VLA image parameters are shown in Table \ref{tab:parameters}. LoTSS contours for each source are at 40, 160, 640, 1280 mJy/beam for 3CR\,18; at 10, 40, 160, 640 mJy/beam for 3CR\,54; at 40, 80, 240, 640, 1200, 1600 mJy/beam for 3CR\,197.1; at 2, 8, 16 mJy/beam for 3CR\,198; at 10, 40, 160, 640 mJy/beam for 3CR\,223.1; at 10, 40, 80, 160, 640 mJy/beam for 3CR\,268.2; at 10, 40, 80, 320 mJy/beam for 3CR\,272; at 10, 20, 80, 320, 500, 800, 1300 mJy/beam for 3CR\,293.1; at 40, 160, 320, 640, 1280 mJy for 3CR\,313; at 10, 20, 40, 80, 160, 320 mJy for 3CR\,332; at 20, 40, 80, 160, 320 mJy for 3CR\,357 and at 100, 200, 800, 1200, 2000 mJy/beam for 3CR\,430, respectively. LoTSS radio maps have beamsizes of 6 arcsec. 150 MHz emission traces better the lobe emission than the GHz emission and, thus, it suggests that the extended X-ray emission coincident with it is most likely due to IC/CMB from lobes.}
\label{fig:lofar}
\end{figure*}

\addtocounter{figure}{-1}
\begin{figure*}[h]
\includegraphics[width=9.cm]{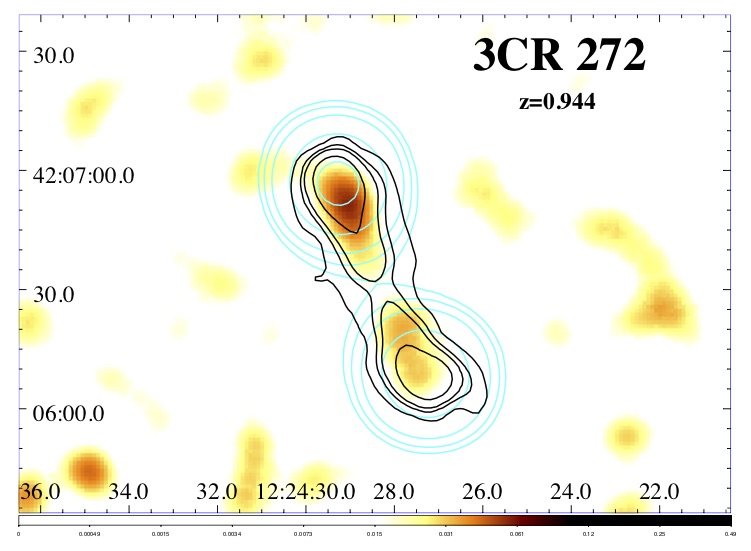}
\includegraphics[width=9.cm]{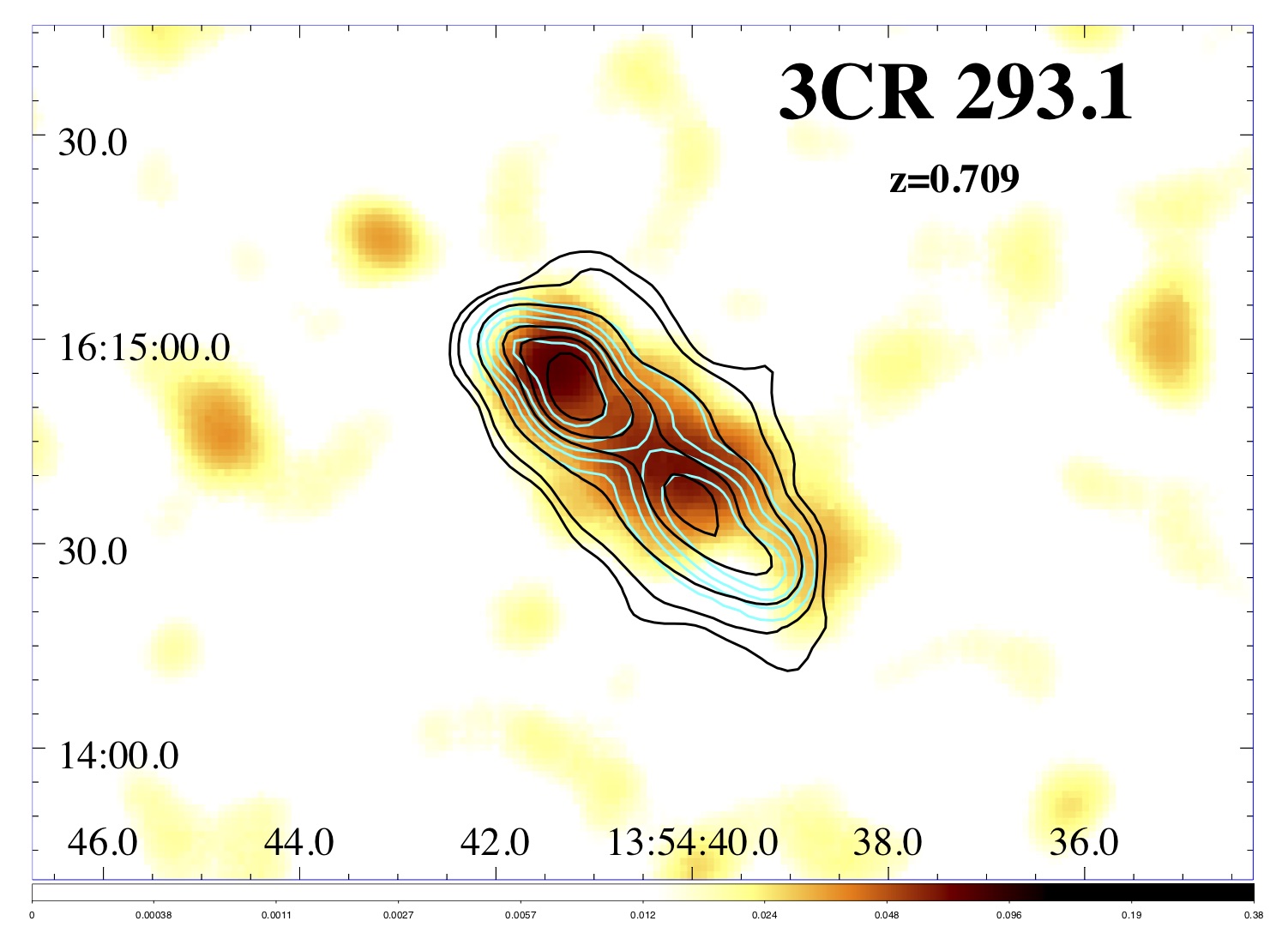}
\includegraphics[width=9.cm]{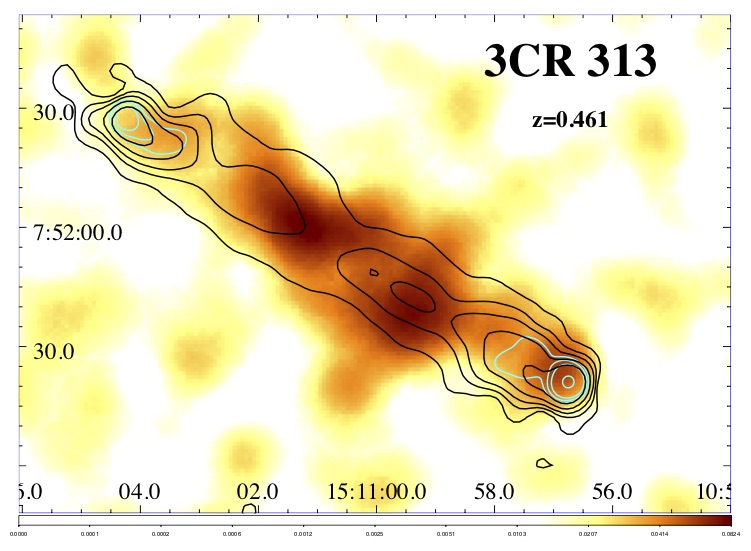}
\includegraphics[width=9.cm]{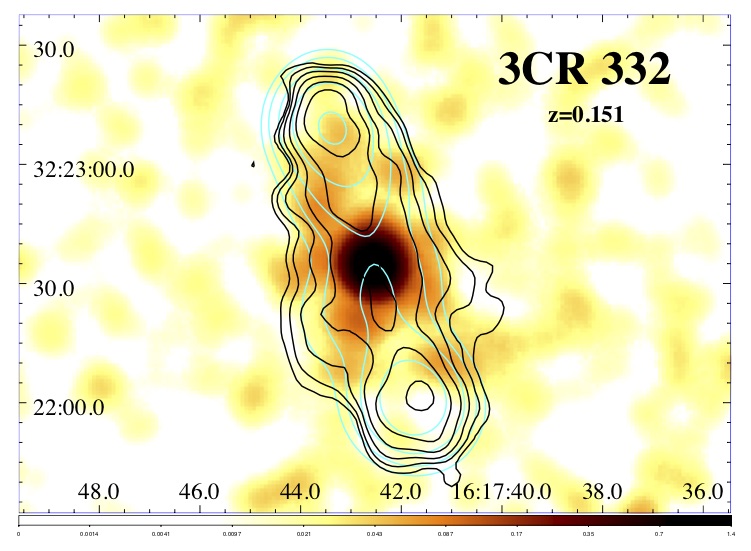}
\includegraphics[width=9.cm]{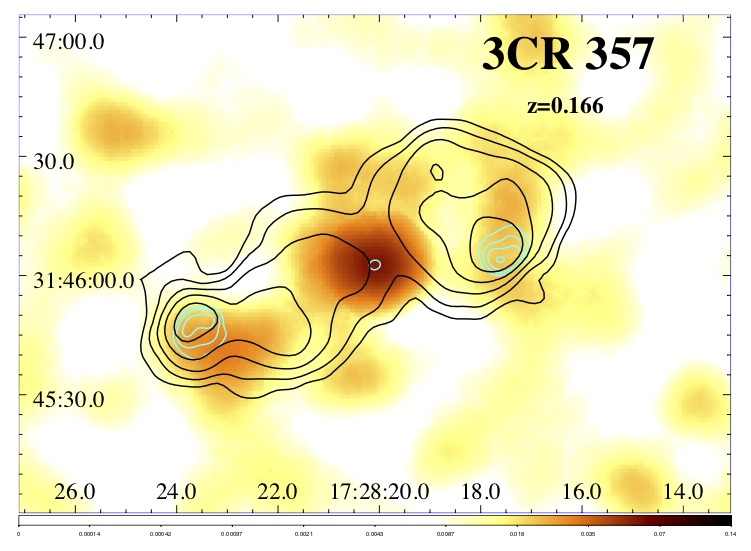}
\includegraphics[width=9.cm]{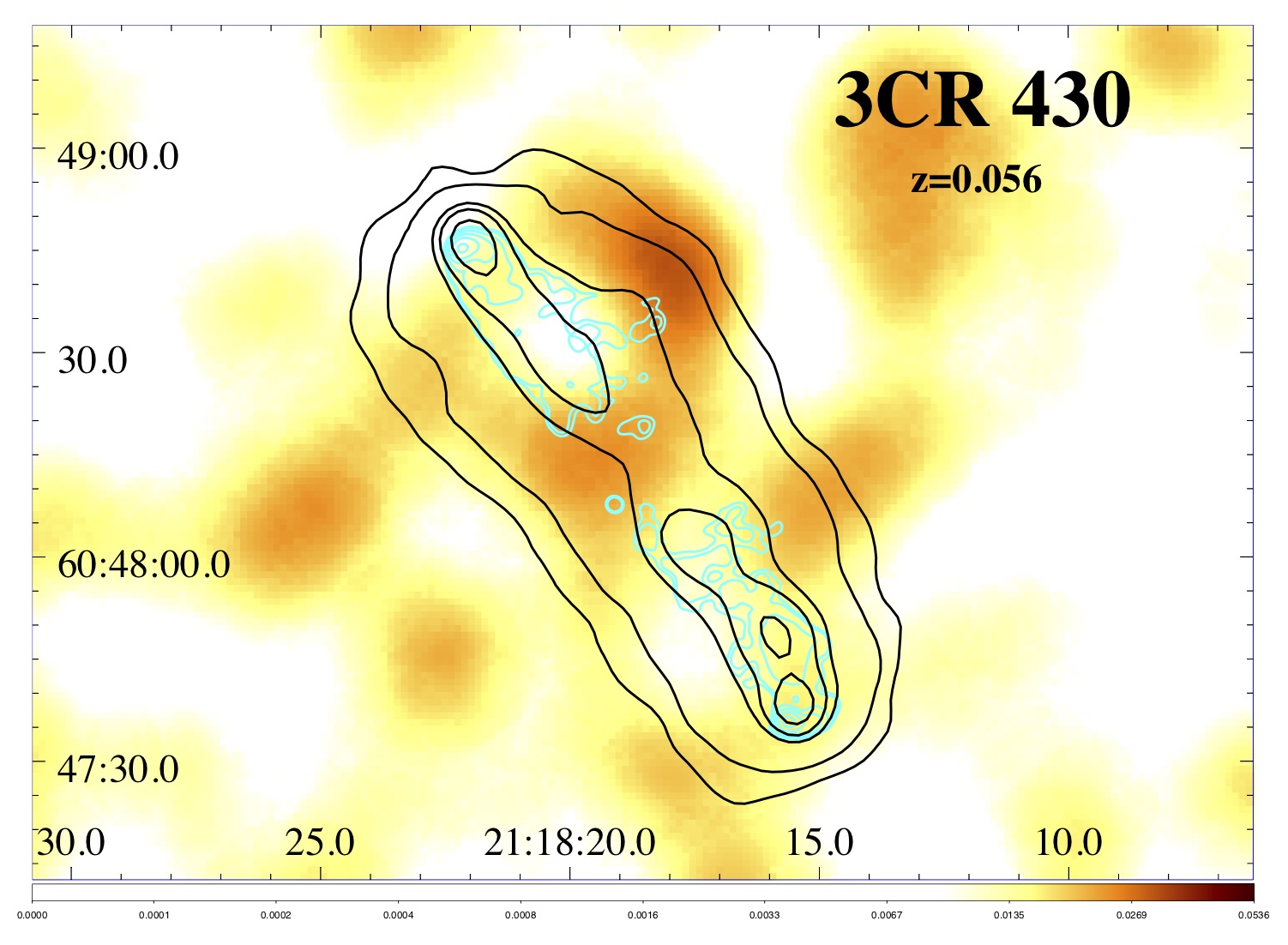}
\caption{(continued)}
\end{figure*}

All sources in our sample but 3CR\,197.1 and 3CR\,434 are extended in TGSS but only 3CR\,44, 3CR\,135, 3CR\,268.2, 3CR\,272, 3CR\,287.1, 3CR\,313, 3CR\,332, 3CR\,357, 3CR\,379.1 and 3CR\,458 present a double morphology at $\sim$150 MHz. We only show the TGSS data of 3CR\,103 (right panel of Fig. \ref{fig:cluster1}), 3CR\,403.1 (Fig. \ref{fig:3c403p1}), 3CR\,430 (right panel of Fig. \ref{fig:cluster2}) and 3CR\,458 (last panel in Fig. \ref{fig:sample}) since we only used it to confirm the sources' double morphology.

\subsection{An optical overview.}

We checked for the presence of ``red sequences'' (i.e., a color-magnitude relation followed by galaxies belonging to the same galaxy cluster; see \citealt{Visvanathan1977} and \citealt{DeLucia2007}) in the fields of some sources in our sample to determine whether these sources belong to galaxy clusters. Sources selected were those for which the ratio of photons along and perpendicular to the radio axis, $\rho<1$ (Table \ref{tab:counts}) as defined in equation \ref{eq:ratio}, meaning that the emission perpendicular to the radio axis is comparable to the emission along the radio axis, which hints at the presence of thermal emission from ICM in galaxy clusters. Magnitudes in the $r$ and the $i$ band (not corrected for Galactic absorption) for sources lying at $> 500$ kpc radius around the positions of the radio galaxies were obtained from Pan-STARRS\footnote{https://catalogs.mast.stsci.edu/panstarrs/}.

We explored the red sequences in the fields of 3CR\,103 (since we detected extended emission in the field with no redshift information that could be related to the radio galaxy), 3CR\,379.1, 3CR\,430. However, we were not able to identify red sequences in the fields of these sources. Nevertheless, spectroscopic follow up observations would be crucial to confirm these results.



\section{Summary and conclusions}
\label{sec:summary}

We carried out a thorough analysis of 35 FR II radio galaxies with angular sizes above 5 arcsec, measured in the GHz radio maps, in the 3CR $Chandra$ Snapshot Survey observed before Cycle 20 and not in the 3CRR catalog (excluding also 3CR\,187, 3CR\,196.1 and 3CR\,320, for which detailed follow-up analyses have already been performed). Our main goals were:
\begin{enumerate}
    \item To study how the presence of this extended emission affects the detection of hotspots.
    \item To investigate the origin of extended X-ray emission in radio galaxies (either non-thermal emission due to IC/CMB from lobes or thermal emission from ICM from galaxy clusters).
\end{enumerate}{}

Throughout this analysis, we used 0.5 - 3 keV, background and point source subtracted, exposure-weighted observations from the $Chandra$ Snapshot Survey, refining the previous analyses.


We found that $\sim90\%$ of galaxies in our sample presented significant ($>3\sigma$ confidence level) extended emission coincident with the radio axis ($\sim70\%$ above a $5\sigma$ confidence level); while $\sim60\%$ presented significant extended emission in the direction perpendicular to the radio axis ($\sim40\%$ above a $5\sigma$ confidence level). For those galaxies for which we detected extended X-ray emission perpendicular to the radio axis, at similar confidence levels as the emission along the radio axis, we tend to favor the scenario where the underlying emission process is thermal radiation from an ICM, while for sources where no significant emission perpendicular to the radio axis is detected we favor the scenario in which non-thermal IC/CMB emission is produced by electrons stored in the radio lobes.

In total, we found that IC/CMB is the most likely dominant emission process in $\sim71\%$ of our sources, while for $\sim17\%$ of them, the dominant emission process is most likely thermal emission from the ICM (see Table \ref{tab:counts}). However, we could not establish which mechanism was responsible for the extended X-ray emission in the remaining $\sim11\%$. Among optically classified LERGs, there is the same number of sources for which the dominant emission comes from non-thermal emission (IC/CMB) and from thermal emission (ICM). For HERGs, non-thermal emission (IC/CMB) dominates in six times more sources than thermal emission (ICM). This suggests that FR\,II-LERGs inhabit galaxy clusters more often than FR\,II-HERGs (see \citealt{massaro19,Massaro2020} for an in-depth analysis of the environments of radio galaxies). A summary of these results can be found in Table \ref{tab:counts}.

We confirmed the detection of 7 hotspots in the 0.5 - 3 keV band above $3\sigma$ confidence level, excluding the possibility of them being fluctuations of the extended X-ray emission that surrounds them. Using the soft energy band and the \textit{local} background, we did not confirm one of the detected hotspots claimed in previous survey analyses (n29 in 3CR\,52) but do claim newly detected hotspots in two sources. This demonstrates the relevance of taking into account the extended emission along the radio axis when detecting hotspots. Additionally, we detected the presence of extended X-ray emission that could arise from the ICM in the fields of 3CR\,103 and 3CR\,430.

Using X-ray surface brightness profiles, we concluded that the X-ray emission is extended up to extent of the radio structure in, at least, four sources. However, deeper X-ray observations are needed to distinguish between thermal and non-thermal emission processes. 

We performed a multifrequency comparison to further constrain the origin of this emission. We compared low radio frequency and X-ray emission for those sources with LOFAR observations (3CR\,18, 3CR\,54, 3CR\,198, 3CR\,223.1, 3CR\,268.2, 3CR\,272, 3CR\,313, 3CR\,332 and 3CR\,357). In all cases, the low radio frequency morphology seems to trace the extended X-ray emission. Thus, in those cases, we tend to favor IC/CMB as the dominant emission process. We performed an optical investigation of the red sequences of sources in the field of 3CR\,103, 3CR\,379.1 and 3CR\,430; however, it is not clear whether they belong to galaxy clusters/groups.

A summary of these results is presented in Table \ref{tab:counts}, where we show the number of photons in each region, with the number of background photons scaled to the size of the region in parenthesis. The parameter $\rho$ defines the ratio of the emission along and perpendicular to the radio axis scaled to the sizes of the regions according to Eq. \ref{eq:ratio}. We also collect in Table \ref{tab:counts} information on sources that have newly detected hotspots, which ones are in known optical clusters and which is the dominant emission process in each case.

\begin{table*}[]
    \centering
    \caption{Radio and X-ray parameters used for the appendix images of Fig. \ref{fig:sample}}
    \begin{tabular}{|l|c|r|r|c|c|r|c|c|c|}
  \hline
  \multicolumn{1}{|c|}{3CR} &
  \multicolumn{1}{c|}{Optical}&
  \multicolumn{1}{c|}{hotspot} &
  \multicolumn{1}{c|}{hotspot} &
  \multicolumn{1}{c|}{along} &
  \multicolumn{1}{c|}{perpendicular} &
  \multicolumn{1}{c|}{$\rho$}&
  \multicolumn{1}{c|}{New hotspots}&
  \multicolumn{1}{c|}{Optical}&
  \multicolumn{1}{c|}{Dominant}\\
  \multicolumn{1}{|c|}{Name} &
  \multicolumn{1}{c|}{Classification}&
  \multicolumn{1}{c|}{1} &
  \multicolumn{1}{c|}{2} &
  \multicolumn{1}{c|}{radio axis} &
  \multicolumn{1}{c|}{radio axis} &
  \multicolumn{1}{c|}{}&
  \multicolumn{1}{c|}{standard, local}&
  \multicolumn{1}{c|}{Cluster}&
  \multicolumn{1}{c|}{Process}\\
\hline
  18    & BLRG & 2.98 (1.90)  & 0.00 (0.14) & 130.70 (13.69)   & 44.80 (9.89)    & 4.4 & Y, -- &           & IC/CMB \\
  44    & HERG & 1.98 (1.56)  & 0.98 (1.14) & 25.10 (8.00)    & 17.20 (8.82)   & 2.1 & Y, -- & \checkmark& IC/CMB \\
  52    & HERG & 0.00 (0.14)  & 0.98 (1.14) & 21.00 (7.65)    & 20.40 (8.82)   & 1.6 & --    & \checkmark& ICM    \\
  54    & HERG & 0.00 (0.14)  & 2.98 (1.90) & 9.90 (5.37)    & 4.40 (5.53)    & 2.4 & N, N  &           & IC/CMB \\
  63    & HERG & 0.00 (0.14)  & 4.98 (2.40) & 16.60 (4.76)    & 9.20 (4.06)    & 3.9 & Y, -- &           & --     \\
  69    & HERG & 0.00 (0.14)  & 0.98 (1.14) & 15.80 (6.86)    & 4.4 (5.89)    & 2.4 & Y, -- &           & IC/CMB \\
  103   &      & 0.00 (0.14)  & 0.98 (1.14) & 20.30 (8.07)    & 3.5 (9.43)   & 3.1 & --    &           & IC/CMB\\
  107   & HERG & 0.00 (0.14)  & 0.98 (1.14) & 6.70 (3.19)     & 1.40 (2.19)     & 7.4 & --    &           & IC/CMB \\
  114   & LERG & 0.00 (0.14)  & 0.00 (0.14) & 12.10 (5.58)    & 10.10 (9.14)   & 2.8  & --    &           & IC/CMB \\
  133   & HERG & 0.00 (0.14)  & 2.98 (1.90) & 11.90 (3.78)    & 6.70 (3.19)     &4.1  & N, -- &           & IC/CMB \\
  135   & HERG & 0.00 (0.14)  & 0.98 (1.14) & 4.80 (8.61)   & 7.00 (15.42)   &1.2  & --    & \checkmark& --     \\
  165   & LERG & 0.00 (0.14)  & 0.98 (1.14) & 39.20 (10.36)   & 54.20 (16.52)   &2.1  & --    &           & ICM    \\
  166   & LERG & 1.98 (1.56)  & 0.00 (0.14) & 28.20 (7.24)    & 13.10 (6.10)    &2.5  & Y, -- &           & IC/CMB\\
  169.1 & HERG & 0.00 (0.14)  & 0.98 (1.14) & 6.30 (4.64)     & 6.50 (6.48)    &1.8  & Y, -- &           & IC/CMB \\
  180   & HERG & 0.00 (0.14)  & 0.00 (0.14) & 6.50 (6.16)    & 16.00 (15.20)   &1.5  & --    &           & IC/CMB \\
  197.1 & HERG & 0.98 (1.14)  & 0.98 (1.14) & 29.60 (5.10)    & 16.60 (5.79)    &7.1  & Y, -- & \checkmark& IC/CMB \\
  198   & HERG & 0.00 (0.14)  & 0.00 (0.14) & 129.60 (33.31) & 124.90 (64.45)& 1.4 & --    & \checkmark& IC/CMB \\
  223.1 & HERG & 0.00 (0.14)  & 0.00 (0.14) & 9.80 (8.90)   & 12.50 (16.69)   &1.4  & --    &           & IC/CMB \\
  268.2 & HERG & 0.98 (1.14)  & 1.98 (1.56) & 25.10 (11.70)   & 1.40 (12.05)   & 2.0 & N, N  &           & IC/CMB \\
  272   & HERG & 0.98 (1.14)  & 0.98 (1.14) & 10.50 (6.98)    & 9.30 (11.87)   & 1.8 & Y, -- &           & IC/CMB \\
  287.1 & HERG & 0.00 (0.14)  & 0.98 (1.14) & 167.20 (16.99)  & 27.20 (12.85)   &7.2  & Y, -- &           & IC/CMB \\
  293.1 & HERG & 0.00 (0.14)  & 0.00 (0.14) & 14.40 (6.50)    & 17.00 (8.85)   &1.7  & --    &           & IC/CMB \\
  306.1 & HERG & 1.98 (1.56)  & 0.98 (1.14) & 24.10 (9.22)   & 9.10 (11.71)   &2.5  & Y, -- & \checkmark& IC/CMB \\
  313   & HERG & 2.98 (1.90)  & 2.98 (1.90) & 93.60 (16.65)  & 41.8 (20.10)  &2.7  & N, N  & \checkmark& IC/CMB \\
  332   & HERG & 1.98 (1.56)  & 0.00 (0.14) & 203.80 (19.17)  & 47.60 (18.83)  & 7.1 & Y, -- & \checkmark& IC/CMB \\
  357   & LERG & 0.00 (0.14)  & 0.00 (0.14) & 31.80 (12.40)   & 15.10 (15.39)   & 1.8 & --    & \checkmark& IC/CMB \\
  379.1 & HERG & 0.00 (0.14)  & 0.00 (0.14) & 2.50 (5.90)    & 12.60 (9.22)   & 0.7 & --    &           & ICM   \\
  403.1 & LERG & 0.00 (0.14)  & 0.00 (0.14) & 15.40 (9.94)   & 62.90 (21.67)  &1.1  & --    & \checkmark& ICM    \\
  411   & HERG & 1.98 (1.56)  & 0.98 (1.14) & 68.80 (9.46)    & 55.30 (10.95)    &1.2  & Y, -- &           & ICM    \\
  430   & LERG & 0.00 (0.14)  & 0.00 (0.14) & 7.80 (7.44)   & 50.20 (16.19)   &0.8  & --    & 
           & ICM    \\
  434   & LERG & 0.00 (0.14)  & 2.98 (1.90) & 4.20 (3.13)     & 3.30 (5.54)     &2.0  & Y, Y  & \checkmark& --     \\
  435A  & BLRG & 0.98 (1.14)  & 1.98 (1.56) & 6.70 (3.19)     & 6.60 (4.01)     & 3.7 & Y, -- &           & IC/CMB\\
  435B  &      & 0.98 (1.14)  & 1.98 (1.56) & 17.70 (5.99)    & 18.60 (7.82)    & 2.4 & Y, -- &           & IC/CMB\\
  456   & HERG & 9.98 (3.30) & 2.98 (1.90) & 2.90 (2.05)     & 14.30 (5.30)    & 2.2 & Y, Y  &           & --    \\
  458   & HERG & 2.98 (1.90)  & 0.00 (0.14) & 39.40 (15.58)   & 20.90 (20.03)  & 1.6 & N, N  & \checkmark& IC/CMB \\
    
    \hline
    \end{tabular}
    
    Column 1: source name. Column 2: optical classification. Columns 3-6: net source photons measured in the hotspots, along the radio axis and perpendicular to the radio axis with the corresponding background photons for those regions in parenthesis. Background here refers to the \textit{standard} background. Column 7: ratio of photons along and perpendicular to the radio axis as defined by Eq. \ref{eq:ratio}. Column 8: flag for sources with new hotspots detected using the \textit{standard} and the \textit{local} background. ``Y" indicates a newly detected hotspot, ``N" indicates a hotspot detected in this work that was previously detected by other works and ``-" indicated a non-detection. Column 9: flag that indicates whether the presence of an optical galaxy cluster was previously reported. Column 10: Dominant emission process for the extended X-ray emission.
\label{tab:counts}
\end{table*}

Lastly, during this analysis, we made the serendipitous discovery that three hotspots in our sample were detected in WISE, namely n22 in 3CR\,69, s24 in 3CR\,166 and n34 in 3CR\,332. Additional details and images can be seen in Appendix \ref{app:wise}. 



\begin{acknowledgements}

We thank the anonymous referee for the useful comments that led to improvements in the paper.

We are grateful to G. Giovannini for providing insight into the radio emission of several radio galaxies.

This paper is based (in part) on data obtained with the International LOFAR Telescope (ILT). LOFAR (\citealt{vanHaarlem2013}) is the Low Frequency Array designed and constructed by ASTRON. It has observing, data processing, and data storage facilities in several countries, that are owned by various parties (each with their own funding sources), and that are collectively operated by the ILT foundation under a joint scientific policy. The ILT resources have benefited from the following recent major funding sources: CNRS-INSU, Observatoire de Paris and Université d’Orléans, France; BMBF, MIWF-NRW, MPG, Germany; Science Foundation Ireland (SFI), Department of Business, Enterprise and Innovation (DBEI), Ireland; NWO, The Netherlands; The Science and Technology Facilities Council, UK; Ministry of Science and Higher Education, Poland.

The Pan-STARRS1 Surveys (PS1) and the PS1 public science archive have been made possible through contributions by the Institute for Astronomy, the University of Hawaii, the Pan-STARRS Project Office, the Max-Planck Society and its participating institutes, the Max Planck Institute for Astronomy, Heidelberg and the Max Planck Institute for Extraterrestrial Physics, Garching, The Johns Hopkins University, Durham University, the University of Edinburgh, the Queen’s University Belfast, the Harvard-Smithsonian Center for Astrophysics, the Las Cumbres Observatory Global Telescope Network Incorporated, the National Central University of Taiwan, the Space Telescope Science Institute, the National Aeronautics and Space Administration under Grant No. NNX08AR22G issued through the Planetary Science Division of the NASA Science Mission Directorate, the National Science Foundation Grant No. AST-1238877, the University of Maryland, Eotvos Lorand University (ELTE), the Los Alamos National Laboratory, and the Gordon and Betty Moore Foundation.

A.J. acknowledges the financial support (MASF\_CONTR\_FIN\_18\_01) from the Italian National Institute of Astrophysics under the agreement with the Instituto de Astrofisica de Canarias for the ``Becas Internacionales para Licenciados y/o Graduados Convocatoria de 2017''.
This work is supported by the ``Departments of Excellence 2018 - 2022'' Grant awarded by the Italian Ministry of Education, University and Research (MIUR) (L. 232/2016).
This research has made use of resources provided by the Compagnia di San Paolo for the grant awarded on the BLENV project (S1618\_L1\_MASF\_01) and by the Ministry of Education, Universities and Research for the grant MASF\_FFABR\_17\_01.
F.M. acknowledges the financial contribution from the agreement ASI-INAF n.2017-14-H.0.
A.P. acknowledges financial support from the Consorzio Interuniversitario per la Fisica Spaziale (CIFS) under the agreement related to the grant MASF\_CONTR\_FIN\_18\_02.
F.R. acknowledges support from FONDECYT Postdoctorado 3180506 and CONICYT project Basal AFB-170002.
C.S. acknowledges support from the ERC-StG DRANOEL, n. 714245.
R.K. , W.F., and C.J. acknowledge support from the Smithsonian Institution and the Chandra High Resolution Camera Project through NASA contract NAS8-03060.

\end{acknowledgements}

\appendix

\section{Tables and images}
\label{app}

Table \ref{tab:parameters} shows parameters chosen for images in Fig. \ref{fig:sample}. These parameters are the radio frequency and contour levels for the radio emission and the pixel size and smoothing factors for the X-ray emission.

Images for all sources in our sample are shown in this section, in Fig. \ref{fig:sample}. We show their point-source subtracted X-ray emission in the 0.5 - 3 keV band (as described in \S~\ref{sec:reduction}) with different pixel sizes and smoothing per source (see Table \ref{tab:parameters}). Contours correspond to the radio emission at frequencies reported in Table \ref{tab:parameters}.

\newpage
\begin{longtable}{|c|c|c|c|c|c|}
\caption[Image parameters for X-ray and radio emission.]{Image parameters for X-ray and radio emission.}
  \\\hline
  \multicolumn{1}{|c|}{3CR} &
  \multicolumn{1}{c|}{Radio freq.} &
  \multicolumn{1}{c|}{Beam size} &
  \multicolumn{1}{c|}{Contour}&
  \multicolumn{1}{c|}{Pixel size}&
  \multicolumn{1}{c|}{Smoothing Gaussian}\\
   \multicolumn{1}{|c|}{Name} &
  \multicolumn{1}{c|}{(GHz)} &
  \multicolumn{1}{c|}{(arcsec)} &
  \multicolumn{1}{c|}{levels (mJy/beam)}&
  \multicolumn{1}{c|}{(arcsec)}&
  \multicolumn{1}{c|}{kernel (arcsec)}\\
\hline
18   &  1.4  & 1.70  & 5, 10, 20, 40, 80, 160, 320, 640 & 0.246 & 2.95\\     
44   &  1.4  & 1.46  & 2, 4, 8, 20, 40, 80              & 0.492   & 5.90\\     
52   &  1.4  & 3.59  & 10, 20, 80, 320, 640             &0.984   & 9.84\\     
54   &  8.0  & 0.40  & 0.2, 0.8, 3.2, 12.8, 51.2        & 0.492    & 6.89\\  
63   &  1.4  & 1.36  & 1, 2, 4, 8, 16, 32, 64           & 1.968   & 23.62\\ 
69   &  4.8  & 5.00  & 2, 4, 8, 16, 32, 64              &0.984   & 9.84\\     
103  &  1.4  & 3.27  & 10, 20, 40, 80, 160, 320, 640    &0.984   & 9.84\\   
107  &  4.8  & 0.46  & 0.2, 0.4, 0.8, 1.6, 3.2, 6.4     & 0.123 & 1.48\\   
114  &  1.4  & 1.27  & 2, 4, 8, 16, 32                  & 0.246 & 3.44\\   
133  &  1.4  & 0.98  & 8, 32, 64, 256, 512              & 0.123 & 1.48\\   
135  &  8.0  & 0.66  & 0.02, 0.08, 0.32, 1.28           & 1.968   & 15.74 \\   
165  &  1.4  & 3.36  & 2, 8, 16, 32, 64                 &0.984   & 7.87 \\   
166  &  1.4  & 1.37  & 4, 8, 16, 32                     & 0.492    & 3.94 \\   
169.1  &  1.4  & 1.37  & 2, 4, 8, 16, 32, 64            &0.984   & 15.74\\   
180  &  8.0  & 0.36  & 0.5, 0.7, 1, 2, 4                & 0.492    & 7.87\\   
197.1  &  4.8  & 1.32  & 2, 3, 4, 5, 6, 8               & 0.492    & 3.94 \\   
198  &  4.8  & 25.9  & 1, 2, 4, 8, 10, 12               & 1.968   & 23.62\\     
223.1  &  1.4  & 1.67  & 2, 4, 8, 20, 40, 80            & 0.492    & 3.94 \\  
268.2  &  1.4  & 1.61  & 2, 4, 8, 20, 40, 80            & 0.984   & 9.84\\
272  &  1.4  & 5.40  & 2, 4, 8, 20, 40, 80              & 0.984   & 9.84\\
287.1  &  1.4  & 5.00  & 2, 4, 8, 20, 40, 80            & 0.492    & 4.92\\
293.1  &  1.4  & 5.40  & 2, 8, 16, 32, 64               & 0.984  & 9.84\\
306.1  &  4.8  & 1.75  & 1, 2, 4, 8, 16                 & 0.984  & 9.84\\
313  &  8.0  & 2.43  & 2, 4, 8, 32, 64, 128             &0.984   & 11.81\\
332  &  1.4  & 4.40  & 2, 4, 8, 16, 32, 64              &0.984   & 7.87 \\
357  &  4.8  & 1.92  & 1, 2, 4, 8, 16                   &0.984   & 13.78\\
379.1  &  4.8  & 1.55  & 1, 2, 4, 8                     &0.984   & 11.81\\
403.1  &  0.3  & 7.03  & 16, 32, 64                     &0.984   & 13.78\\
411  &  1.4  & 1.30  & 4, 8, 16, 32, 64                 & 0.123 & 1.48\\
430  &  4.8  & 1.34  & 1.5, 2, 4, 8, 16, 32, 64         &0.984   & 15.74\\
434  &  1.4  & 1.40  & 4, 8, 16, 32, 64                 & 0.246 & 3.44\\
435  &  4.8  & 1.76  & 1, 2, 4, 8, 16, 32               & 0.246 & 3.44\\
456  &  1.8  & 1.43  & 4, 16, 64, 256                   & 0.123 & 1.72\\
458  &  1.4  & 6.40  & 2, 4, 8, 16, 32, 64              &0.984   & 13.78\\

\hline
\caption{Column description: (1) Source name; (2) radio frequency of contours in GHz; (3) beam size of the radio maps; (4) radio contour levels in mJy/beam; (5) pixel size of the X-ray image; (6) size of the Gaussian Kernel used for smoothing the X-ray image.}
\label{tab:parameters}
\end{longtable}

\newpage
\begin{figure}[h]
\includegraphics[width=9.5cm]{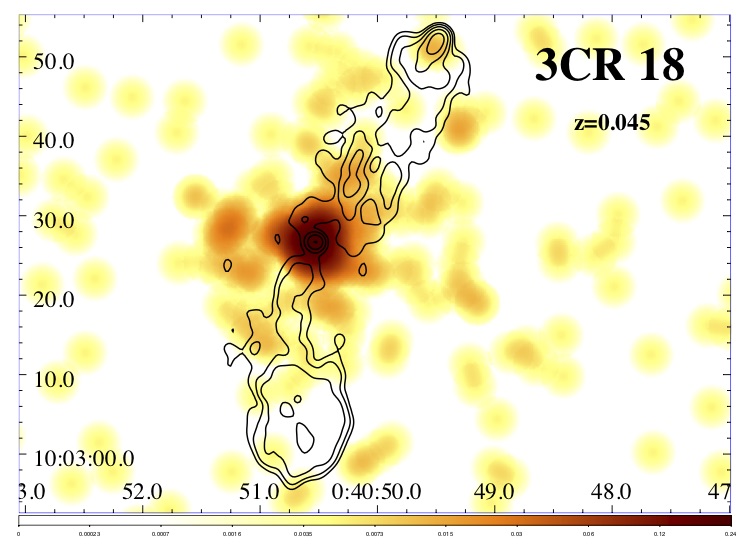}
\includegraphics[width=9.5cm]{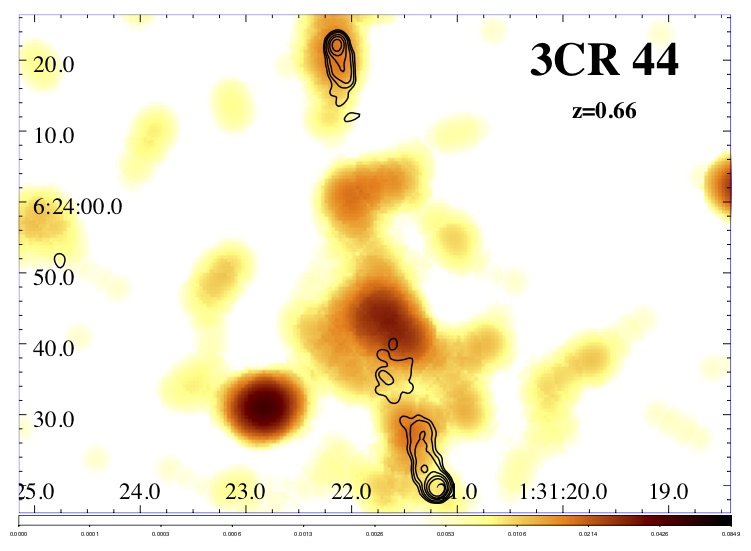}
\includegraphics[width=9.5cm]{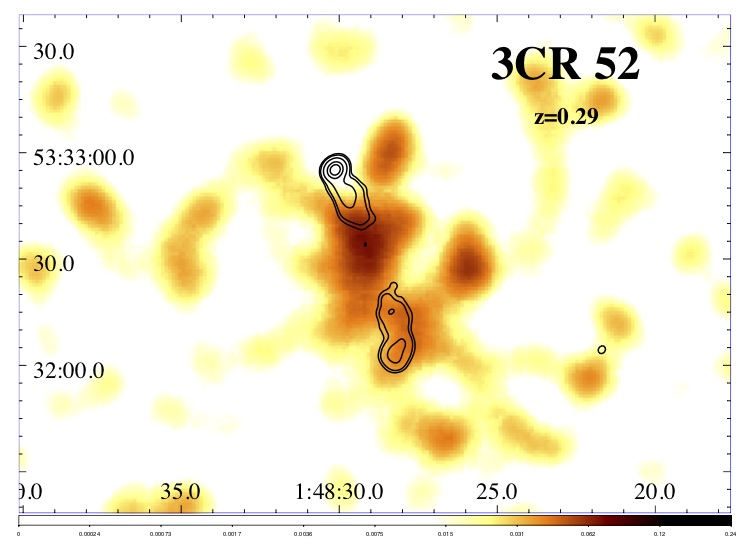}
\includegraphics[width=9.5cm]{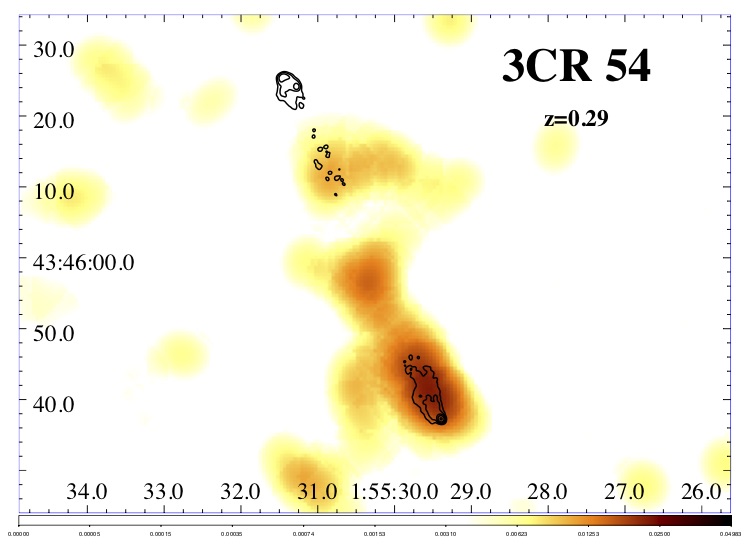}
\includegraphics[width=9.5cm]{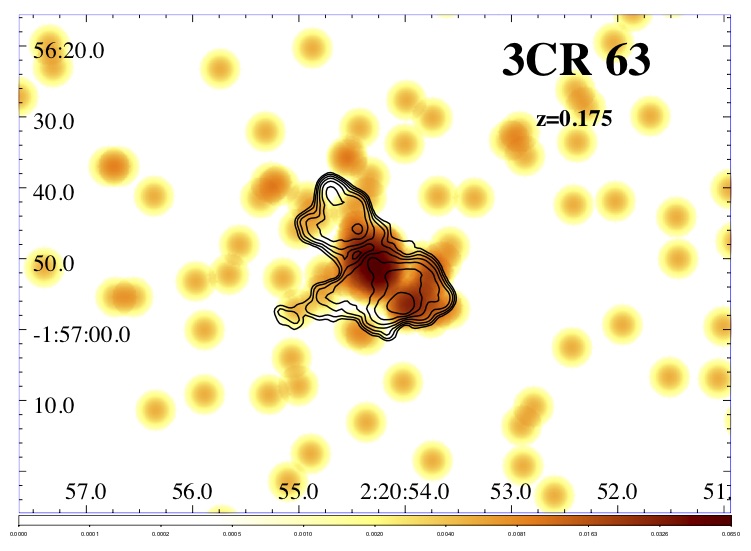}
\includegraphics[width=9.5cm]{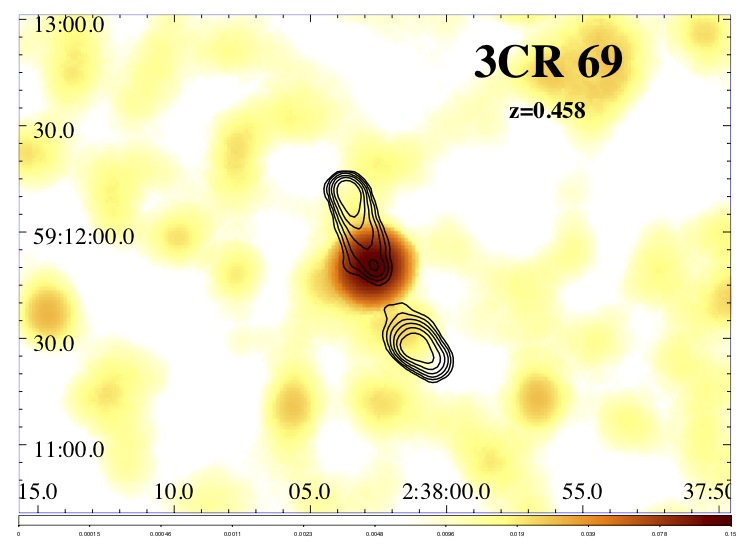}

\caption{X-ray emission, point-source subtracted (see \S~\ref{sec:reduction}), in the 0.5 - 3 keV band with VLA radio contours overlaid for all sources in our sample. For 3CR\,458, we also show the TGSS contours at levels of 40, 160, 640, 2560 mJy/beam. Pixel size of X-ray images, as well as frequency and contour levels for each source are shown in Table \ref{tab:parameters}.}

\label{fig:sample}
\end{figure}

\addtocounter{figure}{-1}
\begin{figure}[h]
\includegraphics[width=9.5cm]{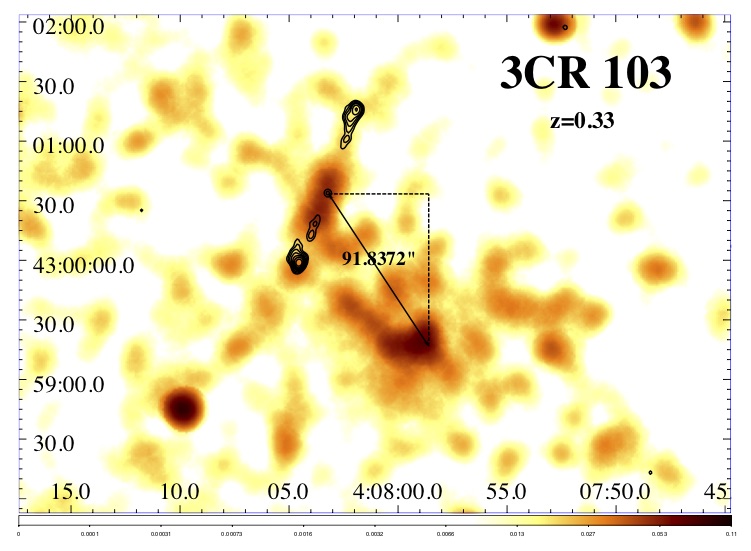}
\includegraphics[width=9.5cm]{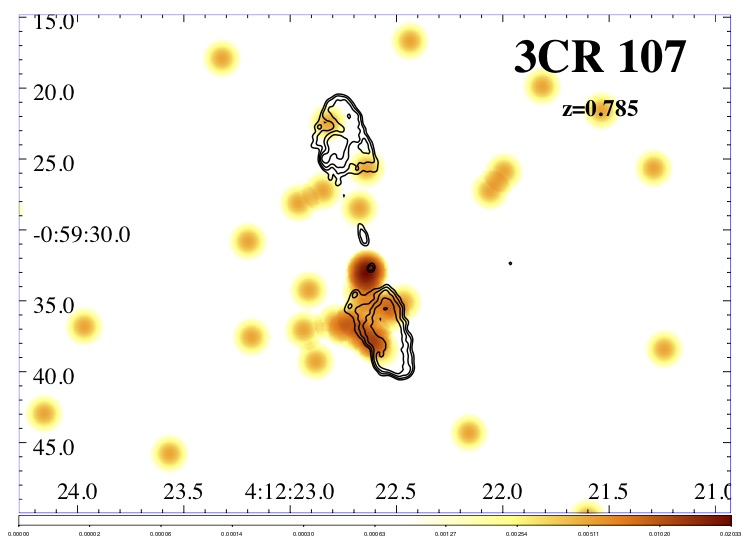}
\includegraphics[width=9.5cm]{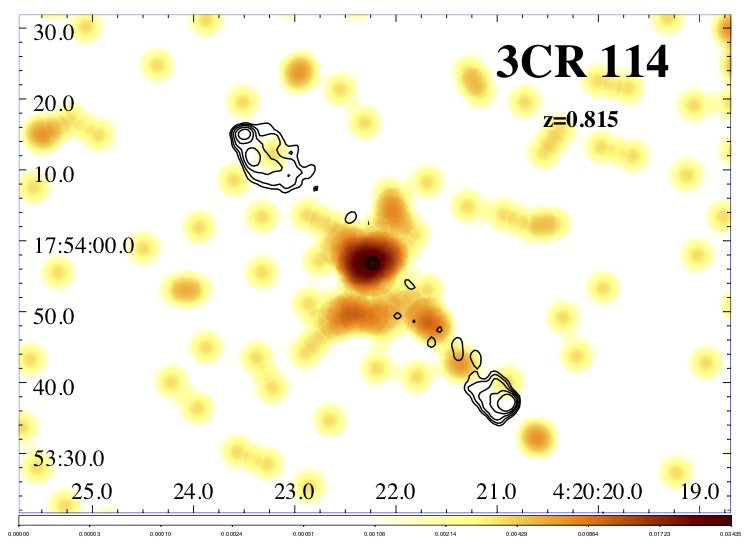}
\includegraphics[width=9.5cm]{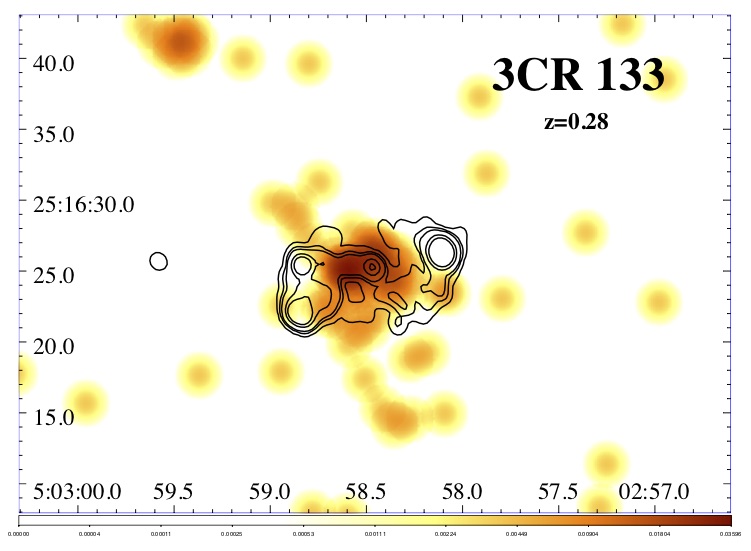}
\includegraphics[width=9.5cm]{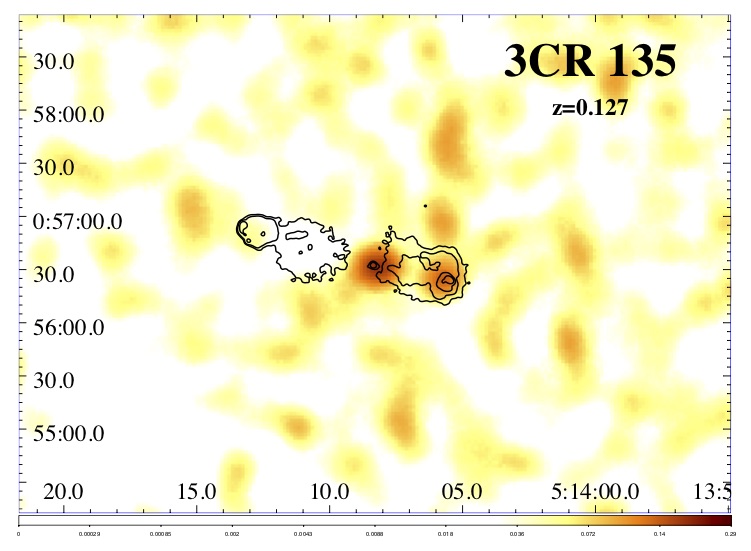}
\includegraphics[width=9.5cm]{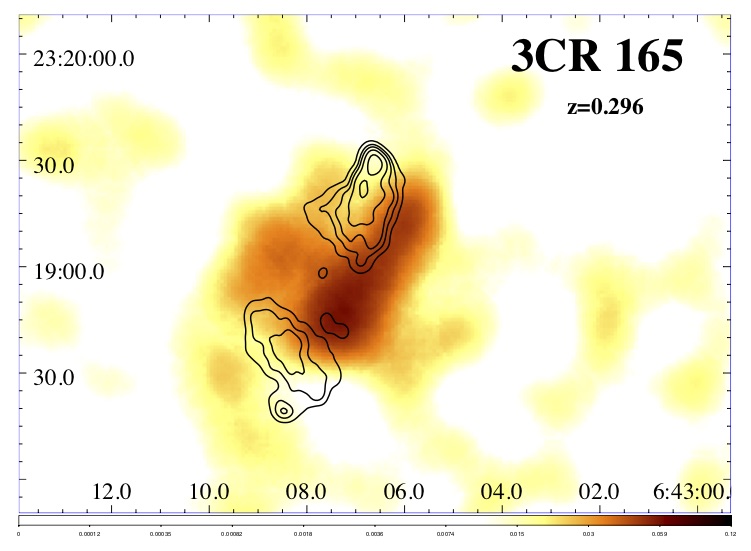}
\caption{(continued)}
\end{figure}

\addtocounter{figure}{-1}
\begin{figure}[h]
\includegraphics[width=9.5cm]{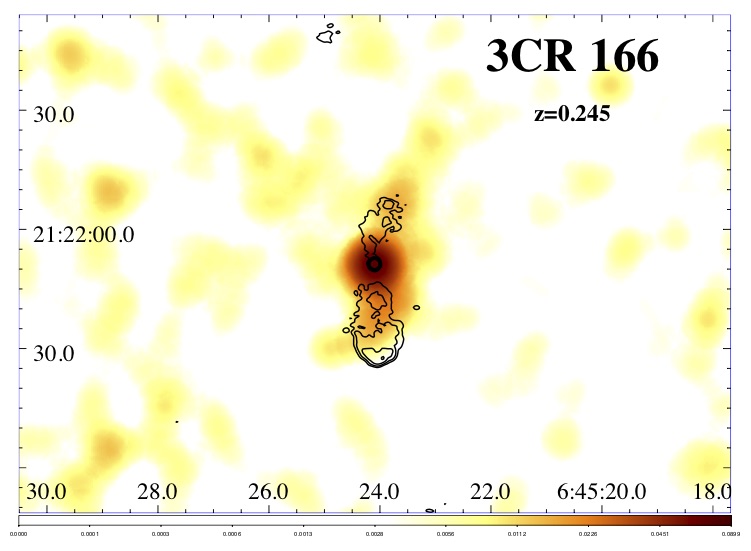}
\includegraphics[width=9.5cm]{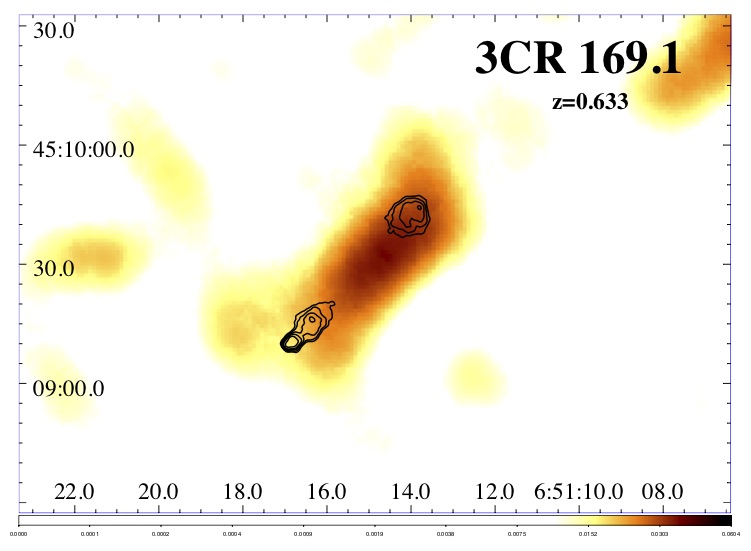}
\includegraphics[width=9.5cm]{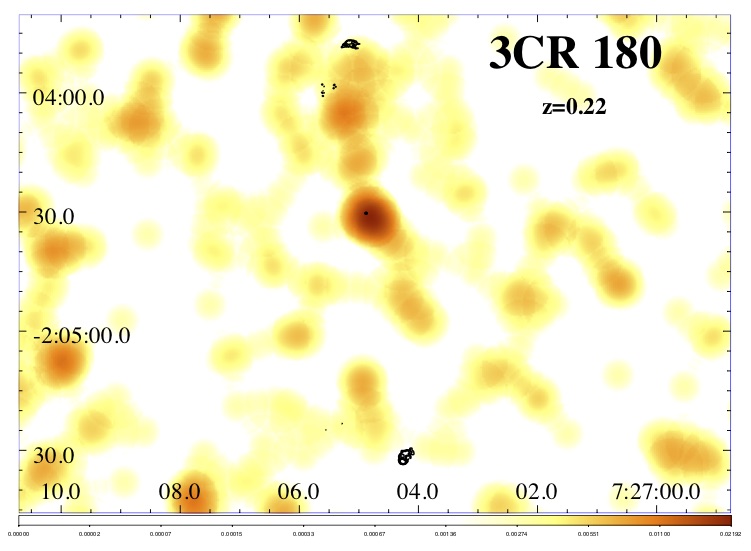}
\includegraphics[width=9.5cm]{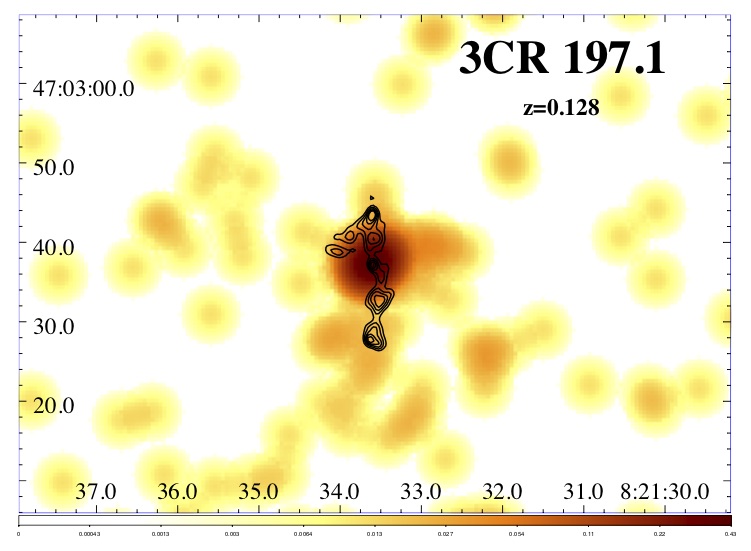}
\includegraphics[width=9.5cm]{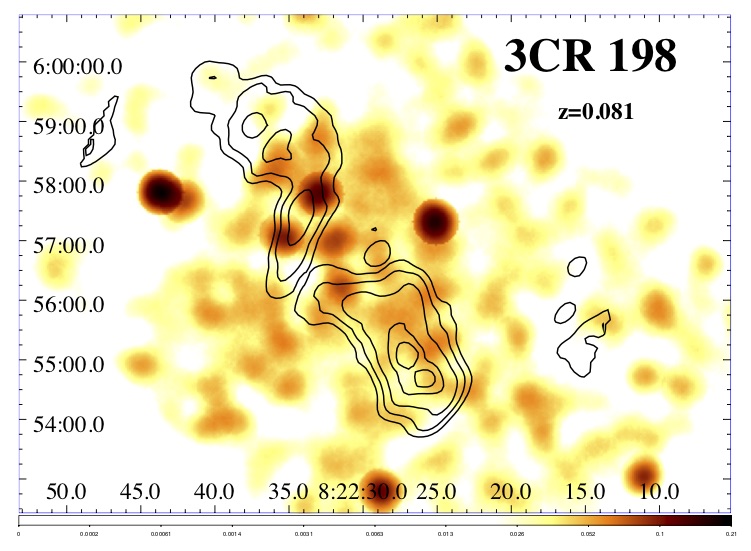}
\includegraphics[width=9.5cm]{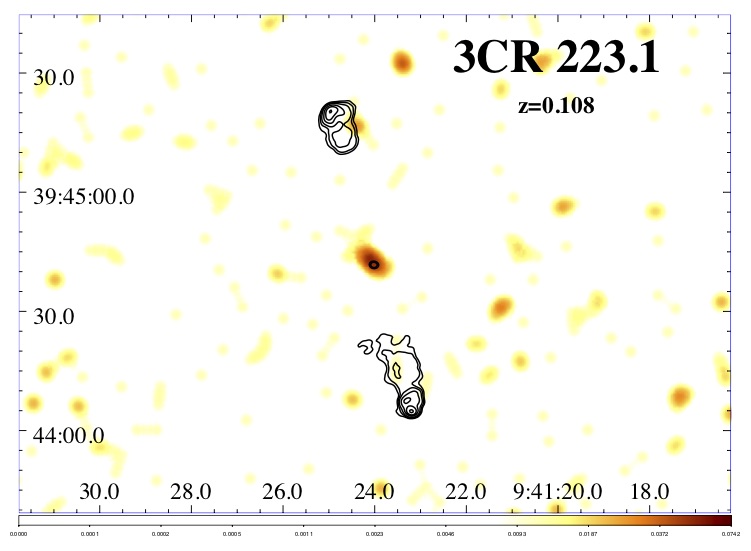}
\caption{(continued)}
\end{figure}

\addtocounter{figure}{-1}
\begin{figure}[h]
\includegraphics[width=9.5cm]{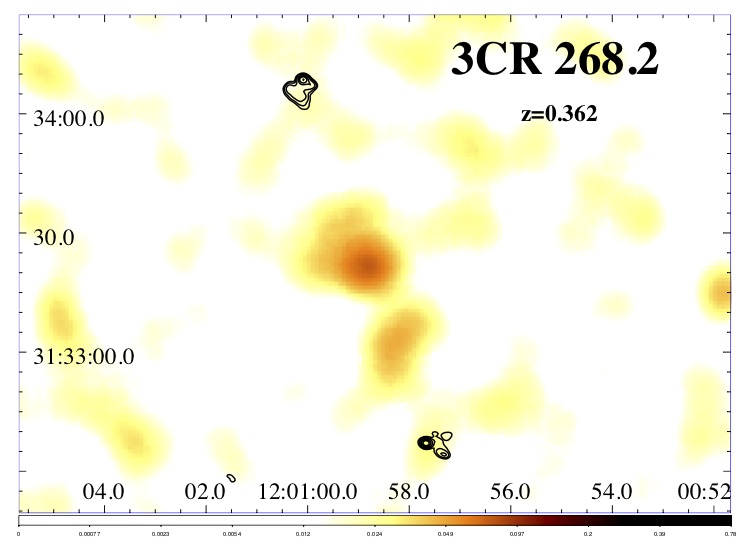}
\includegraphics[width=9.5cm]{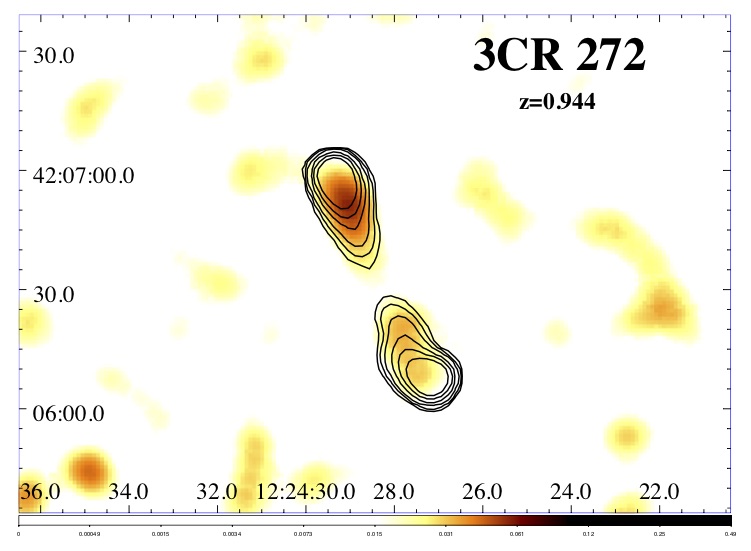}
\includegraphics[width=9.5cm]{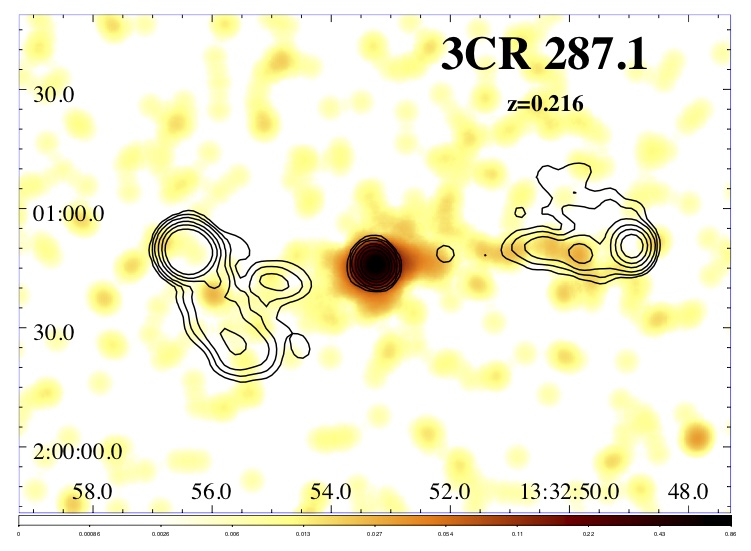}
\includegraphics[width=9.5cm]{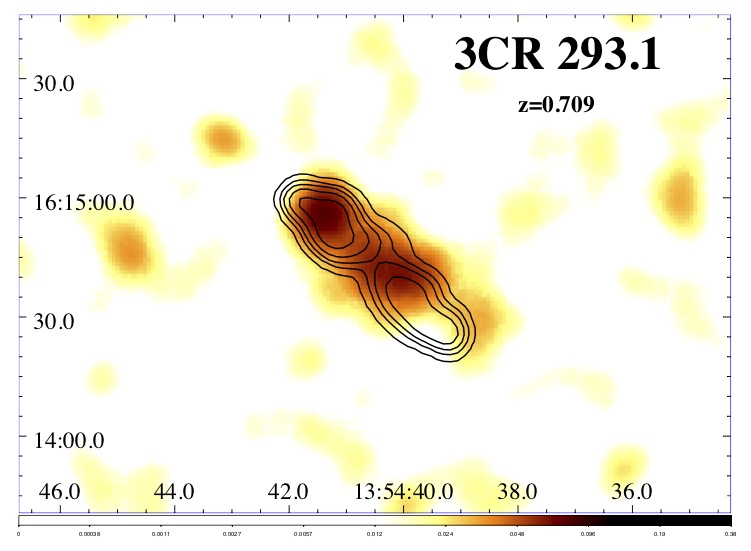}
\includegraphics[width=9.5cm]{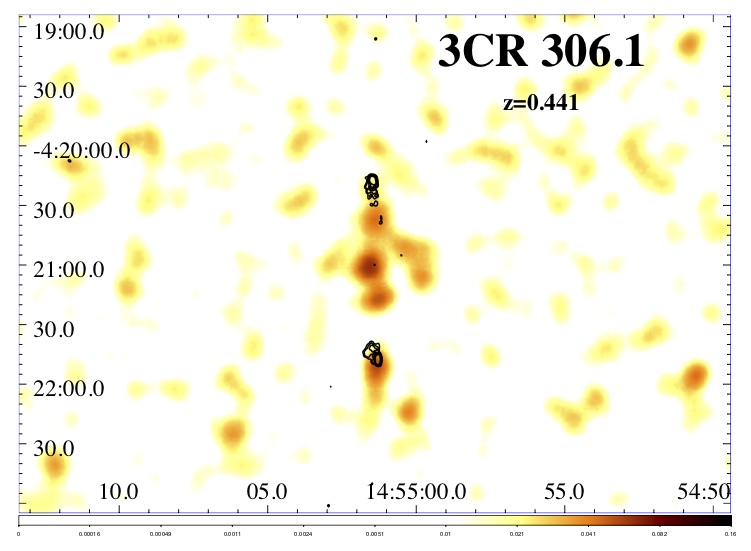}
\includegraphics[width=9.5cm]{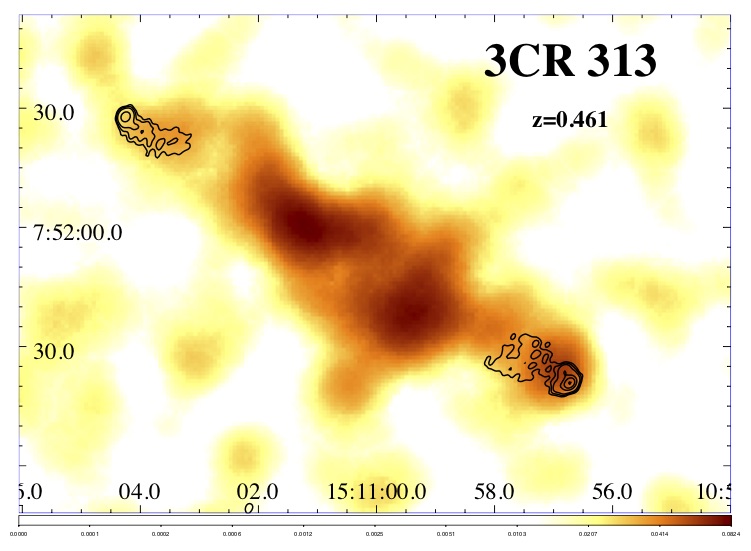}
\caption{(continued)}
\end{figure}

\addtocounter{figure}{-1}
\begin{figure}[h]
\includegraphics[width=9.5cm]{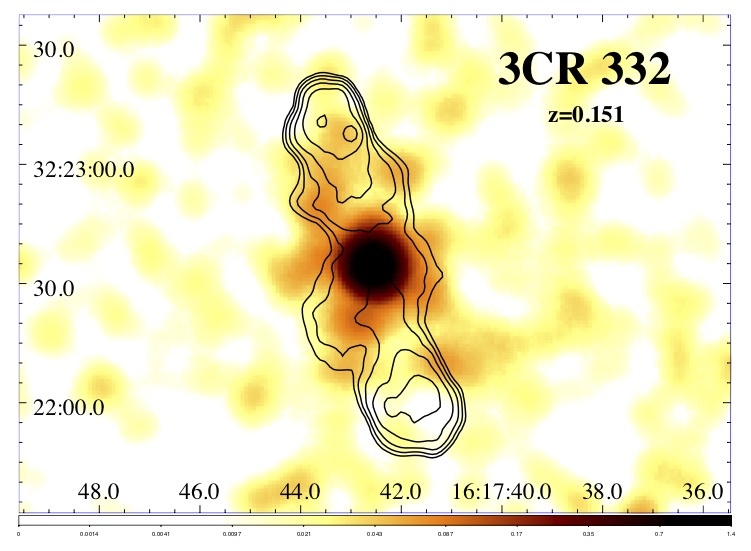}
\includegraphics[width=9.5cm]{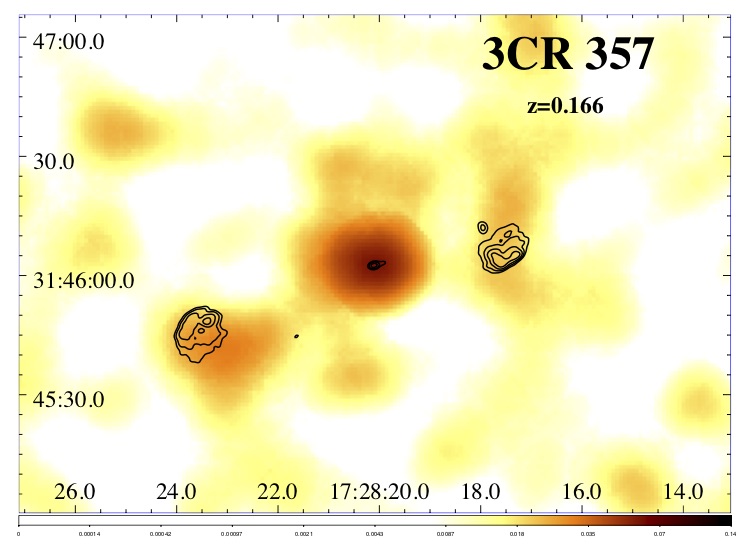}
\includegraphics[width=9.5cm]{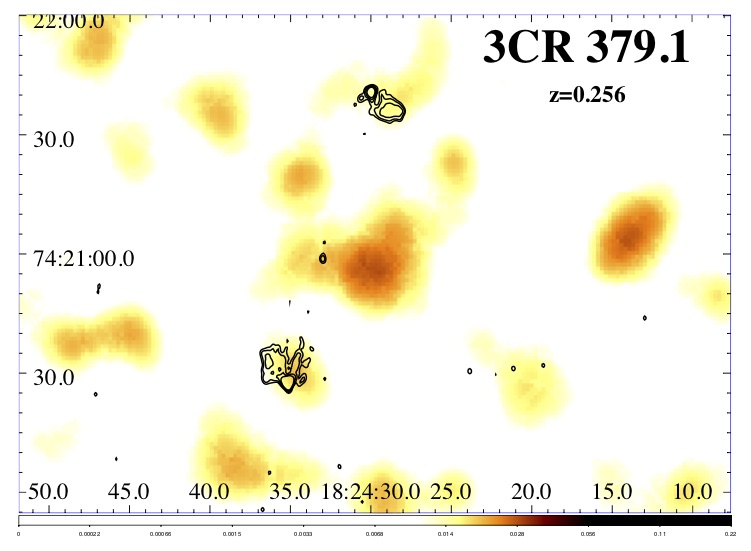}
\includegraphics[width=9.5cm]{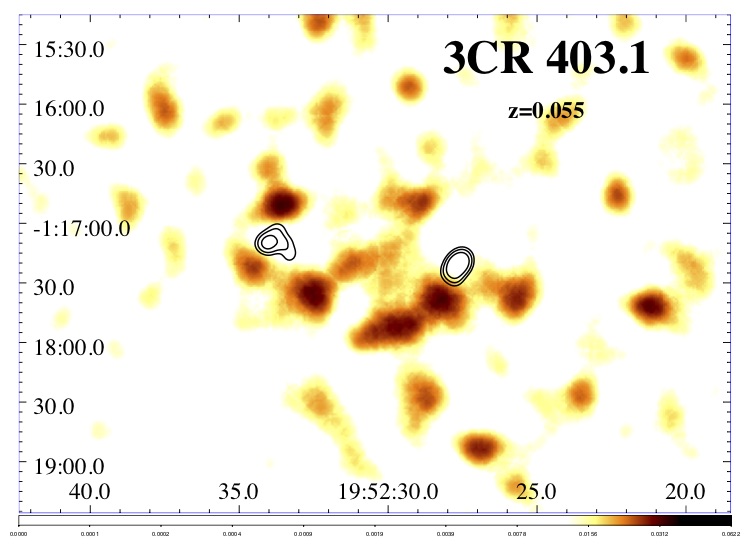}
\includegraphics[width=9.5cm]{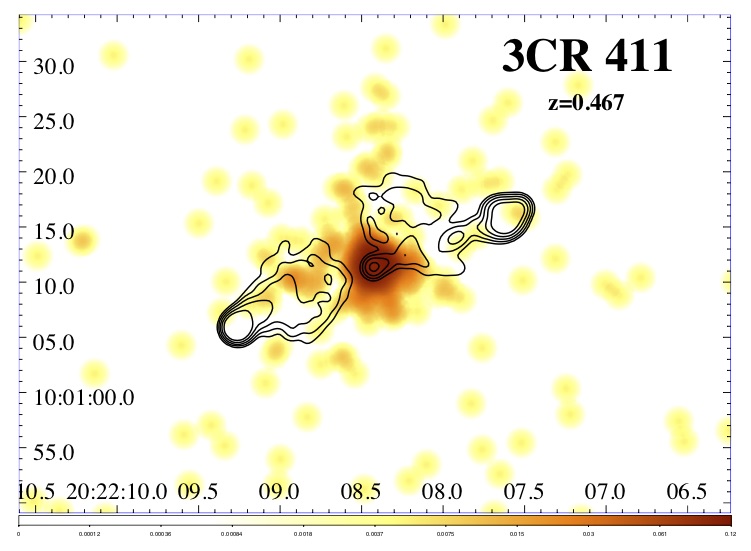}
\includegraphics[width=9.5cm]{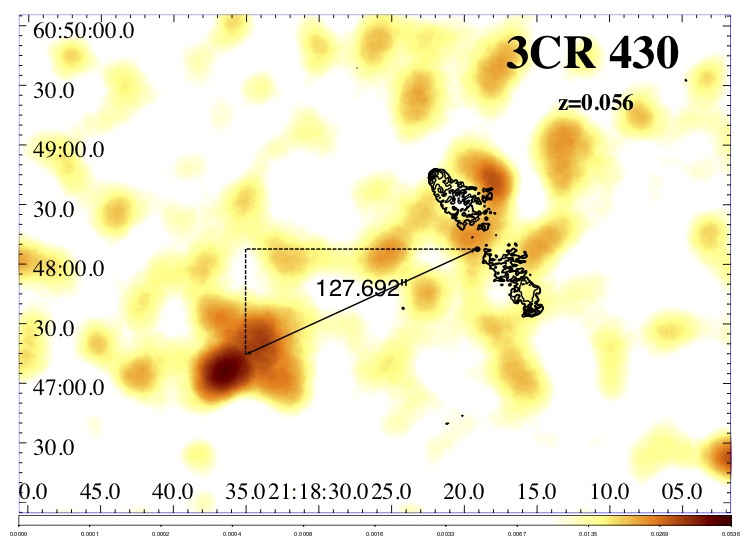}
\caption{(continued)}
\end{figure}

\addtocounter{figure}{-1}
\begin{figure}[h]
\includegraphics[width=9.5cm]{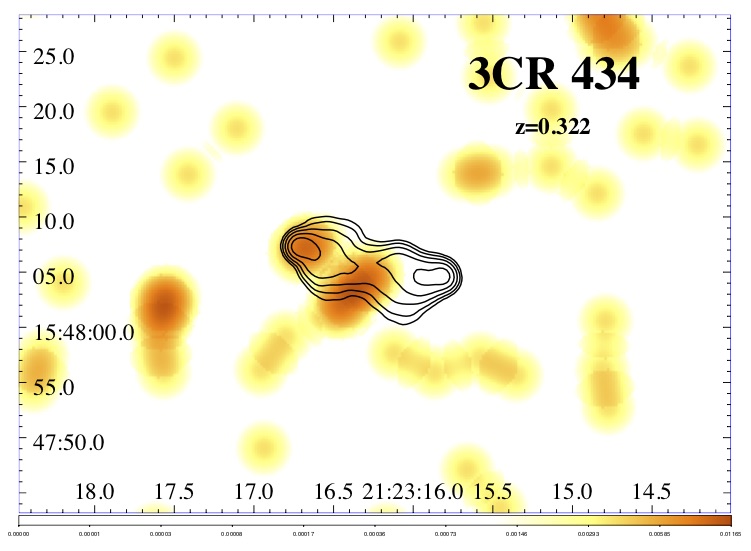}
\includegraphics[width=9.5cm]{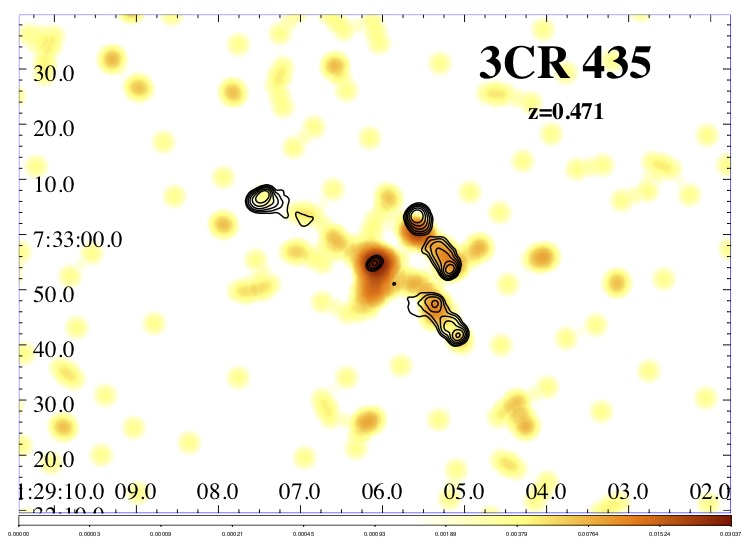}
\includegraphics[width=9.5cm]{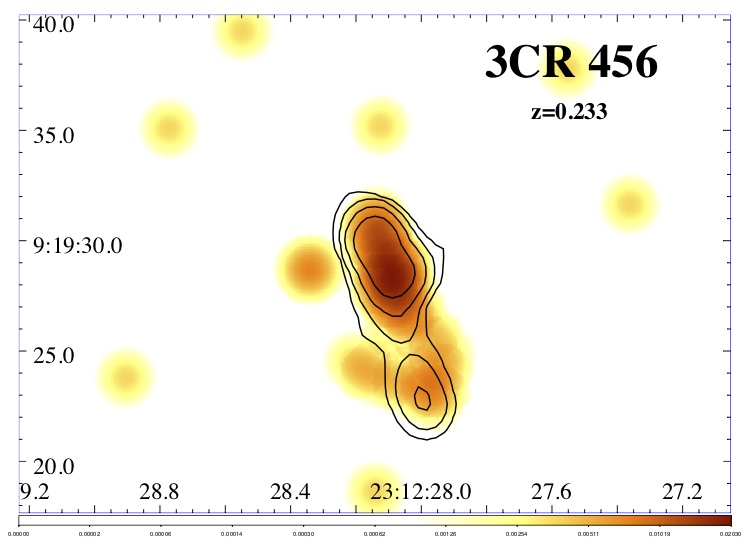}
\includegraphics[width=9.5cm]{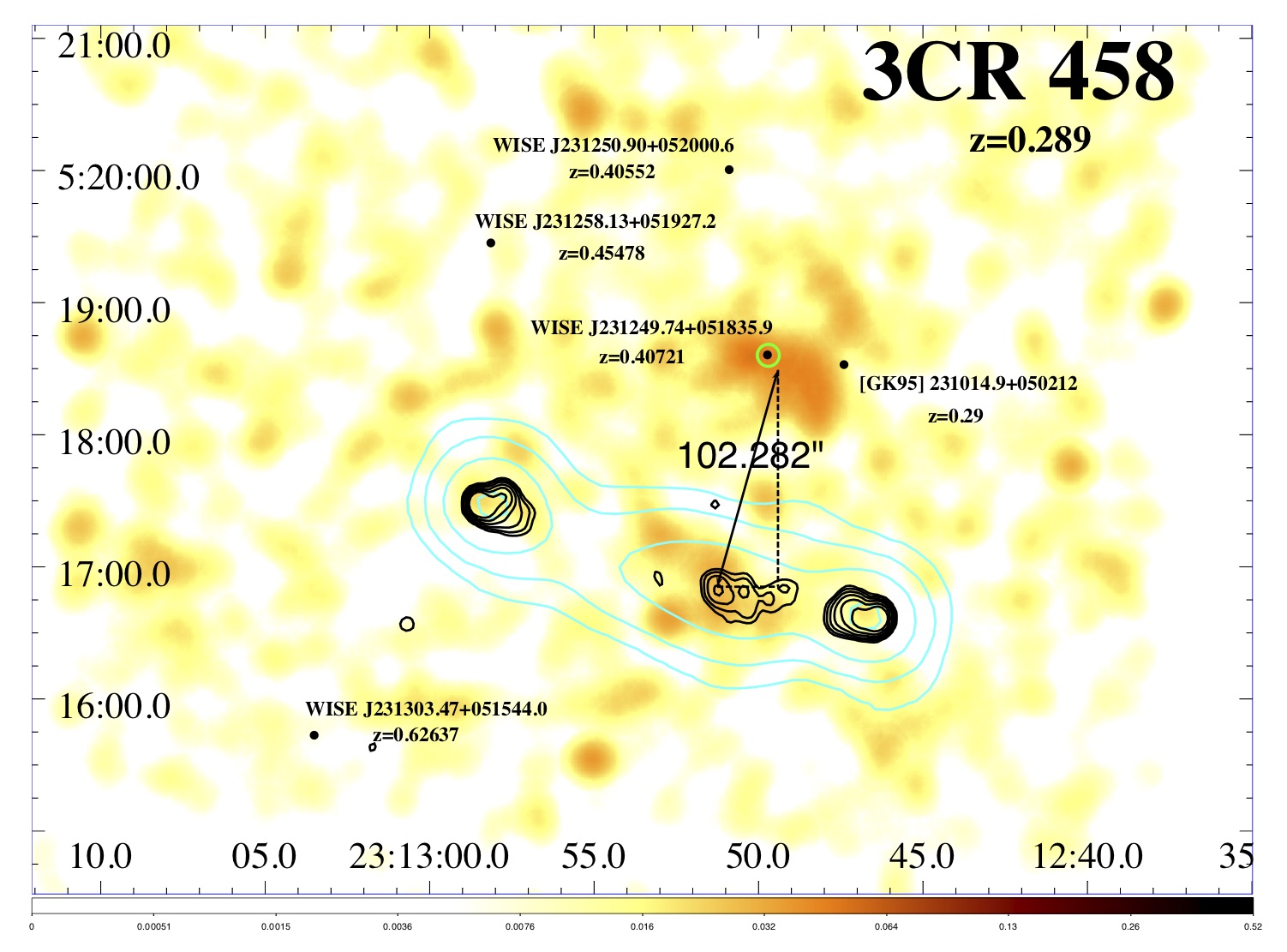}
\caption{(continued)}
\end{figure}

\section{A WISE serendipitous discovery}
\label{app:wise}
We cross-matched the radio position of all hotspots in our radio galaxy sample with the $AllWISE$ catalog aiming at finding their mid-IR counterparts. This was done while performing the test on the mid-IR counterparts of X-ray sources detected using the \textsc{wavdetect} algorithm (see \S~\ref{sec:reduction}).

We adopted a searching radius of 3.3 arcsec, chosen on the basis of our previous analyses (see e.g., \citealt{Dabrusco2013,Dabrusco2014,Massaro2014}), since the chance probability to have spurious cross-matches between radio sources with an average positional uncertainty of the order of $\sim$1 arcsec and their potential $WISE$ counterparts is less than $\sim$1\%, for angular separations below this value (see also \citealt{Massaro2011a, Dabrusco2012, Massaro2013a, Massaro2013b}, for additional information on the method to compute the chance probability).

\begin{figure}

\includegraphics[width=9.cm,height=5.cm]{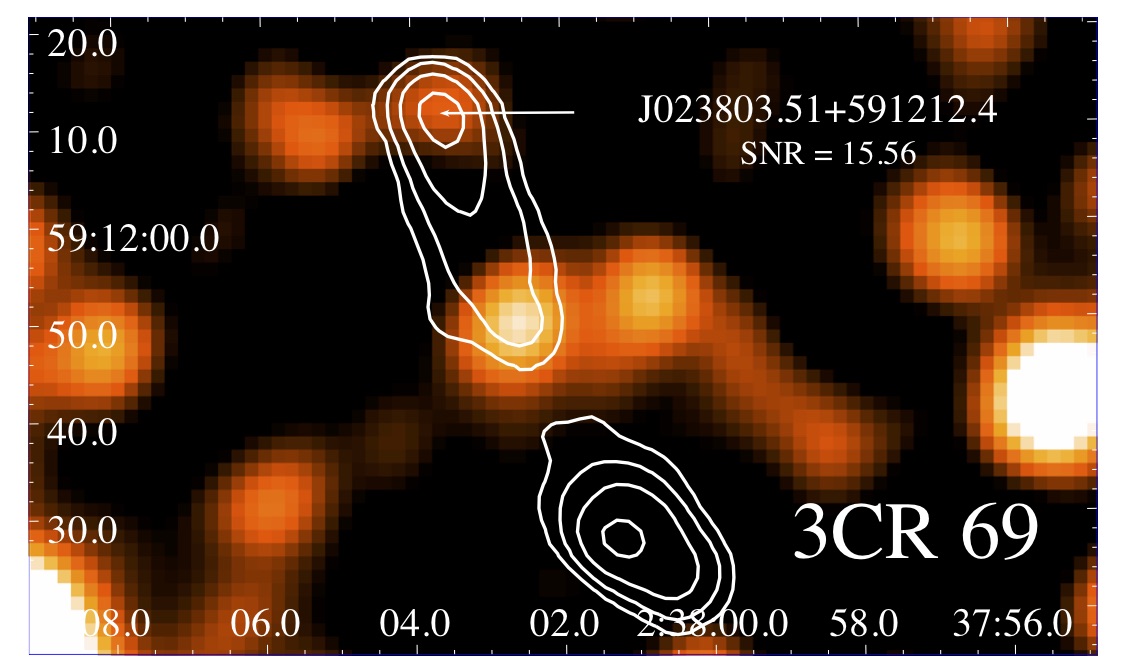}
\includegraphics[width=9.cm,height=5.cm]{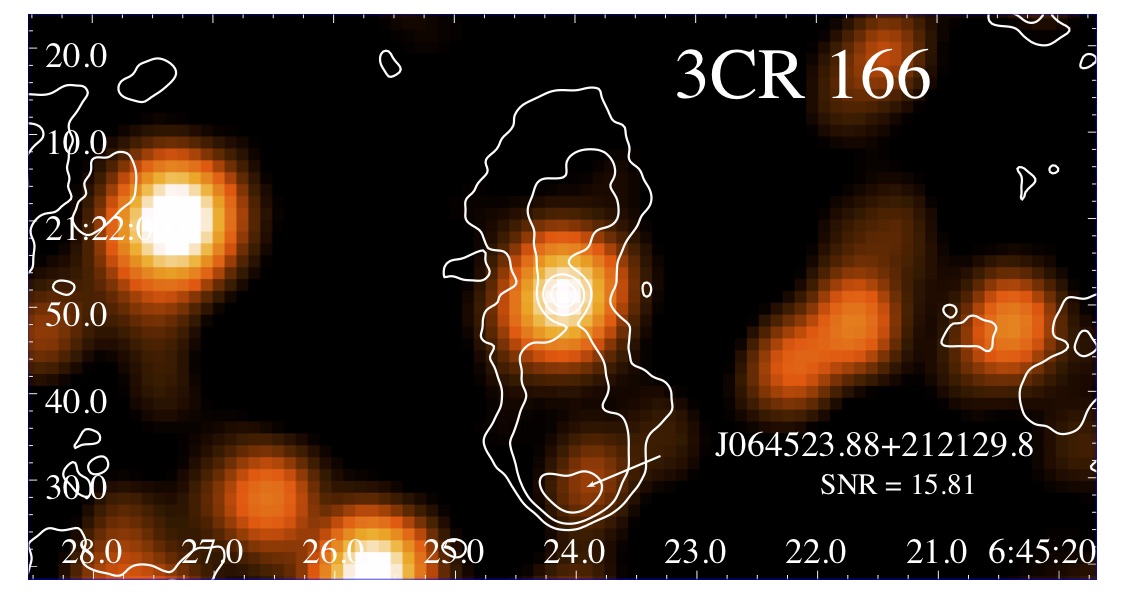}
\includegraphics[width=9.cm,height=5.cm]{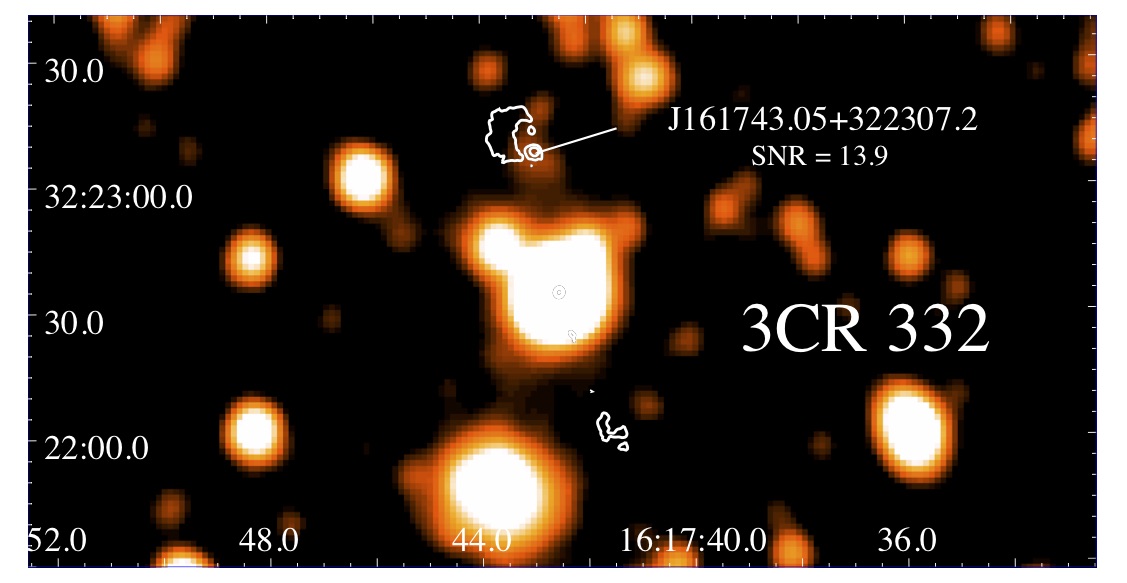}

\caption{Top left: 3.4 $\mu$m $WISE$ image of 3CR\,69 with 4.8 GHz VLA contours overlaid. Radio contours start at 2 mJy/beam and increase by a factor of four. VLA radio map has a beam size of 5 arcsec. Top rigth: 3.4 $\mu$m $WISE$ image of 3CR\,166 with 1.4 GHz VLA contours overlaid. Radio contours start at 3 mJy/beam and increase by a factor of four. VLA radio map has a beam size of 1.37 arcsec. Bottom: 3.4 $\mu$m $WISE$ image of 3CR\,332 with 1.4 GHz VLA contours overlaid. Radio contours start at 1 mJy/beam and increase by a factor of four. VLA radio map has a beam size of 4.4 arcsec.}
\label{fig:wisehot}
\end{figure}

We found three hotspots, namely n22 in 3CR\,69 (with $W1$ magnitude $15.553 \pm 0.05$), s24 in 3CR\,166 ($W1=15.805 \pm 0.057$) and n34 in 3CR\,332 ($W1=14.214 \pm 0.033$), having a clear mid-IR counterpart, all detected at 3.4 $\mu$m and 4.6 $\mu$m in $WISE$ images, as shown in Fig. \ref{fig:wisehot}, respectively. The first two sources have a signal-to-noise ratio (SNR) greater than 15 while the third hotspot in 3CR\,322 a SNR of 13.9, all measured at 3.4 $\mu$m. These SNRs are generally lower than those for example found for mid-IR counterparts of blazars with radio flux density of the order of a few mJy (see e.g. catalogs in \citealt{Dabrusco2014,Dabrusco2019,Demenezes2019}). Thus, although these SNRs are well above the $AllWISE$ catalog threshold\footnote{https://irsa.ipac.caltech.edu/Missions/wise.html,} we also carry out the following test to get a better idea of their significance. Since these hotspots have flux densities of the order of 1 mJy, we cross-matched all FIRST sources of 1 mJy (a total of 2907) with the $AllWISE$ catalog within 3.3 arcsec. We found a total of 1407 cross-matches, all unique. Then we found that 124 out of 1407 (i.e., $\sim$8\%) have SNR between 13 and 16 and $\sim$1/3 of all cross-matches are detected with SNR greater than 15, all computed at 3.4 $\mu$m that is the band with the largest number of constraints to claim a mid-IR detection. Thus we are confident that our claim on the discovery of mid-IR counterparts of three hotspots is reliable. 

It is worth noting that mid-IR counterparts of hotspots and lobes of radio galaxies have already been observed in the past, especially for 3CR sources (\citealt{Werner2012} and references therein). Our serendipitous discovery proves that $WISE$ is also able to detect them with a shallow mid-IR survey. Additionally, one hotspot in our sample was previously detected in the literature in the optical and near-IR, namely the eastern hotspot of 3CR\,133 (\citealt{Madrid2006}). Lastly, \citet{deKoff1996} proposed a possible candidate for the optical counterpart of the western hotspot of 3CR\,287.1.

\end{document}